\documentclass[a4paper,11pt]{article}
\pdfoutput=1 

\usepackage{jcappub} 

\usepackage[T1]{fontenc} 
\usepackage{footnote}

\usepackage{graphicx,ulem}
\usepackage{longtable}
\usepackage{float}
\usepackage{dcolumn}
\usepackage{graphics,epsfig}
\usepackage{amsmath,amssymb,latexsym,mathrsfs}
\usepackage{bm}
\usepackage{color}
\usepackage{color}
\usepackage{subfigure}
\usepackage{multirow}
\usepackage{amsfonts}

\newcommand{\rr}{\mathrm}

\newcommand{\neff}{N_{\rm eff}}
\newcommand{\lcdm}{\Lambda\mathrm{CDM}}

\newcommand{\zbeg}{z_{\rm beg}}
\newcommand{\zend}{z_{\rm end}}
\newcommand{\zreio}{z_{\rm reio}}
\newcommand{\dzreio}{\Delta z_{\rm reio}}

\begin{document}

\title{Exploring Cosmic Origins with CORE:  Cosmological Parameters}

\author[1,2]{Eleonora Di Valentino,}
\author[3]{Thejs Brinckmann,}
\author[4]{Martina Gerbino,}
\author[3,5]{Vivian Poulin,}
\author[1]{François R. Bouchet,}
\author[3]{Julien Lesgourgues,}
\author[6]{Alessandro Melchiorri,}
\author[7]{Jens Chluba,}
\author[3]{S\'ebastien Clesse,}
\author[8]{Jacques Delabrouille,}
% ...

\affiliation[1]{Institut d'Astrophysique de Paris (UMR7095: CNRS \& UPMC-Sorbonne Universités), F-75014, Paris, France}
\affiliation[2]{Sorbonne Universit\'es, Institut Lagrange de Paris (ILP), F-75014, Paris, France}
\affiliation[3]{Institute for Theoretical Particle Physics and Cosmology (TTK), RWTH Aachen University, D-52056 Aachen, Germany.}
\affiliation[4]{The Oskar Klein Centre for Cosmoparticle Physics, Department of Physics, Stockholm University, AlbaNova, SE-106 91 Stockholm, Sweden}
\affiliation[5]{LAPTh, Universit\'e Savoie Mont Blanc \& CNRS, BP 110, F-74941 Annecy-le-Vieux Cedex, France.}
\affiliation[6]{Physics Department and Sezione INFN, University of Rome La Sapienza, Ple Aldo Moro 2, 00185, Rome, Italy}
\affiliation[7]{Jodrell Bank Centre for Astrophysics, School of Physics and Astronomy, The University of Manchester, Oxford Road, Manchester, M13 9PL, U.K.}
\affiliation[8]{APC, AstroParticule et Cosmologie, Universit\'e Paris Diderot, CNRS/IN2P3, CEA/Irfu, 
Observatoire de Paris Sorbonne Paris Cit\'e, 10, rue Alice Domon et
Leonie Duquet, 75205 Paris Cedex 13, France}
% ...

\emailAdd{bouchet@iap.fr,lesgourg@physik.rwth-aachen.de, alessandro.melchiorri@roma1.infn.it}

\abstract{
We forecast the main cosmological parameter constraints achievable with the CORE space mission which is dedicated to mapping the polarisation of the Cosmic Microwave Background (CMB). CORE was recently submitted in response to ESA's fifth call for medium-sized mission proposals (M5). Here we report the results from our pre-submission study of the impact of various instrumental options, in particular the telescope size and sensitivity level, and review the great, transformative potential of the mission as proposed. Specifically, we assess the impact on a broad range of fundamental parameters of our Universe as a function of the expected CMB characteristics, with other papers in the series focusing on controlling astrophysical and instrumental residual systematics. In this paper, we assume that only a few central CORE frequency channels are usable for our purpose, all others being devoted to the cleaning of astrophysical contaminants. On the theoretical side, we assume $\lcdm$ as our general framework and quantify the improvement provided by CORE over the current constraints from the Planck 2015 release. We also study the joint sensitivity of CORE and of future Baryon Acoustic Oscillation and Large Scale Structure experiments like DESI and Euclid. Specific constraints on the physics of inflation are presented in another paper of the series. In addition to the six parameters of the base $\lcdm$, which describe the matter content of a spatially flat universe with adiabatic and scalar primordial fluctuations from inflation, we derive the precision achievable on parameters like those describing curvature, neutrino physics, extra light relics, primordial helium abundance, dark matter annihilation, recombination physics, variation of fundamental constants, dark energy, modified gravity, reionization and cosmic birefringence. In addition to assessing the improvement on the precision of individual parameters, we also forecast the post-CORE overall reduction of the allowed parameter space with figures of merit for various models increasing by as much as $\sim 10^7$ as compared to Planck 2015, and $10^5$ with respect to Planck 2015 + future BAO measurements.}

\maketitle

\flushbottom


\section{Introduction}

In the quarter century since their first firm detection by the COBE satellite \cite{Smoot:1992td}, Cosmic Microwave Background (CMB) anisotropies have revolutionized the field of cosmology with an enormous impact on several branches of astrophysics and particle physics.  From observations made by ground-based experiments such as TOCO \cite{Miller:1999qz},  DASI \cite{Halverson:2001yy} and ACBAR \cite{Kuo:2002ua}, balloon-borne experiments like BOOMERanG \cite{deBernardis:2000sbo1,deBernardis:2000sbo2}, MAXIMA \cite{Hanany:2000qf} and Archeops \cite{Benoit:2002mm}, and satellite experiments such as COBE, WMAP \cite{wmap1,wmap2} and, more recently, Planck \cite{planck2013,planck2015}, a cosmological "concordance" model has emerged, in which the need for new physics beyond the standard model of particles is blatantly evident.
The impressive  experimental progress in detector sensitivity and observational techniques, combined with the accuracy of linear perturbation theory, have clearly identified the CMB as the "sweet spot" from which to accurately constrain cosmological parameters and fundamental physics. Such a fact calls for new and significantly improved measurements of CMB anisotropies, to continue mining their scientific content.

In particular, observations of the CMB angular power spectrum are not only in impressive agreement with the expectations of
the so-called $\lcdm$ model, based on cold dark matter (CDM hereafter), inflation and a cosmological constant, but they now also constrain several parameters with exquisite precision. For example, the cold dark matter density is now constrained to $1.25$\% accuracy using recent Planck measurements, naively yielding an evidence for CDM at about $\sim 80$ standard deviations (see \cite{planck2015}). Cosmology is indeed extremely powerful in identifying CDM, since on cosmological scales the gravitational effect of CDM are cleaner and can be precisely discriminated from those of standard baryonic matter. In this respect, no other cosmological observable aside from the CMB could show, if considered alone, the need for CDM to such a level of significance. Moreover, the cosmological signatures of CDM rely mainly on gravity, 
while astrophysical searches of DM annihilating or decaying into standard model particles depend on the strength of the interaction.
Similarly, a possible signal in underground laboratory experiments depends on the coupling between CDM particles and ordinary matter (nuclei and electrons).
It is possible to construct CDM models that could interact essentially just through gravity, and the current lack of detection of CDM in underground and astrophysics experiments is leaving this possibility open. If this is the case, structure formation on cosmological scales could result in the best observatory we have where to study the CDM properties, and a further improvement from future CMB measurements will clearly play a crucial and complementary role. The CMB even allows to put bounds on the stability and decay time of CDM through purely gravitational effects \cite{Ichiki:2004vi,Lattanzi:2013uza,Audren:2014bca,Poulin:2016nat}.

CMB measurements also provide an extremely stringent constraint on standard baryonic matter. The recent results from Planck constrain the baryonic content with a $0.7 \%$ accuracy, nearly a factor $2$ better than the present constraints derived from primordial deuterium measurements \cite{cooke2016}, obtained assuming standard
Big Bang Nucleosynthesis. In this respect, the experimental uncertainties on nuclear rates like $d(p,\gamma)^3$He that enter in BBN computations are starting to be relevant for accurate estimates of the baryon content from measurements of primordial nuclides. A combination of CMB and primordial deuterium measurements is starting to produce independent bounds on these quantities (see, e.g. \cite{nollett,divalentinobbn}). As a matter of fact, a further improvement in the determination of the baryon density is mainly expected from future CMB anisotropy measurements and could help not only in testing the BBN scenario but also in providing independent constraints on nuclear physics.

In this direction, it is also important to stress that CMB measurements are already so accurate that they are able to constrain some aspects of the physics of hydrogen recombination, such as the $2s-1s$ two photon decay channel transition rate, with a precision higher than current experimental estimates  \cite{planck2015}. New CMB measurements can, therefore, considerably improve our knowledge of the physics of recombination. Since primordial Helium also recombines, albeit at higher redshifts, the CMB is sensitive to the primordial $^4$He abundance which lowers the free electron number density at recombination. The Planck mission already detected the presence of primordial Helium at the level of $\sim 10$ standard deviation  \cite{planck2015}. Next-generation CMB experiments could significantly improve this measurement, reaching a precision comparable with current direct measurements from extragalactic HII regions that may, however, still be plagued by systematics \cite{helium1,helium2}. Constraining the physics of recombination will also bound the possible presence of extra ionizing photons that could be produced by dark matter self annihilation or decay (see e.g. \cite{2004PhRvD..70d3502C,2009PhRvD..80b3505G,dmannihilation3,dmannihilation2,dmannihilation1}). The Planck 2015 data release already produced significant constraints on dark matter annihilation at recombination that are fully complementary to those derived from laboratory and astrophysical experiments \cite{planck2015}.

The CMB is also a powerful probe of the density and properties of "light" particles, i.e. particles with masses below $\sim 1$~eV that become non-relativistic between 
recombination (at redshift $z\sim 1100$, when the primary CMB anisotropies are visible) and today. Such particles may affect primary and secondary CMB anisotropies, as well as structure formation. In particular, this can change the amplitude of gravitational lensing produced by the intervening matter fluctuations (\cite{lensingreview}) and leave clear signatures in the CMB power spectra. Neutrinos are the most natural candidate to leave such an imprint (see e.g.  \cite{ Lesgourgues:1519137,neutrinoreview}).  From neutrino oscillation experiments we indeed know that neutrinos are massive and that their total mass summed over the three eigenstates should be {\it larger} than $M_\nu >60$~meV in the case of a normal hierarchy and of  $M_\nu >100$~meV in the case of an inverted hierarchy (see e.g. \cite{neutrinos1,neutrinos2,neutrinos3} for recent reviews of the current data). The most recent constraints from Planck measurements (temperature, polarization and CMB lensing) bound the total mass to $M_\nu < 140$~meV \cite{planck2016reio} at $95 \%$ c.l. Clearly, an improvement of the constraint towards a sensitivity of $\sigma (M_\nu) \sim 30$~meV will provide a {\it guaranteed discovery} for the neutrino absolute mass scale and for the neutrino mass hierarchy (see e.g. \cite{neutrinocosmo1,neutrinocosmo2,neutrinocosmo3,neutrinocosmo4}). 
Neutrinos are firmly established in the standard model of particle physics and a non-detection of the neutrino mass would cast serious doubts on the $\lcdm$ model, opening the window to new physics in the dark sector, such as, for instance, interactions between neutrinos and new light particles \cite{neutrinoless}. On the other hand, several extensions of the standard model of particle physics feature light relic particles that could produce effects similar to massive neutrinos, and might be detected or strongly constrained by future CMB measurements. Thermal light axions (see e.g. \cite{axions1,axions2,axions3}), for example, can produce very similar effects. Axions change the growth of structure formation after decoupling and increase the energy density in relativistic particles at early times\footnote{The effective neutrino number $N_\mathrm{eff}$ is normally defined at times such that all ``light'' particles (neutrinos, axions, etc.) are still ultra relativistic.},  parametrized by the quantity $N_\mathrm{eff}$. Models of thermal axions will be difficult to accommodate with a value of $N_\mathrm{eff}<3.25$, and a CMB experiment with a sensitivity of $\Delta N_\mathrm{eff}=0.04$ could significantly rule out or confirm their existence. Other possible candidates are light sterile neutrinos and asymmetric dark matter (see e.g. \cite{sterile0,sterile1,sterile2} and \cite{asydm}). More generally, a sensitivity to $\Delta N_\mathrm{eff}=0.04$ could rule out the presence of any thermally-decoupled Goldstone boson that decoupled after the QCD phase transition 
(see e.g. \cite{baumann}). The same sensitivity would also probe non-standard neutrino decoupling (see e.g. \cite{neutrinonon}) and the possibility of a low reheating temperature of the order of ${\cal O}$(MeV)~\cite{treheating}.

In combination with galaxy clustering and type Ia luminosity distances,
CMB measurements from Planck have also provided the tightest constraints on the dark energy equation of state
$w$ \cite{planck2015}. In particular, the current tension between the Planck value and the HST value of the Hubble constant from Riess et al. 2016 \cite{riess2016} could be resolved by invoking an equation of state $w<-1$ \cite{dms}. Planck alone is currently unable to constrain the equation of state $w$ and the Hubble constant $H_0$ independently, due to a "geometrical degeneracy" between the two parameters. An improved measurement of the CMB anisotropies could break this degeneracy, produce two independent constraints on $w$ and $H_0$, and possibly resolve the current tension on the value of the Hubble constant. Moreover, modified gravity models have been proposed that could provide an explanation to the current accelerated expansion of our universe. The CMB can be sensitive to modifications of General Relativity through CMB lensing and the late Integrated Sachs-Wolfe (ISW) effect.  Current Planck measurements are compatible with certain types of departures from GR (and even prefer such models, albeit at small statistical significance, see \cite{planckmg}). Future CMB measurements are, therefore, extremely important in addressing this issue.

In order to further improve current measurements and provide deeper insight on the nature of dark matter and dark energy, a CMB satellite mission is clearly our ultimate goal.
This does, however, raise two fundamental questions. The first one is whether we really need to go to space and launch a new satellite, given that several other ground-based and balloon-borne experiments are under discussion or already under construction (see e.g. \cite{cmbs4}). In fifteen years it is certainly reasonable to assume that these experiments will collect excellent data that could, in principle, constrain cosmological parameters to similar precision. However, there is a fundamental aspect to consider: ground-based experiments have very limited frequency coverage and sample just a portion of the CMB sky. Contaminations from unknown foregrounds can be extremely dangerous for ground-based experiments, and can easily fool us. 
The claimed detection of a primordial Gravitational Waves (GW) background from the BICEP2 experiment \cite{Ade:2014xna} was latter ruled out by Planck observations at high frequencies, showing that contaminations from thermal dust in our Galaxy are far more severe than anticipated. This shows that unprecedented control of systematics and a wide frequency coverage are required, both of which call for a space-based mission.
In fact, future ground-based and satellite experiments must be seen as complementary: while ground-based experiments could provide a first hint for primordial GWs or neutrino masses, a satellite experiment could monitor the frequency dependence of the corresponding signal with the highest possible accuracy, and unambiguously confirm its primordial nature. 

Moreover, most of the future galaxy and cosmic shear surveys will sample several extended regions of the sky. Cross correlations with CMB data in the same sky area will offer a unique opportunity to test for systematics and new physics. It is, therefore, clear that a full sky survey from a satellite will offer much more complete, consistent and homogeneous information than several ground based observations of sky patches. Moreover, an accurate full-sky map of CMB polarisation on large angular scales can provide extremely strong constraints on the reionization optical depth, breaking degeneracies with other parameters such as neutrino masses.

The second fundamental question related to a new CMB satellite proposal arises from the fact that after increasing sensitivity and frequency coverage, one has to deal with the intrinsic limit of cosmic variance. At a certain point, no matter how much we increase the instrumental sensitivity, we reach the cosmic variance limit and stop improving the precision of parameter estimates.
This clearly opens the following issue: how close are we from cosmic variance with current CMB data? The Planck satellite measured the temperature angular spectrum up to the limit of cosmic variance in a wide range of angular scales; however, we are far from this limit when we consider
polarization spectra. But how much can current constraints improve with a future CMB satellite?

This is exactly the question we want to address in this paper. Assuming that foregrounds and systematics are under control, as should be the case with a well-designed satellite mission, we study by how much current constraints can improve, and find whether these improvements are worth the effort.
In this respect, we adopt the proposed baseline experimental configuration of the recent CORE satellite proposal \cite{coreprop}, submitted in response to ESA's call for a Medium-size mission opportunity (M5) as the successor of the Planck satellite. We refer to this experimental configuration (with a $\sim 120$~cm mirror) as CORE-M5 in all the next sections of this paper. We compare the results from CORE-M5 with other possible experimental configurations that range from a minimal and less expensive configuration (LiteCORE-80), with a $\sim 80$~cm mirror, aimed mainly at measuring large and mid-range angular scale polarization, up to a much more ambitious configuration (COrE+), with a $\sim 150$~cm mirror.
Given different experimental configurations, we forecast the achievable constraints assuming a large number of possible models, trying to review most of the science that could be extracted from the CORE data (with the exception of constraints on GWs and on inflation, addressed separately in a companion paper~\cite{ECOinflation}).
After a description of the analysis method in Section II, we start in Section III by providing the constraints achievable under the context of the $\lcdm$ concordance model. We then review the constraints that could be obtained on spatial curvature (Section IV), extra relativistic relics  (Section V), primordial nucleosynthesis and Helium abundance  (Section VI), neutrinos (Section VII), dark energy  (Section VIII), extended parameters spaces (Section IX), recombination  (Section X), Dark Matter annihilation and decay (Section XI), variation of fundamental constants  (Section XII), reionization  (Section XIII),  modified gravity  (Section XIV) and cosmic birefringence  (Section XV).

This work is part of a series of papers that present the science achievable by the CORE space mission and focuses on the constraints on cosmological parameters and fundamental physics that can be derived from future measurements of CMB temperature and polarization angular power spectra and lensing. The constraints on inflationary models are discussed in detail in a companion paper \cite{ECOinflation} while the cosmological constraints from complementary galaxy clusters data provided by CORE are presented in  \cite{ECOclusters}. The impact of CORE on the study of extragalactic sources is presented in \cite{ECOsources}.

\section{Experimental setup and fiducial model\label{sec:setup}}

We run Monte Carlo Markhov Chains (MCMC) forecasts for several possible experimental configurations of the CORE CMB satellite, 
following the commonly used approach described for example in \cite{lensextr} and \cite{shimon}. The method consists in generating mock data according to some fiducial model.
One then postulates a Gaussian likelihood with some instrumental noise level, and fits theoretical predictions for various cosmological models to the mock data, using standard Bayesian extraction techniques.
For the purpose of studying the sensitivity of the experiment to each cosmological parameter, as well as parameter degeneracies and possible parameter extraction biases, it is sufficient to set the mock data spectrum equal to the fiducial spectrum, instead of generating random realisations of the fiducial model.

Unless otherwise specified, we choose a fiducial minimal $\Lambda$CDM
model compatible with the recent Planck 2015 results \cite{planck2016reio}, i.e. with baryon density 
$\Omega_{b}h^2=0.02218$, cold dark matter density 
$\Omega_{c}h^2= 0.1205$, spectral index $n_s=0.9619$, and optical depth $\tau=0.0596$. 
This model also assumes a flat universe with a cosmological constant, 
$3$ neutrinos with effective number $N_\mathrm{eff}=3.046$ (with masses and hierarchy
that change according to the case under study), and standard recombination.

We use publicly available 
Boltzmann codes to calculate the corresponding theoretical angular power
spectra $C_{\ell}^{TT}$, $C_{\ell}^{TE}$, $C_{\ell}^{EE}$ 
for temperature, cross temperature-polarization and  polarization\footnote{Note that we don't consider 
the $B$ mode lensing channel.}. Depending on cases, we use either {\sc CAMB}\footnote{\tt
http://camb.info/}~\cite{Lewis:1999bs} or {\sc CLASS}\footnote{\tt http://class-code.net}~\cite{Lesgourgues:2011re,Blas:2011rf}, 
which are known to agree at a high degree of precision~\cite{Lesgourgues:2011rg,Lesgourgues:2011rh,Lesgourgues:2013bra}.

In the mock likelihoods, the variance of the ``observed'' multipoles $a_{lm}$'s is given by the sum of the fiducial $C_\ell$'s and of
an instrumental noise spectrum given by:

\begin{equation}
N_\ell = w^{-1}\exp(\ell(\ell+1)\theta^2/8\ln2),
\end{equation}

\noindent where $\theta$ is the FWHM of the beam assuming a Gaussian profile
and where $w^{-1}$ is the experimental power noise related to the 
detectors sensitivity $\sigma$ by $w^{-1} = (\theta\sigma)^2$.

As we discussed in the introduction, we adopt as main dataset the one presented for the recent CORE proposal,  
a complete survey of polarised sky emission in 19 frequency bands, with sensitivity and angular resolution requirements 
summarized in Table~\ref{tab:CORE-bands}.

{\small 
\begin{table}[h]
\begin{center}
{\footnotesize
\scalebox{0.93}{\begin{tabular}{|c|c|c|c|c|c|c|c|c|c|c|c|}
\hline 
channel & beam  &  $N_{\rm det}$  &  $\Delta T$  &  $\Delta P$  & $\Delta I$& $\Delta I$  & $\Delta y\times 10^6$  &  PS  ($5\sigma$) \\
GHz     &	arcmin &    &  $\mu K$.arcmin &  $\mu K$.arcmin &  $\mu K_{\rm RJ}$.arcmin  & kJy/sr.arcmin &  $y_{\rm SZ}$.arcmin & mJy  \\
\hline 
\hline 

60      & 17.87   & 48      & 7.5          & 10.6         & 6.81          & 0.75         & -1.5         & 5.0                     \\ 
70      & 15.39   & 48      & 7.1          & 10           & 6.23          & 0.94         & -1.5         & 5.4              \\ 
80          & 13.52   & 48      & 6.8          & 9.6          & 5.76          & 1.13         & -1.5         & 5.7           \\ 
90         & 12.08   & 78      & 5.1          & 7.3          & 4.19          & 1.04         & -1.2         & 4.7            \\ 
100       & 10.92   & 78      & 5.0            & 7.1          & 3.90          & 1.2          & -1.2         & 4.9          \\ 
115       & 9.56    & 76      & 5.0            & 7.0            & 3.58          & 1.45         & -1.3         & 5.2         \\ 
130      & 8.51    & 124     & 3.9          & 5.5          & 2.55          & 1.32         & -1.2         & 4.2         \\ 
145      & 7.68    & 144     & 3.6          & 5.1          & 2.16          & 1.39         & -1.3         & 4.0             \\ 
160      & 7.01    & 144     & 3.7          & 5.2          & 1.98            & 1.55         & -1.6         & 4.1            \\ 
175        & 6.45    & 160     & 3.6          & 5.1          & 1.72          & 1.62         & -2.1         & 3.9         \\ 
195     & 5.84    & 192     & 3.5          & 4.9          & 1.41          & 1.65         & -3.8         & 3.6           \\ 
220      & 5.23    & 192     & 3.8          & 5.4          & 1.24          & 1.85         & -           & 3.6         \\ 
255      & 4.57    & 128     & 5.6          & 7.9          & 1.30          & 2.59         & 3.5          & 4.4         \\ 
295       & 3.99    & 128     & 7.4          & 10.5         & 1.12          & 3.01         & 2.2          & 4.5           \\ 
340      & 3.49    & 128     & 11.1         & 15.7         & 1.01            & 3.57         & 2.0            & 4.7             \\ 
390       & 3.06    & 96      & 22.0           & 31.1         & 1.08          & 5.05         & 2.8          & 5.8          \\ 
450      & 2.65    & 96      & 45.9         & 64.9         & 1.04            & 6.48         & 4.3          & 6.5           \\ 
520      & 2.29    & 96      & 116.6        & 164.8        & 1.03            & 8.56         & 8.3          & 7.4      \\ 
600       & 1.98    & 96      & 358.3        & 506.7        & 1.03            & 11.4         & 20.0           & 8.5            \\ 
\hline
\hline
Array         &              & 2100 & 1.2  & 1.7 &  &  & 0.41 &   \\
\hline 
\end{tabular}}
}
\end{center}
\caption{\small Proposed CORE-M5 frequency channels. 
The sensitivity is calculated assuming $\Delta \nu/\nu=30\%$ bandwidth, 60\% optical efficiency, total noise of twice the expected photon noise from the sky and the optics of the instrument at 40K temperature. This configuration has 2100 detectors, about 45\% of which are located in CMB channels between 130 and 220\,GHz. Those six CMB channels yield an aggregated CMB sensitivity of $2\,\mu$K.arcmin ($1.7\,\mu$K.arcmin for the full array). 
}
\label{tab:CORE-bands}
\end{table}
}

Obviously, data from low (60-115\,GHz) and high frequencies (255-600\,GHz) channels will be mainly used for monitoring foreground
contaminations (and deliver rich related science). In our forecasts we therefore use only the six channels in the frequency range of $130-220$\,GHZ.
As stated in the introduction we will refer to this experimental configuration as {\bf CORE-M5}.

\begin{figure}
	\centering
		\includegraphics[width=5.2cm]{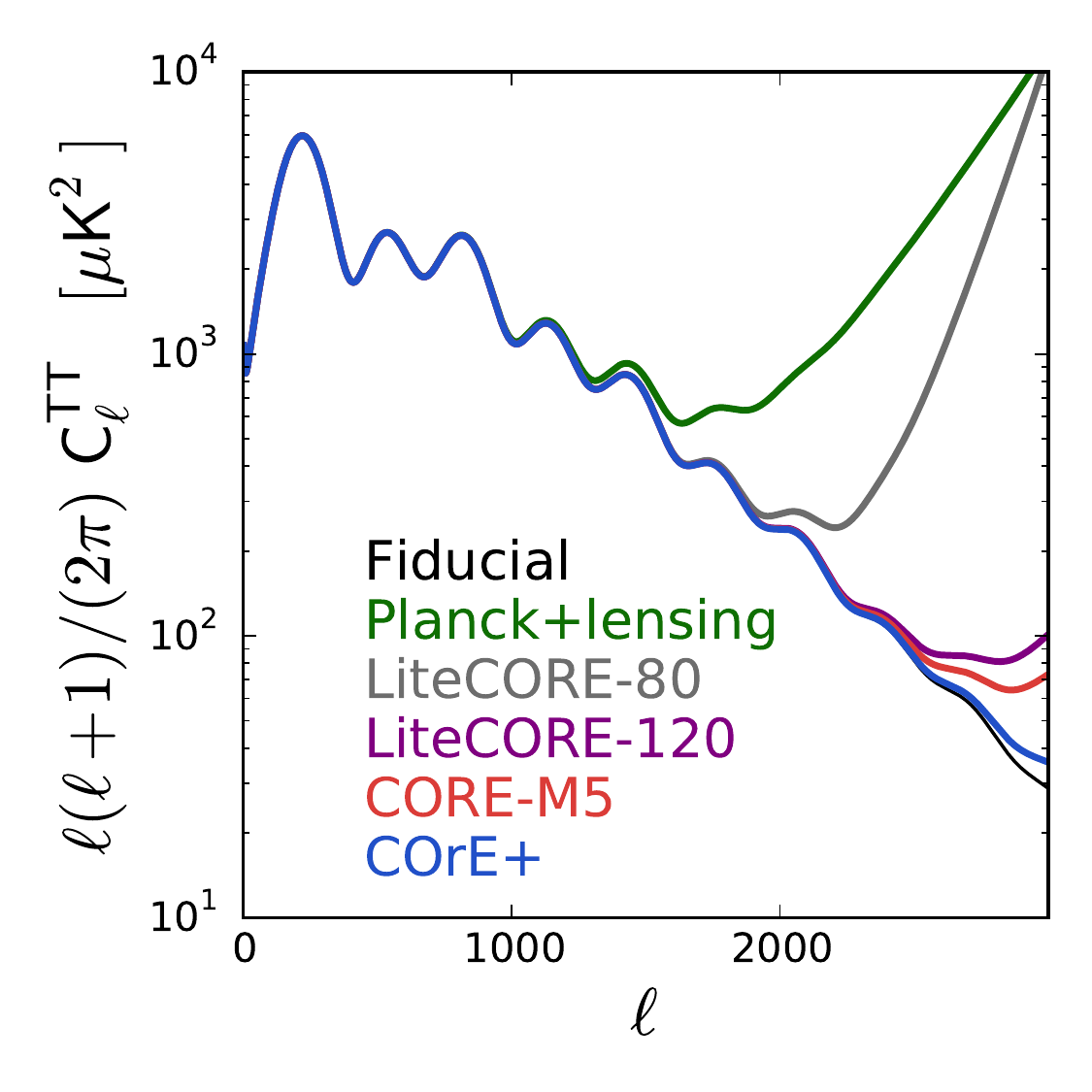} 
		\hspace{-0.2cm}\includegraphics[width=5.2cm]{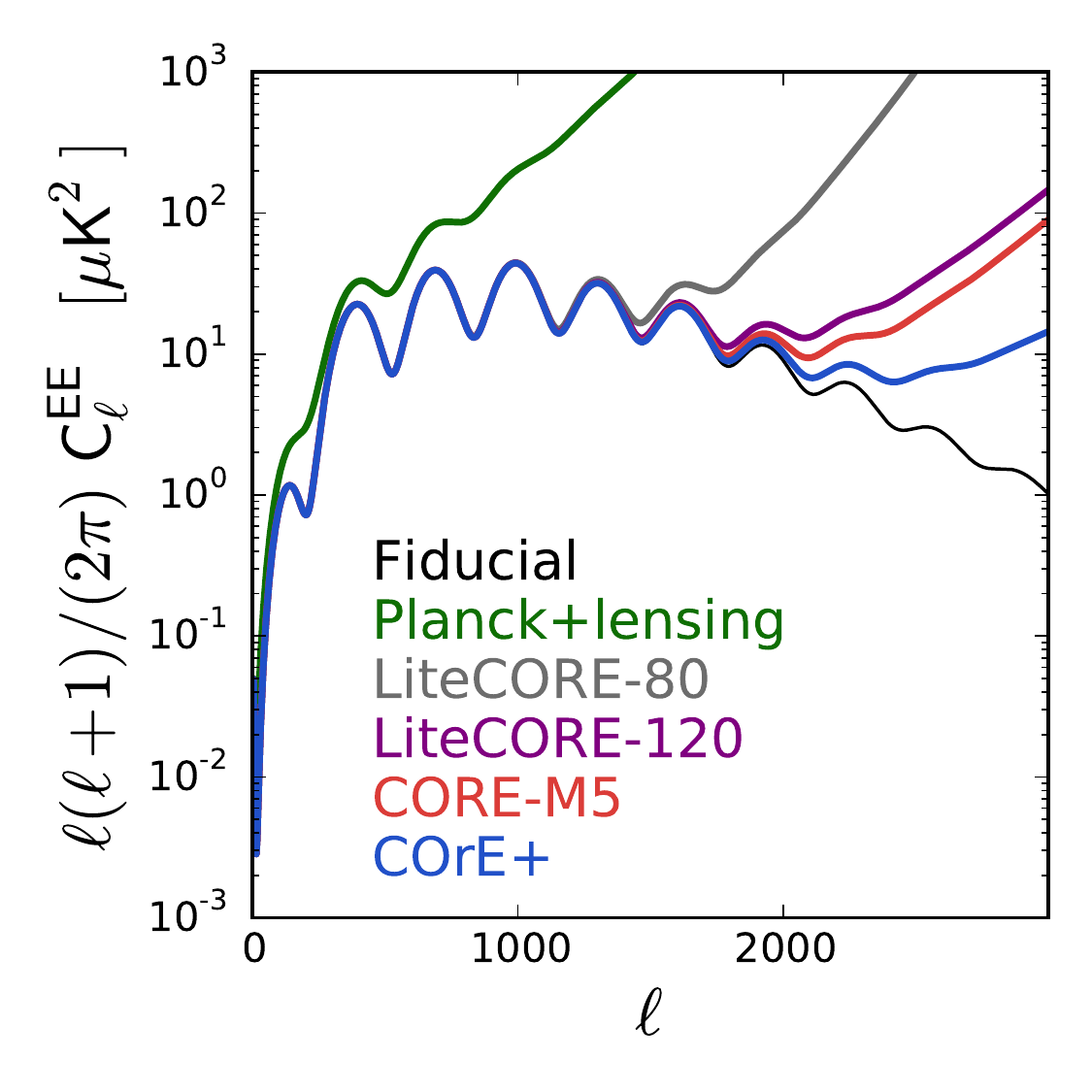}
		\hspace{-0.2cm}\includegraphics[width=5.2cm]{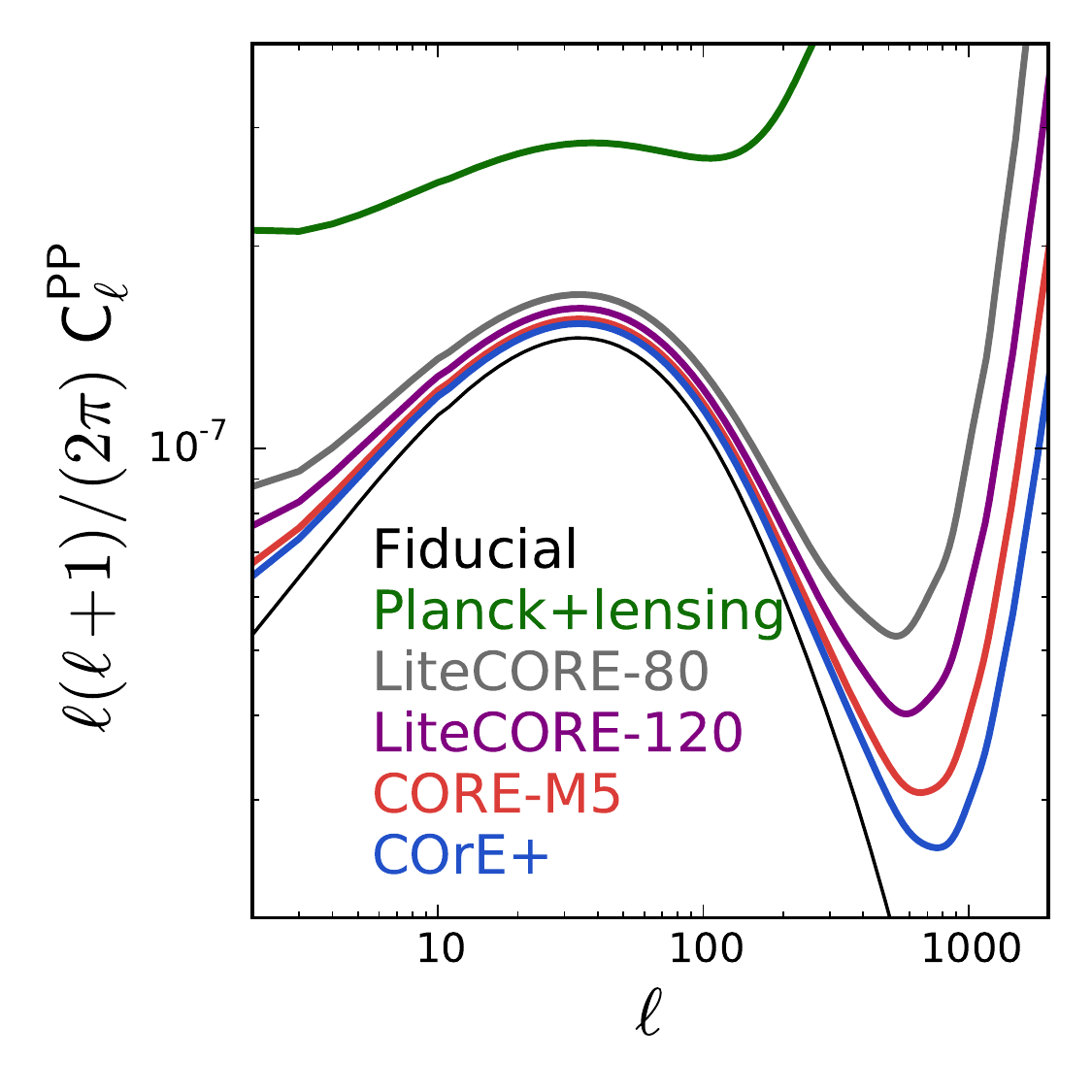}	
	\caption{Fiducial model and variance $C_l+N_l$ of each data point $a_{lm}$, given the sensitivity of each CORE configuration (Planck is also shown for comparison). As long as the variance traces the fiducial model, the data is cosmic variance limited. This happens down to different angular scales for the temperature (left) and E-mode polarisation (middle). For CMB lensing extraction (right), on all scales, there is a substantial difference between the noise level of the different configurations.}
\label{fig:noise}
\end{figure}

In what follows we also compare the baseline CORE-M5 configuration with other four 
possible versions: LiteCORE-80, LiteCORE-120, LiteCORE-150 and COrE+. 
Experimental specifications for these configurations are given in Table \ref{tab:specifications}.
We assume that beam uncertainties are small and that uncertainties due to foreground
removal are smaller than statistical errors.  In Figure~\ref{fig:noise}, for each configuration, we show the variance $C_l+N_l$
compared to the fiducial model $C_l$ for the temperature (left) and polarisation (middle) auto-correlation spectra. The data are cosmic-variance-limited up to the multipole at which this variance departs from the fiducial model.

\begin{table} [h!]
\begin{center}\footnotesize
\begin{tabular}{|c|c|c|c|}
\hline
Channel [GHz] & FWMH [arcmin] & $\Delta T $ [$\mu$K arcmin] & $\Delta P $ [$\mu$K arcmin] \\ 
\hline
\multicolumn{4}{|c|}{ LiteCORE-80,  $l_{\rr{max}} = 2400, f_{\rr{sky}} = 0.7$  }    \\
\hline
 $ 80$ & $20.2$  & $8.8$ & $12.5$ \\
 $ 90$ & $17.8$  & $7.1$ & $10.0$ \\
 $100 $ & $15.8$  & $8.5$ & $12.0$ \\
 $120 $ & $13.2$  & $6.7$ & $9.5$ \\
 $ 140$ & $11.2$  & $5.3 $ & $7.5$ \\
 $ 166$ & $8.5$  & $5.0$ & $7.0$ \\
 $ $195 & $8.1$  & $3.6 $ & $5.0$ \\
\hline
\multicolumn{4}{|c|}{ LiteCORE-120, $l_{\rr{max}} = 3000, f_{\rr{sky}} = 0.7$ }    \\
\hline
 $ 80$ & $13.5$  & $8.8$ & $12.5$ \\
 $ 90$ & $11.9$  & $7.1$ & $10.0$ \\
 $ 100$ & $10.5$  & $8.5$ & $12.0$ \\
 $ 120$ & $8.8$  & $6.7$ & $9.5$ \\
 $ 140$ & $7.4$  & $5.3 $ & $7.5$ \\
 $ 166$ & $6.3 $  & $5.0$ & $7.0$ \\
 $ 195$ & $5.4 $  & $3.6 $ & $5.0$ \\
\hline
\multicolumn{4}{|c|}{ LiteCORE-150, $l_{\rr{max}} = 3000, f_{\rr{sky}} = 0.7$ }    \\
\hline
 $ 80$ & $10.8$  & $8.8$ & $12.5$ \\
 $ 90$ & $9.5$  & $7.1$ & $10.0$ \\
 $ 100$ & $8.4$  & $8.5$ & $12.0$ \\
 $ 120$ & $7.0$  & $6.7$ & $9.5$ \\
 $ 140$ & $5.9$  & $5.3 $ & $7.5$ \\
 $ 166$ & $5.0 $  & $5.0$ & $7.0$ \\
 $ 195$ & $4.3 $  & $3.6 $ & $5.0$ \\
\hline
\multicolumn{4}{|c|}{COrE+, $l_{\rr{max}} = 3000, f_{\rr{sky}} = 0.7$  }   \\
\hline
 $ 100$ & $8.4$  & $6.0  $ & $8.5 $ \\
 $ 115$ & $7.3$  & $5.0  $ & $7.0 $ \\
 $ 130$ & $6.5$  & $4.2  $ & $5.9 $ \\
 $ 145 $ & $5.8$  & $3.6 $ & $5.0 $ \\
 $ 160$ & $5.3 $  & $3.8 $ & $5.4 $ \\
 $175 $ & $4.8 $  & $3.8 $ & $5.3 $ \\
 $195 $ & $4.3 $  & $3.8  $ & $5.3 $ \\
 $220 $ & $3.8 $  & $5.8  $ & $8.1 $ \\
\hline
\end{tabular}
\end{center}
\caption{Experimental specifications for LiteCORE-80, LiteCORE-120, LiteCORE-150 and COrE+:  Frequency channels dedicated to cosmology, beam width, temperature and polarization sensitivities for each channel. }
\label{tab:specifications}
\end{table}

Together with the primary anisotropy signal, we also take into account information
from CMB weak lensing, considering the power spectrum of the 
CMB lensing potential
$C_{\ell}^{PP}$. In what follows we use the quadratic estimator method of Hu \& Okamoto~\cite{okamotohu}, 
that provides an algorithm for estimating the corresponding noise spectrum $N_{\ell}^{PP}$ from
the observed CMB primary anisotropy and noise power spectra. Like in~\cite{Errard:2015cxa}, we use here the noise spectrum 
$N_{\ell}^{PP}$ associated to the $EB$ estimator of lensing, which is the most sensitive one for all CORE configurations (out of all pairs of maps). 
We occasionally repeated the analysis with the actual minimum variance estimator, and found very similar results.
Figure~\ref{fig:noise} shows that the lensing reconstruction noise is different on all scales for the various configurations.

CORE-M5  is clearly sensitive also to  the $BB$ lensing polarization signal, but here we take the conservative
approach to not include it in the forecasts. This leaves open the possibility 
to use this channel for further checks for foregrounds contamination and systematics.
Note that in this work, we consider fiducial models with negligible primordial gravitational waves from inflation.
Otherwise, the $BB$ channel would contain primary signal on large angular scales and could not be neglected.
The sensitivity of CORE-M5 to primordial gravitational waves is studied separately and with a different methodology in a 
companion paper~\cite{ECOinflation}.

We generate fiducial and noise spectra with noise properties as reported in Table \ref{tab:specifications}.
Once a mock dataset is produced we compare a generic theoretical model
through a Gaussian likelihood ${\cal L}$ defined as 
\begin{equation}
 - 2 \ln {\cal L} = \sum_{l} (2l+1) f_{\rm sky} \left(
\frac{D}{|\bar{C}|} + \ln{\frac{|\bar{C}|}{|\hat{C}|}} - 3 \right),
\label{chieff}
\end{equation}
where $\bar{C}_l$ and $\hat{C}_l$ are the fiducial
and theoretical spectra plus noise respectively, 
$|\bar{C}|$, $|\hat{C}|$ denote the determinants of
the theoretical and observed data covariance matrices respectively,
\begin{eqnarray}
|\bar{C}| &=& \bar{C}_\ell^{TT}\bar{C}_\ell^{EE}\bar{C}_\ell^{PP} -
\left(\bar{C}_\ell^{TE}\right)^2\bar{C}_\ell^{PP} -
\left(\bar{C}_\ell^{TP}\right)^2\bar{C}_\ell^{EE} ~, \\
|\hat{C}| &=& \hat{C}_\ell^{TT}\hat{C}_\ell^{EE}\hat{C}_\ell^{PP} -
\left(\hat{C}_\ell^{TE}\right)^2\hat{C}_\ell^{PP} -
\left(\hat{C}_\ell^{TP}\right)^2\hat{C}_\ell^{EE}~,
\end{eqnarray}
$D$ is defined as
\begin{eqnarray}
D  &=&
\hat{C}_\ell^{TT}\bar{C}_\ell^{EE}\bar{C}_V^{PP} +
\bar{C}_\ell^{TT}\hat{C}_\ell^{EE}\bar{C}_\ell^{PP} +
\bar{C}_\ell^{TT}\bar{C}_\ell^{EE}\hat{C}_\ell^{PP} \nonumber\\
&&- \bar{C}_\ell^{TE}\left(\bar{C}_\ell^{TE}\hat{C}_\ell^{PP} +
2\hat{C}_\ell^{TE}\bar{C}_\ell^{PP} \right) \nonumber\\
&&- \bar{C}_\ell^{TP}\left(\bar{C}_\ell^{TP}\hat{C}_\ell^{EE} +
2\hat{C}_\ell^{TP}\bar{C}_\ell^{EE} \right),
\end{eqnarray}
and finally $f_{sky}$ is the sky fraction sampled by the experiment after foregrounds removal.

Note that for temperature and polarization, $\bar{C}_l$ and $\hat{C}_l$ could be defined to include the lensed or unlensed fiducial and theoretical spectra, and in both cases the above likelihood is slightly incorrect. If we use the unlensed spectra, we optimistically assume that we will be able to do a perfect de-lensing of the $T$ and $E$ map, based on the measurement of the lensing map with quadratic estimators. If we use the lensed spectra, we take the risk of double-counting the same information in two observables which are not statistically independent: the lensing spectrum, and the lensing corrections to the $TT$, $EE$ and $TE$ spectra. To deal with this issue, one could adopt a more advanced formalism including non-Gaussian corrections, like in~\cite{BenoitLevy:2012va,Schmittfull:2013uea}. However, we performed dedicated forecasts to compare the two approximate Gaussian likelihoods, and even with the best sensitivity settings of {\it COrE+} we found nearly indistinguishable results (at least for the $\Lambda$CDM+$M_\nu$ model). The reconstructed parameter errors change by negligible amounts between the two cases. The biggest impact is on the error on the sound horizon angular scale $\sigma(\theta_s)$, which is 5\% smaller when using unlensed spectra, because perfect delensing would allow to better identify the primary peak scales. When using the lensed spectra, we do not observe any statistically significant reduction of the error bars, and we conclude that over-counting the lensing information is not important for an experiment with the sensitivity of {\it COrE+}. 
Hence in the rest of this work we choose to always use the version of the Gaussian likelihood that includes lensed $TT$, $EE$ and $TE$ spectra. We will usually refer to our full CMB likelihoods with the acronym ``TEP'', standing for ``Temperature, E-polarisation and lensing Potential data''.

Depending on cases, we derive constraints from simulated data using a modified version of 
the publicly available Markov Chain Monte Carlo
package {\sc CosmoMC}\footnote{\tt http://cosmologist.info}~\cite{Lewis:2002ah}, or with the {\sc MontePython}\footnote{\tt http://baudren.github.io/montepython.html}~\cite{Audren:2012wb} package. With both codes, we normally sample parameters with the Metropolis-Hastings algorithm, with a convergence
diagnostic based on the Gelman and Rubin statistic performed. In exceptional cases, we switch the {\sc MontePython} sampling method
to {\sc MultiNest}~\cite{Feroz:2008xx}.

In what follows we consider temperature and polarization power spectrum data up to $\ell_{\rm max}=3000$,
due to possible unresolved foreground contamination at smaller angular scales and larger multipoles. 
We run {\sc CAMB+CosmoMC} and {\sc CLASS+MontePython} with enhanced accuracy settings\footnote{For CAMB+CosmoMC we checked that:
{\tt accuracy\_setting=1, high\_accuracy\_default = T}
is sufficient. For {\sc CLASS+MontePython} we increased a bunch of precision parameter values with respect to the default of version 2.4.4: \\
{\tt 
tol\_background\_integration = 1.e-3,
tol\_thermo\_integration = 1.e-3,
tol\_perturb\_integration = 1.e-6,
reionization\_optical\_depth\_tol = 1.e-5,
l\_logstep = 1.08,
l\_linstep = 25,
perturb\_sampling\_stepsize = 0.04,
delta\_l\_max = 1000}.
}, including non-linear corrections to the lensing spectrum computed with the latest version of {\sc HaloFit}~\cite{Takahashi:2012em}. We performed several consistency checks proving that the two pipelines produce identical results.

We also include a few external mock data sets in combination with CORE. 
For the BAO scale reconstruction, we included a mock likelihood for a high precision spectroscopic survey like DESI (Dark Energy Spectroscopic Instrument~\cite{Levi:2013gra}). For simplicity, our DESI mock data consists in the measurement of the ``angular diameter distance to sound horizon scale ratio'', $D_A/s$, at 18 redshifts ranging from 0.15 to 1.85, with uncorrelated errors given by the second column of Table V in~\cite{Font-Ribera:2013rwa}.
For the matter power spectrum reconstruction, we simulate data corresponding to the tomographic weak lensing survey of Euclid. We used the public {\tt euclid\_lensing} mock likelihood of {\sc MontePython}, with sensitivity parameters identical to the default settings of version 2.2.2. (matched to the current recommendations of the Euclid science working group). 
Integrals in wavenumber space are conservatively limited to the range $k\leq 0.5h/$Mpc, to avoid propagating systematic errors from deeply non-linear scales.
For simplicity we do not include extra observables from Euclid (galaxy power spectrum, cluster counts, BAO scale...) which would further decrease error bars. Hence we expect our CORE + Euclid forecasts to be very conservative.
\section{$\Lambda$CDM and derived parameters \label{sec:lcdm}}

\subsection{Future constraints from CORE}

Adopting the method presented in the previous section, here we forecast the achievable constraints on cosmological parameters from CORE in four configurations: LiteCORE-80, LiteCORE-120, CORE-M5 and COrE+.  We work in the framework of the $\lcdm$ model, that assumes a flat universe with a cosmological constant, and is based on $6$ parameters: the baryon $\Omega_bh^2$ and cold dark matter  $\Omega_{c}h^2$ densities, the amplitude $A_s$ and spectral index $n_s$ of primordial inflationary perturbations,
the optical depth to reionization $\tau$, and the angular size of the sound horizon at recombination $\theta_s$.
Assuming $\lcdm$, constraints can be subsequently obtained on "derived" parameters (i.e. that are not varied during
the MCMC process) such as the Hubble constant $H_0$ and the r.m.s. amplitude of matter fluctuations on spheres of $8 Mpc^{-1} h$; $\sigma_8$.
The $\Lambda$CDM model has been shown to be in good agreement with current measurements of CMB anisotropies
(see e.g. \cite{planck2015}) and is therefore mandatory to first consider the future possible improvement provided
by a CMB satellite experiment such as CORE on the accuracy of its parameters.

\begin{table}[h]
\begin{center}\footnotesize
\scalebox{0.83}{\begin{tabular}{|c||c|c|c|c|}
\hline
 Parameter                             & LiteCORE-80, TEP     & LiteCORE-120, TEP                         & CORE-M5, TEP	&COrE+, TEP\\
\hline
$ \Omega_b h^2          $ & $ 0.022182\pm 0.000052_{}(2.9)             $& $ 0.022180\pm 0.000041_{} (3.75)             $& $ 0.022182\pm 0.000037_{} (4.0)              $& $ 0.022180\pm 0.000033_{} (4.5)             $\\
$ \Omega_c h^2          $ & $ 0.12047\pm 0.00033_{} (4.1)               $& $ 0.12049\pm 0.00030_{} (4.8)               $& $ 0.12048\pm 0.00026_{} (5.4)               $& $ 0.12048\pm 0.00026_{} (5.4)                $\\
$ 100\theta_{MC}          $ & $ 1.040691\pm 0.000097_{} (3.2)             $& $ 1.040691\pm 0.000082_{} (3.7)              $& $ 1.040691\pm 0.000078_{} (4.0)             $& $ 1.040693\pm 0.000073_{} (4.3)             $\\
$ \tau                             $ & $ 0.0598\pm0.0020_{} (4.1)        $& $ 0.0597\pm 0.0020_{} (4.5)                   $& $ 0.0597\pm 0.0020_{} (4.5)                 $& $ 0.0597\pm 0.0020_{} (4.5)                 $\\
$ n_s                     $ & $ 0.9619\pm 0.0016_{} (2.8)                 $& $ 0.9620\pm 0.0015_{} (3.0)                 $& $ 0.9619\pm 0.0014_{} (3.2)                 $& $ 0.9619\pm 0.0014_{} (3.2)                 $\\
$ ln(10^{10} A_s)        $ & $ 3.0563\pm 0.0037_{} (3.9)                 $& $ 3.0562\pm 0.0035_{} (4.3)                 $& $ 3.0563\pm 0.0035_{} (5.1)                 $& $ 3.0562\pm 0.0034_{} (5.3)                 $\\
\hline
$ H_0                              $[km/s/Mpc] & $ 66.96\pm 0.14_{} (4.4)                     $& $ 66.95\pm 0.12_{} (5.2)                    $& $ 66.96\pm 0.11_{} (5.6)                     $& $ 66.95\pm 0.10_{} (6.2)                     $\\
$ \sigma_8                         $ & $ 0.8173\pm 0.0014_{} (5.8)                $& $ 0.8173\pm 0.0012_{} (7.4)                 $& $ 0.8172\pm{0.0011} _{} (7.8)                 $& $ 0.8173\pm 0.0010_{} (8.6)                 $\\
\hline
\end{tabular}}
\end{center}
\caption{Forecasted constraints at $68 \%$ c.l. on cosmological parameters assuming standard $\Lambda$CDM for the CORE-M5 proposal and for three other possible CORE experimental configurations. The dataset used includes TT, EE, TE angular spectra and information from Planck CMB lensing.
The numbers in parenthesis show the improvement $i=\sigma^{Planck}/\sigma^{CORE}$ with respect to the current constraints coming from the Planck satellite.}
\label{tab:params_lcdm}
\end{table}

Our results are reported in Table \ref{tab:params_lcdm}, where we show the constraints at $68 \%$ c.l. on the cosmological parameters from CORE-M5 and we compare the results with three other possible experimental configurations: LiteCORE-80, LiteCORE-120 and COrE+.  Besides the standard $6$ parameters we also show the constraints obtained on derived parameters such as the Hubble constant $H_0$ and the amplitude of density fluctuations $\sigma_8$.

\subsection{Improvement with respect to the Planck 2015 release}

\begin{figure}
	\centering
		\includegraphics[width=7.5cm]{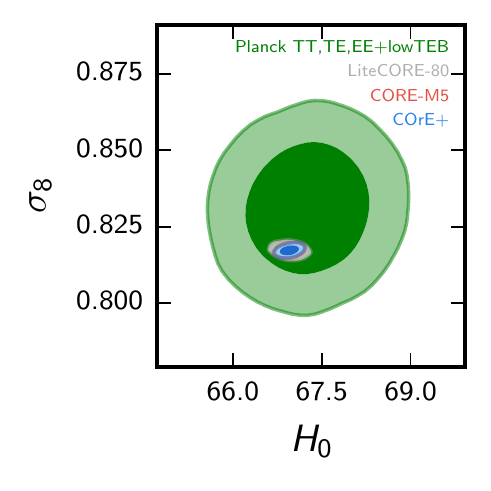}
		\includegraphics[width=7.5cm]{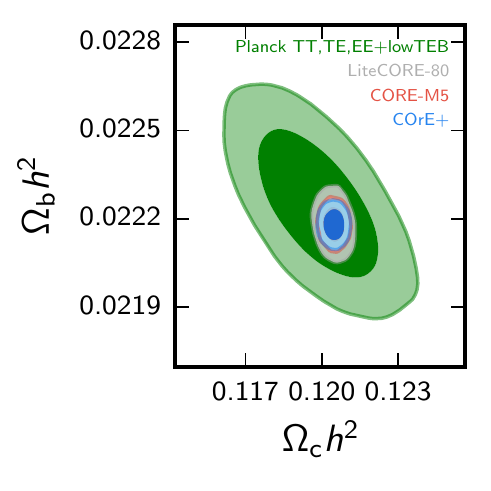}
	\caption{2D posteriors in the $\sigma_8$ vs $H_0$ plane (left panel) and on the $\Omega_bh^2$ vs $\Omega_{c}h^2$ plane (right panel) from the recent Planck 2015 data release (temperature and anisotropy) and from the simulated LiteCORE-80, CORE-M5 and COrE+ experimental configurations. $\Lambda$CDM is assumed for the CORE simulations. The improvement of any CORE configuration in constraining parameters with respect to Planck is clearly visible.}
\label{fig:lcdm_improvement}
\end{figure}

In Table \ref{tab:params_lcdm} we also show the improvement in the accuracy with respect to the most recent constraints coming from the TT, TE and EE angular
spectra data from the Planck satellite \cite{planck2016reio} simply defined as $i=\sigma^{Planck}/\sigma^{CORE}$.
As we can see, even the cheapest configuration of LiteCORE-80 could improve current constraints with respect to Planck by
a factor that ranges between $\sim 3$, for the scalar spectral index $n_s$, and $\sim 6$, for the $\sigma_8$ density fluctuations amplitude.
The most ambitious configuration, COrE+, could lead to even more significant improvements: up to a factor $\sim 8$ in
$\sigma_8$ and up to a factor $\sim 6$ for $H_0$, for example. Similar constraints can be achieved by the proposed CORE-M5 configuration.
The improvement with respect to current Planck measurements is clearly visible in Figure \ref{fig:lcdm_improvement}, where we show the 2D posteriors in the $\sigma_8$ vs $H_0$ plane (left panel) and on the $\Omega_bh^2$ vs $\Omega_{c}h^2$ plane (right panel) from the recent Planck 2015 data release (temperature and polarization) and from the LiteCORE-80, CORE-M5 and COrE+ experimental configurations.
These numbers clearly indicate that there is still a significant amount of information that can be extracted from the CMB angular spectra even after the very 
precise Planck measurements. It is also important to note that the most significant improvements are on two key observables: $\sigma_8$ and the
Hubble constant $H_0$ that can be measured in several other independent ways.
A precise measurement of these parameters, therefore, offers the opportunity for a powerful test of the standard cosmological model. It should indeed also be noticed that the
recent determination of the Hubble constant from observations of luminosity distances of Riess et al. (2016) \cite{riess2016} is in conflict
 at above $3$ standard deviations with respect to the value obtained by Planck (see also \cite{Grandis:2016fwl, Bernal:2016gxb}). 
 A significantly higher value of the Hubble constant has also recently been reported by the H0LiCOW collaboration \cite{Bonvin:2016crt}, from a  joint analysis of three multiply-imaged quasar systems with measured gravitational time delays. Furthermore, values of $\sigma_8$ inferred from cosmic shear galaxy surveys such as CFHTLenS \cite{cfhtlens} and KiDS \cite{kids} are in tension above two standard deviations with Planck. While systematics can clearly play a role, new physics has been invoked to explain these tensions 
(see e.g. \cite{dms,Qing-Guo:2016ykt,Ko:2016uft,Karwal:2016vyq,Kumar:2016zpg,Ko:2016fcd,Prilepina:2016rlq}) and future and improved CMB determinations of $H_0$ and $\sigma_8$ are crucial in testing this possibility.

\subsection{Comparison between the different CORE configurations}

\begin{figure}
	\centering
		\includegraphics[width=12cm]{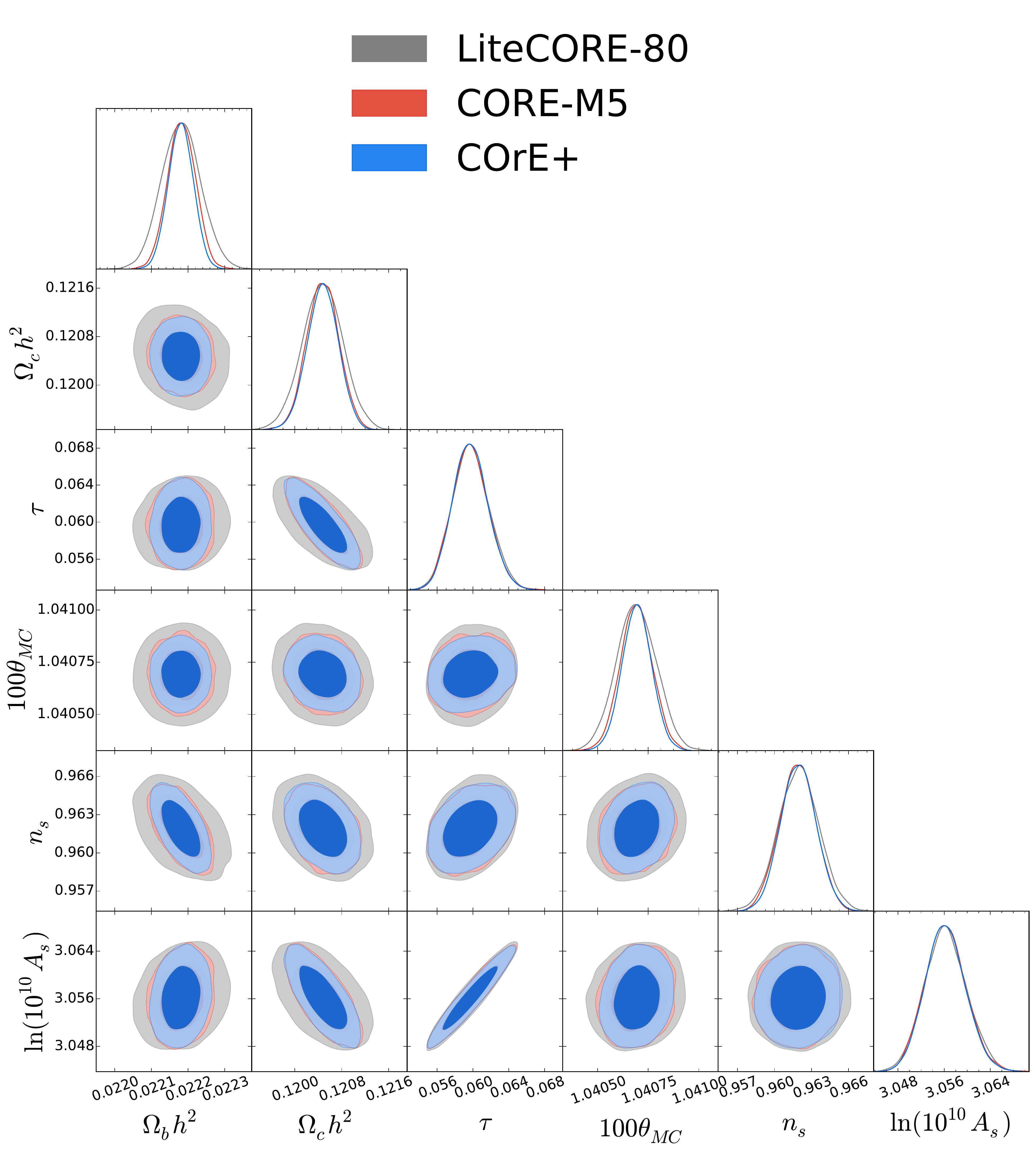}
	\caption{2D posteriors for several combinations of parameters for the LiteCORE-80, CORE-M5 and COrE+ experimental configurations. $\Lambda$CDM is assumed as the underlying fiducial model.}
\label{fig:params_lcdm}
\end{figure}

It is interesting to compare the results between the different experimental configurations as reported in
Table \ref{tab:params_lcdm} and as we can also visually see in Figure \ref{fig:params_lcdm}, where
we show a triangular plot for the 2D posteriors from LiteCORE-80, CORE-M5 and
COrE+.

We find four main conclusions from this comparison:

\begin{itemize}

\item When we move from LiteCORE-80 to COrE+ we notice an improvement of a factor $\sim 1.6$ on the
determination of the baryon density $\Omega_bh^2$, and an improvement of a factor  $\sim 1.4$ on the
determination of the Hubble constant $H_0$ and the amplitude of matter fluctuations $\sigma_8$.
COrE+ is clearly the best experimental configuration in terms of constraints on these cosmological
parameters. However, the CORE-M5 setup provides very similar bounds on these parameters
as COrE+, with a degradation in the accuracy at the level of $\sim 10-12 \%$.

\item Moderate improvements are also present for the CDM density (of about $\sim 1.3$) and
the spectral index ($\sim 1.14$). The constraints from CORE-M5 and COrE+ are almost identical
on these parameters.

\item The constraints on the optical depth are identical for all four experimental configurations considered.
This should not come as a surprise, since $\tau$ is mainly determined by the large angular scale polarization that
is measured with almost the same accuracy with all the versions of CORE.

\item Moving from COrE+ to CORE-M5 the maximum degradation on the
constraints is about $12 \%$ (for the baryon density).
\end{itemize}

From these results, and considering also the contour plots in Figure \ref{fig:lcdm_improvement} and Figure \ref{fig:params_lcdm} that are almost identical
between CORE-M5 and COrE+, we can conclude that CORE-M5, despite having a mirror of smaller size, will produce essentially the same constraints
on the parameters with respect to COrE+ with, at worst, a degradation in the accuracy of just $\sim 12\%$.

\subsection{Constraints from CORE-M5 and future BAO datasets}

\begin{figure}
	\centering
		\includegraphics[width=7.5cm]{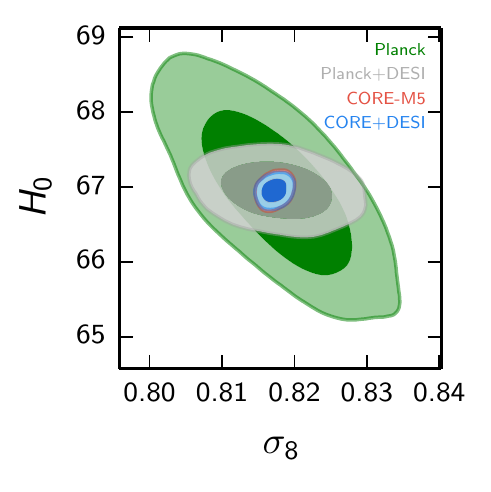}
		\includegraphics[width=7.5cm]{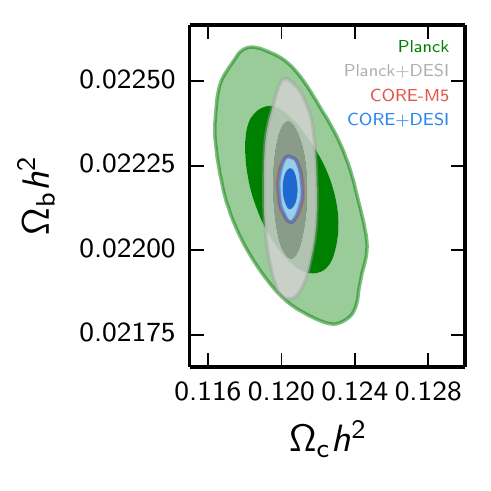}
	\caption{2D posteriors in the $H_0$ vs $\sigma_8$ (left panel) and $\Omega_bh^2$ vs $\Omega_ch^2$ (right panel) planes
	from Planck (simulated), CORE-M5, and future BAO dataset from the DESI survey. $\Lambda$CDM is assumed as the underlying fiducial model.}
\label{fig:params_lcdm_bao}
\end{figure}

We have also considered the constraints achievable by a combination of the CORE-M5 data with information from Baryonic Acoustic
Oscillation derived from a future galaxy survey as DESI. We found that the inclusion of this dataset
will have minimal effect on the CORE-M5 constraints on $\Lambda$CDM parameters. This can clearly be seen in Figure \ref{fig:params_lcdm_bao},
where we plot the 2D posteriors in the $H_0$ vs $\sigma_8$ (left panel) and $\Omega_bh^2$ vs $\Omega_ch^2$ (right panel) planes.
The CORE-M5 and the CORE+DESI contours are indeed almost identical.

It is also interesting to investigate whether the Planck dataset, when combined with future BAO
datasets, could reach a precision on the $\Lambda$CDM parameters comparable with the one obtained by CORE-M5.
To answer to this question we have simulated
the Planck dataset with a noise consistent with the one reported in the 2015 release and combined it with our simulated DESI dataset.
The 2D posteriors are reported in Figure \ref{fig:params_lcdm_bao}: as we can see, while the inclusion of the DESI dataset with Planck
will certainly help in constraining some of  $\Lambda$CDM parameters, such as $H_0$ and the CDM density, the final accuracy will not be
competitive with the one reachable by CORE-M5. In particular, there will be no significant improvement in the determination of
$\sigma_8$ and the baryon density.

\section{Constraints on curvature}

\subsection{Future constraints from CORE}

Measuring the spatial curvature of the Universe is one of the most important goals of modern cosmology, since
flatness is a key prediction of inflation. A precise measurement
of the spatial curvature could, therefore, highly constrain some classes of inflationary
models (see e.g. \cite{inflationcurvature1,inflationcurvature2,inflationcurvature3}).
For example, inflationary models with positive curvature have been proposed in \cite{inflationcurvature1}, while 
models with negative spatial curvature have been proposed in \cite{openinf1,openinf2,openinf3,openinf4,openinf5,openinf6}.
Interestingly, the most recent constraint coming from the Planck 2015 angular power spectra 
data marginally prefers a universe with positive spatial curvature, with curvature density parameter 
$\Omega_k=-0.040_{-0.016}^{+0.024}$ at $68 \%$ CL \cite{planck2015}, suggesting a closed universe 
at about two standard deviations. Moreover, including curvature in the analysis strongly weakens the Planck constraints
on the Hubble constant, due to the well know geometric degeneracy (see e.g. \cite{geometricdegeneracy1,geometricdegeneracy2,geometricdegeneracy3}).
When $\Omega_k$ is varied, the Planck 2015 dataset gives 
$H_0=55_{-5.0}^{+4.3}$ km/s/Mpc at $68 \% $ c.l., i.e. a constraint weaker by nearly one order of
magnitude with respect to the flat case ($H_0=67.59\pm0.73$ km/s/Mpc at $68 \%$ CL \cite{planck2015}).

As shown in \cite{planck2015}, the compatibility with a flat universe is restored when 
the Planck data is combined with the Planck CMB lensing dataset, yielding 
$\Omega_k=-0.0037_{-0.0069}^{+0.0084}$ at $68 \%$ c.l.. However, the inclusion of
the CMB lensing dataset still provides a quite weak constraint on the Hubble constant of 
$H_0=66.1\pm3.1$ km/s/Mpc at $68 \%$ c.l..
It is, therefore, quite important to understand what level
of precision can be reached by future CMB data alone on $\Omega_k$ and, subsequently, on the
Hubble constant, $H_0$.

\begin{table}[h]
\begin{center}\footnotesize
\scalebox{0.83}{\begin{tabular}{|c||c|c|c|c|c|}
\hline 
Parameter         & Planck + lensing & LiteCORE 80, TEP& LiteCORE120, TEP& CORE-M5, TEP& COrE+, TEP  \\            
\hline
$\Omega_k$ &      $-0.0037\,^{+0.0084}_{-0.0069}$&$0.0000\pm0.0021$& $0.0000\pm0.0019$ & $ 0.0000\pm0.0019$ & $ 0.0000\,_{-0.0017}^{+0.0020}$ \\
$\Omega_bh^2$ &    $0.02226\pm0.00016$&  $0.022182\pm0.000059$& $0.022182\,_{-0.000046}^{+0.000041}$ & $ 0.022183\pm0.000038$ & $ 0.022182\pm0.000035$ \\
$\Omega_ch^2$ &      $0.1192\pm0.0015$& $0.12050\pm0.00074$& $0.12046\pm0.00068$ & $ 0.12049\pm0.00066$ & $ 0.12050\,_{-0.00064}^{+0.00071}$ \\
$100\theta_{MC}$ &   $1.04087\pm0.00032$&   $1.04069\pm0.00011$& $1.040685\pm0.000090$ & $ 1.040686\pm0.000085$ & $ 1.040688\pm0.000080$ \\
$\tau$ &    $0.055\,\pm0.019$&  $0.0597\,_{-0.0023}^{+0.0020}$& $0.0598\pm0.0021$ & $ 0.0596\pm0.0020$ & $ 0.0596\pm0.0020$ \\
$n_s$ &   $0.9658\pm0.0048$&   $0.9620\pm0.0021$& $0.9619\pm0.0019$ & $ 0.9620\pm0.0019$ & $ 0.9619\pm0.0019$ \\
$ln(10^{10}A_s)$ &   $3.043\pm0.037$&   $3.0562\pm0.0044$& $3.0563\pm0.0044$ & $ 3.0561\pm0.0044$ & $ 3.0561\pm0.0043$ \\
\hline
$H_0$ [km/s/Mpc]&  $66.1\pm3.1$&    $66.98\pm0.75$& $66.96\pm0.68$ & $ 66.97\pm0.66$ & $ 66.97\,_{-0.62}^{+0.69}$ \\
$\sigma_8$ &   $0.806\pm0.019$&   $0.8174\pm0.0044$& $0.8172\pm0.0040$ & $ 0.8173\pm0.0040$& $ 0.8173\pm0.0039$ \\
\hline
\end{tabular}}
\end{center}
\caption{$68\%$~CL future constraints on cosmological parameters in the $\Lambda$CDM + $\Omega_k$ model for 
four different CORE experimental configurations. A flat universe is assumed as fiducial model.
Current constraints from the Planck 2015 release (temperature,
polarization and lensing) are also reported in the second column for comparison.}\label{tab:omegak}
\end{table}

In Table \ref{tab:omegak} we report the results from our forecasts using CMB data only from four experimental
configurations: LiteCORE-80, LiteCORE-120, CORE-M5 and COrE+.
As we can see, all configurations are able to constrain curvature with similar accuracy, which is anyway always
about a factor $8$ better than current constraints coming from Planck angular spectra data (about a factor $4$ when compared
with Planck+CMB lensing). Future CMB data can, therefore, improve
the Planck 2015 constraint on curvature by nearly one order of magnitude. The current best fit Planck value of $\Omega_k=-0.033$
(see e.g. \cite{planck2015}) can be tested (and falsified) at the level of $\sim 16$ standard deviations.
Constraints on the Hubble constant are also significantly improved: a future CORE mission can
provide constraints on the Hubble constant with a $1 \sigma$ accuracy  better than $\sim 1$ km/s/Mpc 
independently from the assumption of a flat universe. The 2D posteriors on the $\Omega_k$ vs $H_0$ plane are reported in 
Figure \ref{fig:omegak_desi} (left panel).

\subsection{Future constraints from CORE+DESI}

\begin{table}[h]
\begin{center}\footnotesize
\scalebox{0.83}{\begin{tabular}{|c||c|c|c|c|c|}
\hline 
Parameter         & Planck + lensing & LiteCORE 80, TEP& LiteCORE120, TEP& CORE-M5, TEP& COrE+, TEP  \\          
        &+DESI&+DESI&+DESI&+DESI&+DESI \\
\hline
$\Omega_k$ &     $-0.0000\pm0.0016$& $-0.00001\pm0.00081$& $-0.0005\pm0.00078$ & $ 0.00002\pm0.00075$ & $ 0.00002\pm0.00074$ \\
$\Omega_bh^2$ &      $0.02219\pm0.00015$&$0.022181\pm0.000055$& $0.022181\pm0.000041$ & $ 0.022184\pm0.000037$ & $ 0.022181\pm0.000036$ \\
$\Omega_ch^2$ &      $0.1204\pm0.0017$&$0.12048\,_{-0.00047}^{+0.00055}$& $0.12044\pm0.00047$ & $ 0.12050\, ^{+0.00044}_{-0.00039}$ & $ 0.12050\pm0.00041$ \\
$100\theta_{MC}$ &      $1.04068\pm0.00036$&$1.04069\,_{-0.00010}^{+0.00012}$& $1.040688\,_{-0.000079}^{+0.000092}$ & $ 1.040685\pm0.000080$ & $ 1.040686\pm0.000073$ \\
$\tau$ &      $0.0605\,_{-0.0061}^{+0.0052}$&$0.0597\,_{-0.0022}^{+0.0019}$& $0.0598\pm0.0021$ & $ 0.0596\pm0.0019$ & $ 0.0595\pm0.0020$ \\
$n_s$ &      $0.9620\pm0.0042$&$0.9621\pm0.0018$& $0.9620\pm0.0016$ & $ 0.9619\,^{+0.0015}_{-0.0017}$ & $ 0.9619\pm0.0016$ \\
$ln(10^{10}A_s)$ &      $3.058\,^{+0.010}_{-0.012}$&$3.0561\pm0.0040$& $3.0564\,_{-0.0037}^{+0.0033}$ & $ 3.0560\pm0.0034$ & $ 3.0558\pm0.0035$ \\
\hline
$H_0$[km/s/Mpc] &      $66.96\pm0.26$&$66.95\pm0.25$& $66.95\pm0.25$ & $ 66.96\pm0.24$ & $ 66.95\pm0.24$ \\
$\sigma_8$ &      $0.8177\pm0.0077$&$0.8173\pm0.0023$& $0.8172\pm0.0021$ & $ 0.8173\pm0.0018$ & $ 0.8172\pm0.0018$ \\
\hline
\end{tabular}}
\end{center}
\caption{$68\%$~CL future constraints on cosmological parameters in the $\Lambda$CDM + $\Omega_k$ model for
four CORE experimental configurations combined with simulated data of the DESI BAO survey.
In the second column, for comparison, we also report the constraints from a simulated Planck+DESI dataset.
A flat universe is assumed in the simulated data.}\label{tab:omegak_desi}
\end{table}

Stronger constraints on curvature can be obtained by combining the Planck 2015 data with a combination of
BAO measurements. In this case, the constraint is $\Omega_k=0.0002\pm0.0021$ at $68 \%$ c.l., and also
the Hubble constant is well constrained with $H_0= 67.58\pm0.70$ km/s/Mpc. The precision of these constraints is very close
to the one expected by CMB data alone from CORE and reported in 
Table \ref{tab:omegak}. It is, therefore, interesting to investigate if a future CORE mission can improve the constraints 
on $\Omega_k$ with respect to current Planck+BAO constraints. 

In Table \ref{tab:omegak_desi} we indeed  present the constraints on $\Omega_k$ including future BAO simulated data assuming the
experimental specification of the DESI survey. As we can see, including DESI data significantly shrinks the model space, leading to constraints that 
are now a factor $\sim 2.5$ stronger than the constraints from CORE alone and $2.8$ times more stringent than current Planck+BAO constraints. 
While, as we saw in the previous section, there is little advantage in combining CORE with future BAO survey in constraining the $\Lambda$CDM 
parameters, a significant improvement is expected on extensions such as $\Omega_k$.

\begin{figure}
	\centering
		\includegraphics[width=7.5cm]{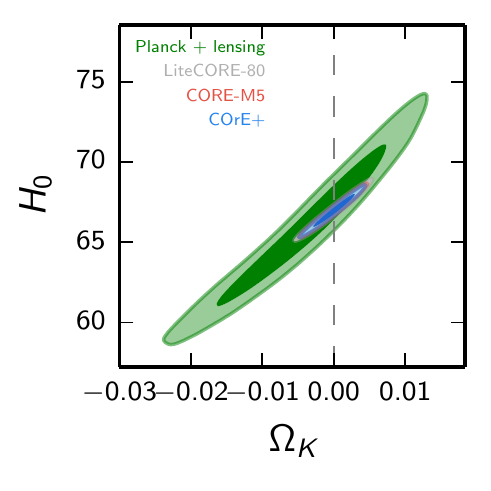}	
		\includegraphics[width=7.5cm]{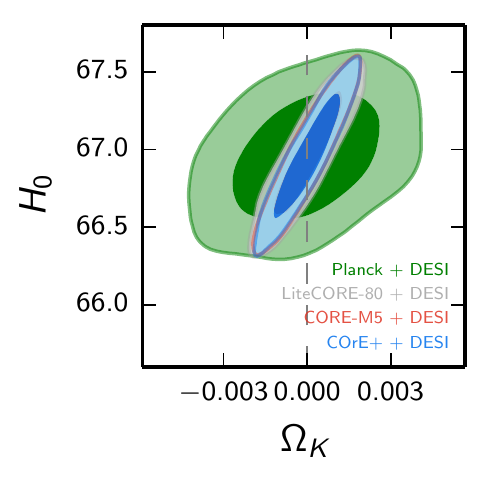}	
	\caption{Left Panel: Constraints on the $H_0$ vs $\Omega_k$ plane from different CORE configurations. Current constraints from Planck+CMB lensing are
	reported for comparison. Right Panel: Constraints on the $H_0$ vs $\Omega_k$ plane for the following (simulated) datasets: Planck+DESI, LiteCORE80+DESI,
	CORE-M5+DESI, COrE++DESI. A flat universe is assumed in the simulated data.}
\label{fig:omegak_desi}
\end{figure}

We can also see that, once the DESI dataset is included, there is little difference in the constraints on $\Omega_k$ between the
CORE configurations. The constraints on the   $H_0$ vs $\Omega_k$ plane from COrE+ and DESI 
are reported in Figure \ref{fig:omegak_desi} (right panel).

\section{Extra relativistic relics\label{sec:relativistic}}

The minimal cosmological scenario predicts that, at least after the time of nucleosynthesis, the density of relativistic particles is given by the contribution of CMB photons plus that of active neutrino species, until they become non-relativistic due to their small mass. This assumption is summarized by the standard value of the effective neutrino number $N_{\rm eff} = 3.046$~\cite{Mangano:2005cc} (see \cite{Dolgov:1998sf} and \cite {Lopez:1998aq} for pioneering work and \cite{Dolgov:2002wy} for a review of the subject). 
A more recent calculation beased on the latest data on neutrino physics finds $N_{\rm eff} = 3.045$~\cite{deSalas:2016ztq}, but at the precision level of CORE the difference is irrelevant, and we will keep 3.046 as our baseline assumption.
However, there are many simple theoretical motivations for relaxing this assumption. We know that the standard model of particle physics is incomplete (e.g. because it does not explain dark matter), and many of its extensions would lead to the existence of extra light or massless particles; depending on their interactions and decoupling time the latter could also contribute to $N_{\rm eff}$. Depending on the context, these extra particles are usually called extra relativistic relics, dark radiation or axion-like particles in more specific cases. In the particular case of particles that were in thermal equilibrium at some point, the enhancement of $N_{\rm eff}$ can be predicted as a function of the decoupling temperature~\cite{Baumann:2016wac}. Even in absence of a significant density of such relics, ordinary neutrinos could have an unexpected density due to non-standard interactions~\cite{neutrinonon}, non-thermal production after decoupling~\cite{Tomas:2001dh}, or low-temperature reheating~\cite{treheating}, leading to a value of $N_{\rm eff}$ larger or smaller than 3.046.
There are 
additional motivations to consider $N_{\rm eff}$ as a free parameter (background of gravitational waves produced by a phase transition, modified gravity, extra dimensions, etc. -- see~\cite{Archidiacono:2011gq} for a review).

Over the last years the extended $\Lambda$CDM + $N_{\rm eff}$ has received a lot of attention within the cosmology community. Assuming $N_{\rm eff}>3.046$ has the potential to solve tensions in observational data: for instance, internal tensions in pre-Planck CMB data, which have now disappeared ($N_{\rm eff} = 2.99 \pm0.20$~(68\%CL) for Planck 2015 TT,TE,EE+lowP~\cite{planck2015}); or tensions between CMB data and direct measurements of $H_0$~\cite{Riess:2016jrr} (however, solving this problem by increasing $N_{\rm eff}$ requires a higher value of $\sigma_8$, which brings further tensions with other datasets~\cite{planck2015}). In any case, the community is particularly eager to measure $N_{\rm eff}$ with better sensitivity in the future, in order to: (i) test the existence of extra relics and probe extensions of the standard model of particle physics; (ii) get a window on precision neutrino physics (since the contribution of neutrinos to $N_{\rm eff}$ depends on the details of neutrino decoupling); and (iii) check whether the tensions in cosmological data are related to the relativistic density or not.

\begin{figure}
	\centering
		\includegraphics[width=7cm]{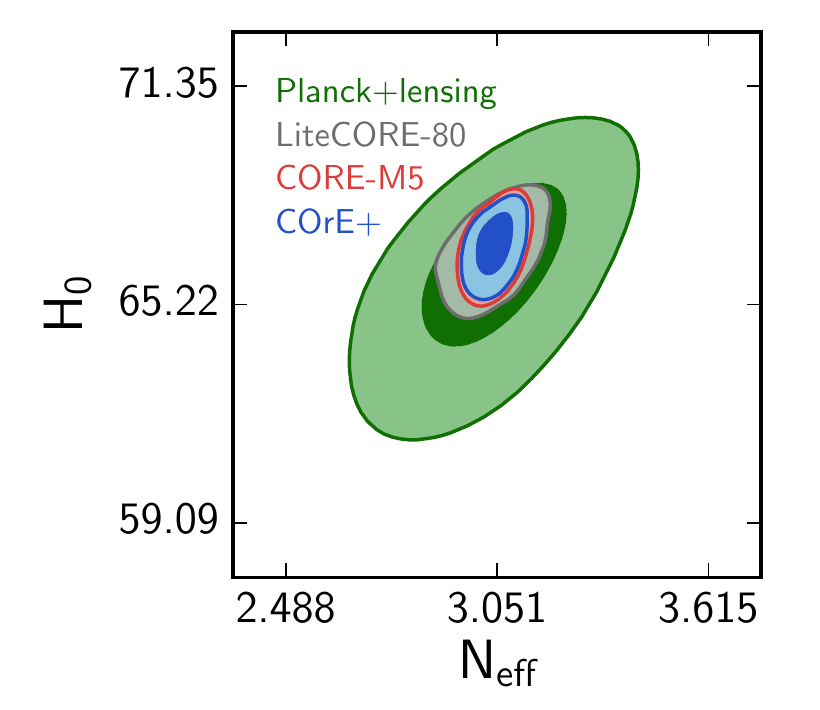} 
		\includegraphics[width=7cm]{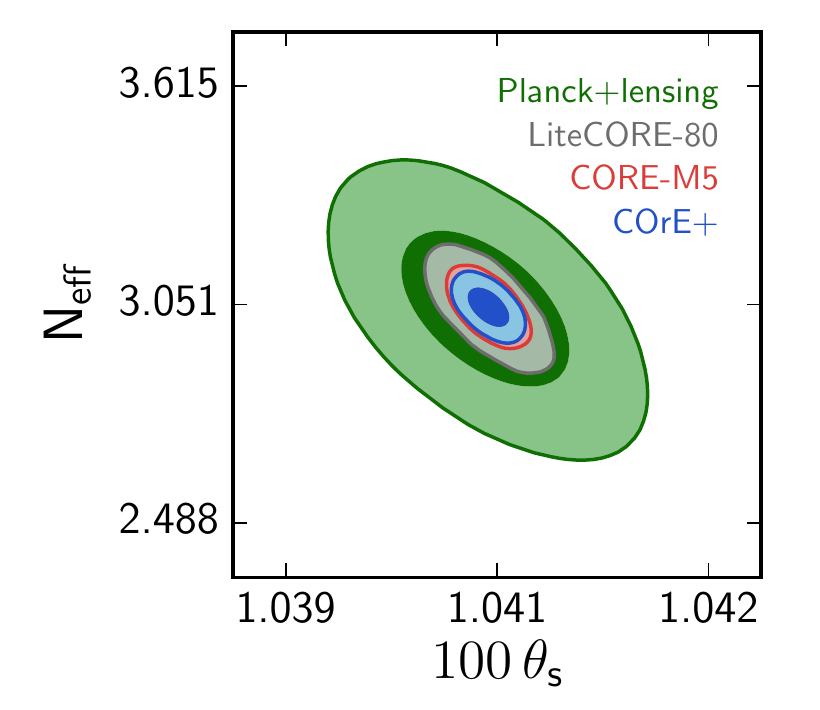}
	\caption{Parameter degeneracy between $N_{\rm eff}$ and $H_0$ or $\theta_s$, assuming the extended model ``DEG+Neff'', with three experimental settings for CORE or with a fake Planck likelihood mimicking the sensitivity of the real experiment (always using all CMB information from TT,TE,EE + lensing extraction). The correlations observed in the Planck case are explained in the text. The degeneracy with $H_0$ is almost entirely resolved by CORE, while that with $\theta_s$ is limited to a much smaller range.}
\label{fig:neff}
\end{figure}

Since CMB data accurately determines the redshift of equality~$z_\mathrm{eq}$, the impact of $N_{\rm eff}$ on CMB observables is usually discussed at fixed $z_\mathrm{eq}$~\cite{Bashinsky:2003tk,Hou:2011ec,Lesgourgues:1519137}. The time of equality can be kept fixed by simultaneously increasing  $N_{\rm eff}$ and the dark matter density $\omega_\mathrm{cdm}$ (or, depending on the choice of parameter basis, $N_{\rm eff}$ and $H_0$). The impact on the CMB is then minimal, which explains the well known ($N_{\rm eff}$, $\omega_\mathrm{cdm}$) or ($N_{\rm eff}$, $H_0$) degeneracy: the latter is clearly visible with Planck data in Figure~\ref{fig:neff} (left plot). However, this transformation does not preserve the angular scale of the photon damping scale on the last scattering surface: hence the best probe of $N_{\rm eff}$ comes from accurate measurements of the exponential tail of the temperature and polarisation spectra at high-$\ell$. Hence the accuracy with which CMB experiments can measure $N_{\rm eff}$ is directly related to their sensitivity and angular resolution, as confirmed by the following forecasts. Increasing $N_{\rm eff}$ has other effects on the CMB coming from gravitational interactions between photons and neutrinos before decoupling: a smoothing of the acoustic peaks (however, very small, and below the per-cent level for variations of the order of $\Delta N_{\rm eff} \sim 0.1$), and a shift of the peaks towards larger angles caused by the ``neutrino drag'' effect~\cite{Bashinsky:2003tk,Hou:2011ec,Lesgourgues:1519137}. This means that in order to keep a fixed CMB peak scale, one should decrease the angular size of the sound horizon $\theta_s$ while increasing $N_{\rm eff}$: this implies an anti-correlation between $\theta_s$ and $N_{\rm eff}$ that can be observed in Figure~\ref{fig:neff} (right plot). Therefore, by accurately measuring $N_{\rm eff}$, we could get a more robust and model-independent measurement of the sound horizon scale, which would in turn be very useful for constraining the expansion history with BAO data.

Since the parameter $N_{\rm eff}$ is closely related to neutrino properties, and since we know that neutrinos have a small mass, we forecast the sensitivity of different experimental set-ups to $N_{\rm eff}$ while varying simultaneously the summed neutrino mass $M_\nu$. This leads to more robust predictions than if we had fixed the mass (although a posteriori we find no significant correlation between $N_{\rm eff}$ and $M_\nu$). We investigate the CORE sensitivity to $N_{\rm eff}$ within two distinct models:
\begin{itemize}
\item The model ``$\Lambda$CDM + $M_\nu$ +$\Delta N_{\rm eff}^\mathrm{massless}$'' has 3 massive degenerate and thermalised neutrino species, plus extra massless relics contributing as $\Delta N_{\rm eff}^\mathrm{massless}>0$. It is motivated by scenarios with standard active neutrinos and extra massless relics (or very light relics with $m \ll 10$~meV).
\item The model ``$\Lambda$CDM + $M_\nu$ + $N_{\rm eff}^\mathrm{massive}$'' only has 3 massive degenerate neutrino species, with fixed temperature, but with a rescaled density. During radiation domination they contribute to the effective neutrino number as $N_{\rm eff}^\mathrm{massive}$, which could be greater or smaller than 3.046. This model provides a rough first-order approximation to specific scenarios in which neutrinos would be either enhanced (e.g. by the decay of other particles) or suppressed (e.g. in case of low-temperature reheating).
\end{itemize}
Our forecasts consist in fitting these models to mock data, with a choice of fiducial parameters slightly different from the previous section\footnote{The new choice of fiducial parameters is 
$\Omega_b h^2=0.022256$,
$\Omega_c h^2=0.11976$,
$100\theta_s=1.0408$,
$\tau=0.06017$,
$n_s=0.96447$,
$ln(10^{10} A_s)=3.0943$,
$M_\nu=60$~meV, with neutrino masses ordered like in the Normal Hierarchy (NH) scenario. \label{foot:dec_fid}}, including in particular neutrino masses summing up to $M_\nu=60$~meV.

\begin{table}[h]
\begin{center}\footnotesize
\scalebox{0.81}{\begin{tabular}{|c||c|c|c|c|c|}
\hline
Parameter & Planck, TEP & LiteCORE-80, TEP & LiteCORE-120, TEP & CORE-M5, TEP & COrE+, TEP\\
\hline
$\Delta N_\mathrm{eff}^\mathrm{massless}$ & $<0.19$ (68\%CL)& $<0.062$ (68\%CL)& $<0.045$ (68\%CL)& $<0.040$ (68\%CL)& $<0.036$ (68\%CL)\\
$M_\nu$ (meV) & $<310$ (68\%CL)& $77^{+37}_{-59}$& $72^{+34}_{-56}$& $71^{+34}_{-54}$& $70^{+35}_{-53}$\\
$\Omega_b h^2$ & $0.02208\pm0.00025$& $0.022305\pm0.000070$& $0.022293\pm0.000052$& $0.022289\pm0.000047$& $0.022284\pm0.000041$\\
$\Omega_c h^2$ & $0.1184\pm0.0030$& $0.12056^{+0.00066}_{-0.00096}$& $0.12030^{+0.00057}_{-0.00079}$& $0.12023^{+0.00052}_{-0.00074}$& $0.12015^{+0.00051}_{-0.00071}$\\
$100\theta_s$ & $1.04087\pm0.00046$& $1.04070\pm0.00013$& $1.04070\pm0.00010$& $1.040700\pm0.000094$& $1.040800\pm0.000085$\\
$\tau$ & $0.071\pm0.018$& $0.0605\pm0.0020$& $0.0606\pm0.0021$& $0.0605\pm0.0021$& $0.0606\pm0.0021$\\
$n_s$ & $0.9589\pm0.0095$& $0.9665\pm0.0026$& $0.9663\pm0.0023$& $0.9661\pm0.0023$& $0.9661\pm0.0022$\\
$ln(10^{10} A_s)$ & $3.071\pm0.037$& $3.0970\pm0.0044$& $3.0964\pm0.0043$& $3.0961\pm0.0042$& $3.0960\pm0.0042$\\
\hline
$H_0$ (km/s/Mpc) & $64.8\pm2.3$& $67.15^{+0.80}_{-0.58}$& $67.13^{+0.74}_{-0.51}$& $67.12^{+0.71}_{-0.50}$& $67.11^{+0.71}_{-0.47}$\\
$\sigma_8$ & $0.778^{+0.038}_{-0.024}$& $0.831^{+0.011}_{-0.006}$& $0.831^{+0.010}_{-0.006}$& $0.8308^{+0.0097}_{-0.0059}$& $0.8307^{+0.0097}_{-0.0055}$\\

\hline
\end{tabular}}
\scalebox{0.81}{\begin{tabular}{|c||c|c|c|c|c|}
\hline
Parameter & Planck, TEP & LiteCORE-80, TEP & LiteCORE-120, TEP & CORE-M5, TEP & COrE+, TEP\\
 & + DESI & + DESI & + DESI & + DESI & + DESI\\ 
\hline
$\Delta N_\mathrm{eff}^\mathrm{massless}$ & $<0.15$ (68\%CL)& $<0.061$ (68\%CL)& $<0.042$ (68\%CL)& $<0.038$ (68\%CL)& $<0.033$ (68\%CL)\\
$M_\nu$ (meV) & $85^{+41}_{-50}$& $72\pm24$& $71^{+23}_{-20}$& $70^{+23}_{-20}$& $65^{+22}_{-20}$\\
$\Omega_b h^2$ & $0.02237\pm0.00015$& $0.022310\pm0.000065$& $0.022293\pm0.000050$& $0.022289\pm0.000045$& $0.022279\pm0.000038$\\
$\Omega_c h^2$ & $0.1216^{+0.0012}_{-0.0020}$& $0.12046^{+0.00048}_{-0.00081}$& $0.12023^{+0.00040}_{-0.00059}$& $0.12017^{+0.00036}_{-0.00054}$& $0.12045^{+0.00034}_{-0.00046}$\\
$100\theta_s$ & $1.04050\pm0.00036$& $1.04070\pm0.00013$& $1.04070\pm0.00010$& $1.040700\pm0.000091$& $1.040700\pm0.000080$\\
$\tau$ & $0.0614\pm0.0046$& $0.0605\pm0.0021$& $0.0606\pm0.0021$& $0.0605\pm0.0021$& $0.0605\pm0.0018$\\
$n_s$ & $0.9695^{+0.0037}_{-0.0053}$& $0.9667^{+0.0020}_{-0.0026}$& $0.9662\pm0.0020$& $0.9661\pm0.0020$& $0.9653^{+0.0016}_{-0.0020}$\\
$ln(10^{10} A_s)$ & $3.102\pm0.010$& $3.0967\pm0.0044$& $3.0962\pm0.0043$& $3.0959\pm0.0040$& $3.0966\pm0.0036$\\
\hline
$H_0$ (km/s/Mpc) & $67.56^{+0.42}_{-0.65}$& $67.23\pm0.33$& $67.15\pm0.29$& $67.13\pm0.29$& $67.13\pm0.28$\\
$\sigma_8$ & $0.833\pm0.011$& $0.8316\pm0.0044$& $0.8311\pm0.0040$& $0.8309\pm0.0038$& $0.8309\pm0.0037$\\

\hline
\end{tabular}}
\scalebox{0.81}{\begin{tabular}{|c||c|c|c|c|c|}
\hline
Parameter & Planck, TEP & LiteCORE-80, TEP & LiteCORE-120, TEP & CORE-M5, TEP & COrE+, TEP\\
 & + DESI + Euclid & + DESI + Euclid & + DESI + Euclid & + DESI + Euclid  & + DESI + Euclid \\ 
\hline
$\Delta N_\mathrm{eff}^\mathrm{massless}$ & $<0.111$ (68\%CL)& $<0.054$ (68\%CL)& $<0.040$ (68\%CL)& $<0.038$ (68\%CL)& $<0.032$ (68\%CL)\\
$M_\nu$ (meV) & $84^{+25}_{-28}$& $71^{+16}_{-18}$& $68^{+15}_{-18}$& $68^{+15}_{-17}$& $67^{+14}_{-17}$\\
$\Omega_b h^2$ & $0.02234\pm0.00013$& $0.022301\pm0.000061$& $0.022290\pm0.000048$& $0.022289\pm0.000045$& $0.022282\pm0.000038$\\
$\Omega_c h^2$ & $0.1211^{+0.0007}_{-0.0013}$& $0.12043^{+0.00034}_{-0.00065}$& $0.12026^{+0.00029}_{-0.00050}$& $0.12023^{+0.00028}_{-0.00046}$& $0.12017^{+0.00027}_{-0.00040}$\\
$100\theta_s$ & $1.04060\pm0.00034$& $1.04070\pm0.00012$& $1.040700\pm0.000095$& $1.040700\pm0.000089$& $1.040800\pm0.000080$\\
$\tau$ & $0.0611\pm0.0046$& $0.0605\pm0.0021$& $0.0604\pm0.0021$& $0.0605\pm0.0021$& $0.0597\pm0.0020$\\
$n_s$ & $0.9678^{+0.0031}_{-0.0040}$& $0.9662\pm0.0021$& $0.9660\pm0.0019$& $0.9659\pm0.0018$& $0.9658\pm0.0017$\\
$ln(10^{10} A_s)$ & $3.100^{+0.008}_{-0.011}$& $3.0967\pm0.0043$& $3.0960\pm0.0041$& $3.0961\pm0.0041$& $3.0958\pm0.0039$\\
\hline
$H_0$ (km/s/Mpc) & $67.37^{+0.28}_{-0.42}$& $67.18\pm0.23$& $67.14\pm0.20$& $67.12\pm0.19$& $67.10\pm0.19$\\
$\sigma_8$ & $0.8314^{+0.0037}_{-0.0030}$& $0.8319^{+0.0034}_{-0.0026}$& $0.8318^{+0.0032}_{-0.0024}$& $0.8317^{+0.0032}_{-0.0023}$& $0.8318^{+0.0030}_{-0.0022}$\\

\hline
\end{tabular}}
\end{center}
\caption{$68\%$~CL constraints on cosmological parameters in the $\Lambda$CDM + $M_\nu$ + $ \Delta N_\mathrm{eff}^\mathrm{massless}$ model (accounting for standard massive neutrino plus extra massless relics, with $ \Delta N_\mathrm{eff}^\mathrm{massless}>0$) from the different CORE experimental specifications and with or without external data sets (DESI BAOs, Euclid cosmic shear). For Planck alone, we quote the results from the 2015 data release, while for combinations of Planck with future surveys, we fit mock data with a fake Planck likelihood mimicking the sensitivity of the real experiment (although a bit more constraining). }
\label{table:deltaneff}
\end{table}

\begin{table}[h]
\begin{center}\footnotesize
\scalebox{0.81}{\begin{tabular}{|c||c|c|c|c|c|}
\hline
Parameter & Planck, TEP & LiteCORE-80, TEP & LiteCORE-120, TEP & CORE-M5, TEP & COrE+, TEP\\
\hline
$N_\mathrm{eff}^\mathrm{massive}$ & $2.93\pm0.19$& $3.045\pm0.063$& $3.047\pm0.045$& $3.045\pm0.041$& $3.045\pm0.036$\\
$M_\nu$ (meV) & $<310$ (68\%CL)& $<110$ (68\%CL)& $73^{+37}_{-53}$& $73^{+37}_{-52}$& $72^{+37}_{-49}$\\
$\Omega_b h^2$ & $0.02208\pm0.00025$& $0.022250\pm0.000089$& $0.022255\pm0.000066$& $0.022254\pm0.000060$& $0.022255\pm0.000051$\\
$\Omega_c h^2$ & $0.1184\pm0.0030$& $0.1198\pm0.0011$& $0.11983\pm0.00082$& $0.11981\pm0.00077$& $0.11979\pm0.00071$\\
$100\theta_s$ & $1.04087\pm0.00046$& $1.04080\pm0.00016$& $1.04080\pm0.00012$& $1.04080\pm0.00011$& $1.04080\pm0.00010$\\
$\tau$ & $0.071\pm0.018$& $0.0604\pm0.0021$& $0.0604\pm0.0021$& $0.0604\pm0.0021$& $0.0603\pm0.0021$\\
$n_s$ & $0.9589\pm0.0095$& $0.9642\pm0.0036$& $0.9644\pm0.0031$& $0.9643\pm0.0030$& $0.9643\pm0.0028$\\
$ln(10^{10} A_s)$ & $3.071\pm0.037$& $3.0950\pm0.0048$& $3.0950\pm0.0045$& $3.0950\pm0.0045$& $3.0948\pm0.0043$\\
\hline
$H_0$ (km/s/Mpc) & $64.8\pm2.3$& $66.81^{+0.88}_{-0.71}$& $66.86^{+0.78}_{-0.59}$& $66.85^{+0.76}_{-0.58}$& $66.86^{+0.70}_{-0.55}$\\
$\sigma_8$ & $0.778^{+0.038}_{-0.024}$& $0.829^{+0.011}_{-0.007}$& $0.8291^{+0.0098}_{-0.0065}$& $0.8289^{+0.0094}_{-0.0066}$& $0.8291^{+0.0090}_{-0.0063}$\\

\hline
\end{tabular}}
\scalebox{0.81}{\begin{tabular}{|c||c|c|c|c|c|}
\hline
Parameter & Planck, TEP & LiteCORE-80, TEP & LiteCORE-120, TEP & CORE-M5, TEP & COrE+, TEP\\
 & + DESI & + DESI & + DESI & + DESI & + DESI\\ 
\hline
$N_\mathrm{eff}^\mathrm{massive}$ & $3.07\pm0.15$& $3.044\pm0.061$& $3.047\pm0.045$& $3.046\pm0.040$& $3.044\pm0.035$\\
$M_\nu$ (meV) & $74^{+35}_{-54}$& $65\pm25$& $66^{+24}_{-22}$& $65^{+24}_{-21}$& $61\pm21$\\
$\Omega_b h^2$ & $0.02228\pm0.00018$& $0.022257\pm0.000082$& $0.022258\pm0.000062$& $0.022257\pm0.000057$& $0.022251\pm0.000048$\\
$\Omega_c h^2$ & $0.1200\pm0.0025$& $0.1197\pm0.0010$& $0.11973\pm0.00075$& $0.11970\pm0.00068$& $0.12002\pm0.00059$\\
$100\theta_s$ & $1.04080\pm0.00045$& $1.04080\pm0.00016$& $1.04080\pm0.00012$& $1.04080\pm0.00011$& $1.040800\pm0.000097$\\
$\tau$ & $0.0608\pm0.0045$& $0.0603\pm0.0021$& $0.0603\pm0.0021$& $0.0603\pm0.0021$& $0.0604\pm0.0018$\\
$n_s$ & $0.9655\pm0.0065$& $0.9644\pm0.0032$& $0.9646\pm0.0028$& $0.9645\pm0.0026$& $0.9637\pm0.0024$\\
$ln(10^{10} A_s)$ & $3.096\pm0.012$& $3.0944\pm0.0049$& $3.0944\pm0.0045$& $3.0944\pm0.0044$& $3.0953\pm0.0038$\\
\hline
$H_0$ (km/s/Mpc) & $67.05\pm0.82$& $66.96\pm0.42$& $66.97\pm0.35$& $66.97\pm0.33$& $66.97\pm0.32$\\
$\sigma_8$ & $0.830\pm0.012$& $0.8307\pm0.0045$& $0.8305\pm0.0041$& $0.8305\pm0.0039$& $0.8304\pm0.0037$\\

\hline
\end{tabular}}
\scalebox{0.81}{\begin{tabular}{|c||c|c|c|c|c|}
\hline
Parameter & Planck, TEP & LiteCORE-80, TEP & LiteCORE-120, TEP & CORE-M5, TEP & COrE+, TEP\\
 & + DESI + Euclid & + DESI + Euclid & + DESI + Euclid & + DESI + Euclid  & + DESI + Euclid \\ 
\hline
$N_\mathrm{eff}^\mathrm{massive}$ & $3.05\pm0.11$& $3.044\pm0.057$& $3.046\pm0.042$& $3.046\pm0.039$& $3.045\pm0.034$\\
$M_\nu$ (meV) & $66^{+31}_{-35}$& $62\pm20$& $62\pm18$& $62\pm17$& $62^{+15}_{-17}$\\
$\Omega_b h^2$ & $0.02225\pm0.00016$& $0.022253\pm0.000081$& $0.022258\pm0.000062$& $0.022256\pm0.000055$& $0.022253\pm0.000047$\\
$\Omega_c h^2$ & $0.1198\pm0.0017$& $0.11976\pm0.00089$& $0.11978\pm0.00067$& $0.11978\pm0.00062$& $0.11977\pm0.00054$\\
$100\theta_s$ & $1.04080\pm0.00038$& $1.04080\pm0.00016$& $1.04080\pm0.00012$& $1.04080\pm0.00011$& $1.040800\pm0.000092$\\
$\tau$ & $0.0607\pm0.0045$& $0.0602\pm0.0021$& $0.0603\pm0.0021$& $0.0602\pm0.0021$& $0.0595\pm0.0020$\\
$n_s$ & $0.9646\pm0.0049$& $0.9644\pm0.0029$& $0.9644\pm0.0025$& $0.9644\pm0.0024$& $0.9644\pm0.0023$\\
$ln(10^{10} A_s)$ & $3.095\pm0.010$& $3.0944\pm0.0048$& $3.0946\pm0.0044$& $3.0945\pm0.0043$& $3.0944\pm0.0041$\\
\hline
$H_0$ (km/s/Mpc) & $66.97\pm0.54$& $66.96\pm0.32$& $66.98\pm0.27$& $66.98\pm0.25$& $66.97\pm0.23$\\
$\sigma_8$ & $0.8313^{+0.0039}_{-0.0029}$& $0.8316^{+0.0035}_{-0.0026}$& $0.8315^{+0.0033}_{-0.0026}$& $0.8315\pm0.0028$& $0.8315^{+0.0030}_{-0.0024}$\\

\hline
\end{tabular}}
\end{center}
\caption{Same as previous table, but for the $\Lambda$CDM + $M_\nu$ + $N_\mathrm{eff}^\mathrm{massive}$ model (accounting for non-thermalised active neutrinos degenerate in mass).}

\label{table:neff}
\end{table}

The results of our MCMC forecasts are shown in Tables~\ref{table:deltaneff}, \ref{table:neff}, and Figure~\ref{fig:neff}.  Since the determination of $N_{\rm eff}$ depends mainly on observations of the exponential tail in the CMB spectra, our results for $\sigma(N_{\rm eff})$ vary a lot with the  sensitivity/resolution assumed for CORE, and are only marginally affected by the inclusion of extra datasets like BAOs and cosmic shear surveys. The value $l_\mathrm{max}$ at which the signal-to-noise blows up in the temperature or polarisation spectrum varies a lot between the different experimental settings, as can be seen in Figure~\ref{fig:noise}. Thus there is a dramatic improvement in $\sigma(N_{\rm eff})$ between Planck and LiteCORE-80 (factor 3), and still a substantial one between LiteCORE-80 and COrE+ (factor 1.7). However, stepping back to the design of CORE-M5, one maintains a very good sensitivity, $\sigma(N_{\rm eff})=0.041$, only 10\% worse than what could be achieved with the better angular resolution of the COrE+ mission. Instead, LiteCORE-120 would be 25\% worse than COrE+. Hence CORE-M5 appears as a good compromise for the purpose of measuring $N_{\rm eff}$.

By achieving  $\sigma(N_{\rm eff})=0.041$ with CORE-M5 alone, or $\sigma(N_{\rm eff})=0.039$ in combination with future BAO data from DESI and/or cosmic shear data from Euclid, we could set very strong bounds on extra relics, neutrino properties, the temperature of reheating, etc., especially compared to Planck + DESI BAOs, which would only yield $\sigma(N_{\rm eff})=0.15$. To be more specific, let us consider the case of early decoupled thermal relics, like in Ref.~\cite{Baumann:2016wac}. Assuming that the last-decoupled relics leave thermal equilibrium at a temperature $T_\mathrm{F}$, and that the subsequent number of relativistic degrees of freedom is entirely accounted for by standard model particles, we notice that there are many well-motivated scenarios predicting a value of $\Delta N_{\rm eff}$ ranging from 0.05 to 0.3, because this corresponds to particles decoupling during the QCD phase transition. In case of a non-detection of extra relics by CORE, the 95\% exclusion bound from CORE + BAOs, $\Delta N_{\rm eff}< 0.076$, would exclude most of this range, while Planck + BAOs would not even touch it. 

A sensitivity of $\sigma(N_{\rm eff})=0.041$ would also have crucial implications for the determination of other important cosmological parameters, through a considerable reduction of parameter degeneracies.
For instance, without making assumptions on $N_{\rm eff}$, Planck + DESI BAOs would measure $H_0$ with 1.2\% uncertainty, and $\omega_\mathrm{cdm}$ with 2\% uncertainty. Figure~\ref{fig:neff} (left plot) shows that CORE-M5 would almost completely resolve the  ($N_{\rm eff}$, $H_0$) degeneracy, such that CORE + DESI BAOs would pinpoint both $H_0$  and $\omega_\mathrm{cdm}$ with 0.5\% uncertainty. This would have repercussions on several other parameters, and would allow to fully exploit the synergy between different types of cosmological data. Also, the determination of $N_{\rm eff}$ based on the observation of the CMB damping tails would reduce the uncertainty on the sound horizon angular scale, from $\sigma(\theta_s)=0.00046$ for Planck to $\sigma(\theta_s)=0.00011$ for CORE: hence the calibration of the sound horizon scale in future BAO data would be much more accurate, and the scientific impact of these observations (for instance, on Dark Energy models) would be enhanced.

\section{Constraints on the primordial Helium abundance \label{sec:Yp}}

In the framework of standard Big Bang Nucleosynthesis (BBN), the abundances of light elements can be 
calculated as a function of the baryon-to-photon ratio $\eta_\mathrm{b} \equiv n_\mathrm{b}/n_\gamma$, of the effective number of relativistic species $N_\mathrm{eff}$, 
and of the chemical potential of electron neutrinos (in the following, the latter is assumed to be zero). This is in particular the case
of the primordial abundance of $^4$He, that, changing the density of free electrons between helium and hydrogen recombination,
has a direct impact on CMB observables and in particular on the damping tail of CMB anisotropies (see e.g. 
\cite{Trotta:2003xg,Galli:2010it,Ichikawa:2007js,Ichikawa:2006dt,Hamann:2007sb}). In the other sections of this paper,
the $^4$He abundance, parameterized by $Y^\mathrm{BBN}_\mathrm{P}  \equiv 4 n_\mathrm{He}/n_\mathrm{b}$,
is calculated consistently as a function of the physical baryon density $\Omega_b h^2$ (that can be translated to $\eta_\mathrm{b}$ by fixing the photon temperature
and neglecting the uncertainty associated to the helium fraction itself) and $N_\mathrm{eff}$, using 
approximate analytical formulas based on the \texttt{PArthENoPE} code \cite{Pisanti:2007hk,Iocco:2008va}.
However, since the CMB is directly sensitive to $Y^\mathrm{BBN}_\mathrm{P}$, it is possible to drop the assumption of standard BBN
and obtain model-independent constraints on the abundance of $^4$He. This is the goal of this section, where we show the constraints
that can be obtained on $Y^\mathrm{BBN}_\mathrm{P}$ with different CORE configurations, in the framework of a minimal extension
of the standard $\Lambda$CDM model, as well as in the case where $N_\mathrm{eff}$ is also allowed to vary.
The fiducial model is the one described in Sec. \ref{sec:setup}, that assumes standard BBN (and vanishing neutrino chemical potential).
Thus the fiducial values $\omega_b = 0.022256$ and $N_\mathrm{eff} = 3.046$ imply $Y^\mathrm{BBN}_\mathrm{P} = 0.24669$.

\subsection{Sensitivity to the helium abundance in a minimal extension of $\Lambda$CDM}

In this section we present constraints assuming standard $\Lambda$CDM but without assuming standard BBN, so in addition
to the six $\Lambda$CDM we let also the primordial Helium
abundance parameter $Y^\mathrm{BBN}_\mathrm{P}$ to vary. The constraints on cosmological parameters for the different COrE experimental configurations are
reported in Table \ref{tab:params_yp}.

\begin{table}[h] 
\begin{center}
\scalebox{0.83}{\begin{tabular}{|c||c|c|c|c|}
\hline
Parameter    & LiteCORE-80, TEP    & LiteCORE-120, TEP    & CORE-M5, TEP    & COrE+, TEP \\
\hline
$Y_{\mathrm{P}}^{\mathrm{BBN}}$ &    $0.2466 \pm 0.0040$ &    $0.2466\pm0.0029$ &    $0.2466^{+0.0029}_{-0.0027}$ &    $0.2466\pm0.0025$ \\
 $\Omega_b h^2$ &    $0.022180\pm0.000078$ &    $0.022181\pm0.000059$ &    $0.022180\pm0.000055$ &    $0.022180\pm0.000047$ \\
 $\Omega_c h^2$ &    $0.12049\pm0.00032$ &    $0.12049\pm0.00029$ &    $0.12048^{+0.00030}_{-0.00027}$ &    $0.12048\pm0.00026$ \\
 $100 \theta_{MC}$ &    $1.04069\pm0.00016$ &    $1.04069\pm0.00012$ &    $1.04069\pm0.00011$ &    $1.04069\pm0.00011$ \\ 
 $\tau$ &    $0.0598\pm0.0020$ &    $0.0597\pm0.0020$ &    $0.0597^{+0.0019}_{-0.0022}$ &    $0.0597^{+0.0019}_{-0.0021}$ \\
 $n_s$ &    $0.9619\pm0.0028$ &    $0.9619\pm0.0025$ &    $0.9620\pm0.0024$ &    $0.9619\pm0.0023$ \\ 
 $\ln(10^{10}A_s)$ &    $3.0564\pm0.0039$ &    $3.0562\pm0.0036$ &    $3.0562\pm0.0036$ &    $3.0562\pm0.0035$ \\
 \hline
$H_0$ [km/s/Mpc]&    $66.95\pm0.16$ &    $66.95\pm0.14$ &    $66.95^{+0.13}_{-0.14}$ &    $66.95^{+0.12}_{-0.11}$ \\
 $\sigma_8$ &    $0.8174\pm0.0018$ &    $0.8172\pm0.0015$ &    $0.8173\pm0.0014$ &    $0.8172 \pm 0.0013$\\
 \hline
\end{tabular}}
\end{center}
\caption{Parameter constraints for $\Lambda$CDM + $Y^\mathrm{BBN}_\mathrm{P}$ (68\% CL uncertainties), for different CORE experimental configurations. }
\label{tab:params_yp}
\end{table}

\begin{figure}
	\centering
		\includegraphics[width=7cm]{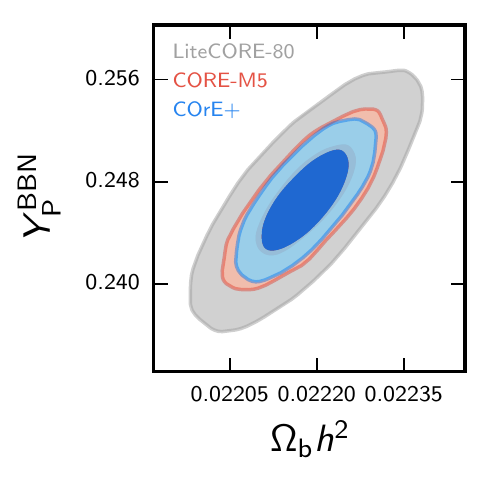}
		\includegraphics[width=7cm]{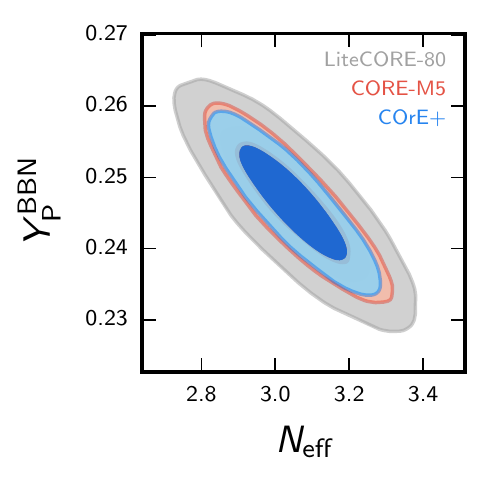}				
	\caption{Left Panel: Two-dimensional credible regions at 68\% and 95\% in the $Y^\mathrm{BBN}_\mathrm{P}$ vs $\omega_b$ plane for the different CORE configurations. The standard value for $N_\mathrm{eff}=3.046$ is assumed. Right Panel: Two-dimensional credible regions at 68\% and 95\% in the $Y^\mathrm{BBN}_\mathrm{p}$ vs $N_\mathrm{eff}$ plane for the different CORE configurations. $N_\mathrm{eff}$ is allowed to vary. }
\label{fig:params_yp_ob}
\end{figure}

It can be seen that the primordial abundance of $^4$He can be constrained with an uncertainty
$\sigma(Y^\mathrm{BBN}_\mathrm{P})=2.5\times 10^{-3}$ by the COrE+ configuration. This gets slightly degraded, by a factor
$\sim 1.2$ for the LiteCORE-120 and CORE-M5 configurations (the two configurations yield very similar results), but is significantly worse by $\sim 60 \%$ for LiteCORE-80. 
These numbers should be compared with the present bound from \textit{Planck} TT+lowP+BAO of $Y^\mathrm{BBN}_\mathrm{P} = 0.255^{+0.036}_{-0.038}$ 
(at 95\% CL) \cite{planck2015}. 
We find in all cases a dramatic improvement over the sensitivity to $Y^\mathrm{BBN}_\mathrm{P}$ from this combination of \textit{Planck} and BAO, 
gaining a factor of 4.6, 6.6 or 7.4 on this parameter, for LiteCORE-80, CORE-M5 or  COrE+, respectively.
Quite remarkably, the uncertainty on $Y^\mathrm{BBN}_\mathrm{P}$ for these CORE experimental configurations is at least two times smaller than the present observational error
in astrophysical determination of the same quantity: Ref. \cite{Aver:2013wba} reports $Y^\mathrm{BBN}_\mathrm{P} = 0.2465 \pm 0.0097$ (at 68\% CL) from a compilation of helium data. In Figure \ref{fig:params_yp_ob} (left panel) we show the 2D constraints in the $Y^\mathrm{BBN}_\mathrm{P}$ vs $\omega_b$ plane.

\subsection{Sensitivity to the helium abundance in $\Lambda$CDM+ $N_\mathrm{eff}$}

\begin{table}[h] 
\begin{center} 
\scalebox{0.83}{\begin{tabular}{|c||c|c|c|c|}
\hline
Parameter    & LiteCORE-80, TEP    & LiteCORE-120, TEP    & CORE-M5, TEP    & COrE+, TEP \\ 
\hline
$N_\mathrm{eff}$ &    $3.06\pm0.12$ &    $3.05\pm0.10$ &    $3.05\pm0.10$ &    $3.048\pm0.092$ \\ 
 $Y_{\mathrm{P}}^{\mathrm{BBN}}$ &    $0.2463\pm0.0069$ &    $0.2463\pm0.0057$ &    $0.2464\pm0.0056$ &    $0.2465\pm0.0052$ \\ 
 $\Omega_b h^2$ &    $0.022185^{+0.000085}_{-0.000084}$ &    $0.022182\pm0.000063$ &    $0.022181\pm0.000058$ &    $0.022183^{+0.000048}_{-0.000050}$ \\ 
 $\Omega_c h^2$ &    $0.1206\pm0.0018$ &    $0.1206\pm0.0015$ &    $0.1206\pm0.0014$ &    $0.1205\pm0.0013$ \\ 
 $100 \theta_{MC}$ &    $1.04067\pm0.00044$ &    $1.04068\pm0.00034$ &    $1.04068\pm0.00033$ &    $1.04068^{+0.00029}_{-0.00030}$ \\ 
 $\tau$ &    $0.0597 \pm 0.0020$ &    $0.0597\pm0.0020$ &    $0.0597\pm0.0020$ &    $0.0598 \pm 0.0020$ \\ 
 $n_s$ &    $0.9620^{+0.0034}_{-0.0033}$ &    $0.9619\pm0.0030$ &    $0.9620\pm0.0027$ &    $0.9619\pm0.0025$ \\ 
 $\ln(10^{10}A_s)$ &    $3.0565^{+0.0051}_{-0.0047}$ &    $3.0563\pm0.0042$ &    $3.0562\pm0.0042$ &    $3.0563^{+0.0040}_{-0.0043}$ \\ 
 \hline
$H_0$  [km/s/Mpc]&    $67.00^{+0.74}_{-0.75}$ &    $66.99^{+0.62}_{-0.64}$ &    $66.99\pm0.62$ &    $66.96\pm0.56$ \\ 
 $\sigma_8$ &    $0.8175\pm0.0037$ &    $0.8174^{+0.0030}_{-0.0031}$ &    $0.8174\pm0.0028$ &    $0.8173\pm0.0025$ \\ 
 \hline 
\end{tabular}}
\end{center}
\caption{Parameter constraints for $\Lambda$CDM + $Y^\mathrm{BBN}_\mathrm{P}$ + $N_\mathrm{eff}$ (68\% CL uncertainties), for different CORE experimental configurations. \label{tab:params_yp_neff}}
\end{table}

It is well known that there is a strong parameter degeneracy between $Y^\mathrm{BBN}_\mathrm{P}$ and $N_\mathrm{eff}$ 
(see e.g. \cite{planck2015} and references therein).
For this reason, in this section we present constraints on primordial $^4$He varying the six $\Lambda$CDM parameters, and letting also 
$Y^\mathrm{BBN}_\mathrm{P}$ and $N_\mathrm{eff}$ to vary.
The constraints on cosmological parameters for the different CORE experimental configurations are
reported in Table \ref{tab:params_yp_neff}.
As we can see, including variations in $N_\mathrm{eff}$ opens a degeneracy. The constraints
on $Y^\mathrm{BBN}_\mathrm{P}$ gets worse by roughly a factor of 2 for COrE+, CORE-M5 and LiteCORE-120, and slightly less than that
for LiteCORE-80, with respect to those obtained by fixing $N_\mathrm{eff}=3.046$. The improvement on $Y^\mathrm{BBN}_\mathrm{P}$ moving from
LiteCORE-80 to COrE+ is now a factor $\sim 1.4$, while it is about $\sim 1.1$ in moving from LiteCORE-120 or CORE-M5
to COrE+. These constraints are however still significantly stronger than those presently available from \textit{Planck}  \cite{planck2015}
or from astrophysical observations \cite{Aver:2013wba}. We note that the degeneracy also affects estimates of the effective number of relativistic species,
by greatly enlarging by nearly a factor two the uncertainty on this parameter when  $Y^\mathrm{BBN}_\mathrm{P}$ is left free to vary, as it can be seen
by comparing the numbers in Tables \ref{table:neff} and \ref{tab:params_yp_neff}.
This degeneracy is clearly seen in Fig. \ref{fig:params_yp_ob} (right panel) where we plot the two-dimensional credible regions in the
$Y^\mathrm{BBN}_\mathrm{p}$ vs $N_\mathrm{eff}$ plane. 

\subsection{Constraints on the neutron lifetime}

\begin{table}[h]
\scalebox{0.87}{\begin{tabular}{|c||c|c||c|c|}
\hline
 Parameter                             &    CORE-M5, TEP	& CORE-M5, TEP &   CORE-M5,TEP	& CORE-M5,TEP \\
                             &    	& + DESI &   	& + DESI \\

\hline
$ \Omega_{\mathrm{b}} h^2          $ & $ 0.022180\pm 0.000054             $& $0.022179  \pm 0.000053     $& $ 0.022180 \pm 0.000057             $&  $0.022179 \pm 0.000055$\\
$N_{\text{eff}}$ &$3.046$ &$3.046$ & $ 3.05 \pm 0.10$ & $3.049 \pm 0.084$\\
$Y_{\mathrm{p}}^{\mathrm{BBN}}$ & $0.2466 _{-0.0027} ^{+0.0029}$ & $0.2466 _{-0.0027}^{+0.0029}$& $0.2464 \pm 0.0056 $ & $0.2465 \pm 0.0050$\\ 
$ \tau_{\mathrm{n}} \, [\mathrm{sec}]                             $ & $ 880 \pm 13       $&$880 \pm 13$ & $  879 \pm 33  $ & $880 _{-29}^{+30}$\\
\hline
\end{tabular}}
\caption{Constraints on the neutron lifetime from CORE-M5 and CORE-M5+DESI under the assumption of BBN. The constraints reported in columns two and three 
have been derived under the assumption of the standard value $N_{\rm eff}=3.046$, while the results reported in the last two columns have been obtained 
assuming $N_{\rm eff}$ free.}
\label{tab:neutron}
\end{table}

Given the CORE-M5 constraints on parameters as the baryon density $\Omega_b h^2$, the Helium abundance $Y_p$ and the neutrino number of relativistic
relics $N_{\rm eff}$, it is possible to constrain the neutron lifetime under the assumption of BBN (\cite{Salvati:2015wxa}).
CMB data can indeed offer a completely independent determination of $\tau_n$, useful also for checking the validity of the cosmological scenario. 
In Table \ref{tab:neutron} we report the constraints on $\tau_n$ assuming BBN for CORE-M5 and CORE-M5+DESI in the case of $N_{\rm eff}=3.046$ and
$N_{\rm eff}$ free. As we can see, when $N_{\rm eff}=3.046$ CORE-M5 will constrain $\tau_n$ with an uncertainty of about  $\sim 1.5 \%$. Adding 
DESI will not improve significantly this bound. When $N_{\rm eff}$ is let free to vary, the CORE-M5 constraint will relax to about
$\sim 3 \%$ uncertainty, with a small improvement when the DESI dataset is included. Current laboratory data constrain the neutron lifetime
with a precision of $\sim 1 -2 s$ but with a $\sim 4.5$ $\sigma$ tension between different experiments with a difference of $\sim 9$s (see discussion in \cite{Salvati:2015wxa}). 
Future data from CORE could therefore help in clarifying the issue.

\section{Neutrino physics\label{sec:neutrinos}}

Neutrino oscillation data show that neutrinos must be massive, but the
data are insensitive to the absolute neutrino mass scale. For a normal hierarchy of masses ($m_1,
m_2 \ll m_3$), the mass summed over all eigenstates is approximately at least 
$60\,\mathrm{meV}$, while for an inverted hierarchy ($m_3 \ll m_1, m_2$) the
minimal summed mass is approximately $100\,\mathrm{meV}$~\cite{neutrinos1,neutrinos2,neutrinos3}. The individual neutrino masses in these
hierarchical limits are below the detection limit of current and future
laboratory $\beta$-decay experiments, but they can remarkably be probed by cosmology~\cite{Hu:1997mj,neutrinoreview,neutrinocosmo1,neutrinocosmo2,neutrinocosmo3,neutrinocosmo4,Hamann:2012fe,Lesgourgues:1519137,2015APh....63...66A}. 
The detection of the neutrino mass scale is even considered as one of the safest and most rewarding targets
of future cosmological surveys, since we know that these masses are non-zero, that they have a significant impact on structure formation, and that their measurement will bring an essential clue for particle physicists to decipher the neutrino sector puzzle (origin of masses, leptogenesis and baryogenesis, etc.). Even the unlikely case of a non-detection would be interesting, since it would force us to revise fundamental assumptions in particle physics and/or cosmology, see e.g.~\cite{neutrinoless}.

For individual neutrino masses below 600~meV, the non-relativistic transition of neutrinos takes place after photon decoupling. After that time 
the neutrino density scales like matter instead of radiation, with an impact on the late expansion history of the universe. This is important for
calculating the angular diameter distance to recombination, which determines the position of all CMB spectrum patterns in multipole space.
At the time of the non-relativistic transition, metric fluctuations experience a non-trivial evolution which can potentially impact the observed CMB spectrum in the range $50<\ell<200$ due to the early ISW effect~\cite{Lesgourgues:2012uu,Hou:2012xq,Lesgourgues:1519137}. However, for individual neutrino masses below 100 meV, the non-relativistic transition happens at $z<190$, hence too late to significantly affect the early ISW contribution. Finally, 
massive neutrinos slow down gravitational clustering on scales below the horizon size at
the non-relativistic transition, leaving a clear signature on the matter power spectrum~\cite{Bond:1980ha,Hu:1997mj,Lesgourgues:1519137}. The magnitude of this effect is controlled mainly by the summed neutrino mass $M_\nu$. Roughly speaking, the suppression occurs on wavenumbers $k \geq 0.01h/$Mpc (which means that even relatively large wavelengths are affected), and saturates for $k \geq 1h/$Mpc. Above this wavenumber and at redshift zero, the suppression factor is given in first approximation by $(M_\nu/10\,\mathrm{meV})$\%, i.e. at least 6\% even for minimal normal hierarchy~\cite{Hu:1997mj,Bird:2011rb,Lesgourgues:1519137}. CMB lensing is expected to be a particularly clean probe of this effect~\cite{Kaplinghat:2003bh,Lesgourgues:2005yv,Hall:2012kg,lensingreview}.

\subsection{Neutrino mass splitting\label{sec:splitting}}

Cosmology is mainly sensitive to the summed neutrino mass $M_\nu$, but the mass splitting does play a small role, since the free-streaming length of each neutrino mass eigenstate is determined by the individual masses~\cite{Lesgourgues:2004ps,neutrinoreview,neutrinocosmo3,Lesgourgues:1519137,2015APh....63...66A,neutrinocosmo4}. Hence, before doing forecasts for future high-precision experiments, it is worth checking the impact of making different assumptions of the mass splitting (for fixed total mass) on the results of a parameter extraction. If this impact is found to be small, we can perform generic forecasts sticking to one mass splitting scheme. Otherwise, several different cases should be considered separately.

We know from particle physics that there are two realistic neutrino mass schemes, NH and IH, both tending to a nearly-degenerate situation in the limit of large $M_\nu$, but that limit is already contradicting current bounds ($M_\nu < 210$~meV from Planck 2015 TT+lowP+BAO~\cite{planck2015}, $M_\nu < 140$~meV when including the latest Planck polarisation data~\cite{Aghanim:2016yuo}, $M_\nu < 130$~meV with recent BAO+galaxy survey data~\cite{Cuesta:2015iho} and $M_\nu < 120$~meV with BOSS Lyman-$\alpha$ data~\cite{Palanque-Delabrouille:2015pga}, all at 95\%CL). On top of NH and IH, the cosmological literature often discusses three unrealistic models (for the purpose of speeding up Boltzmann codes and integrating only one set of massive neutrino equations): the degenerate case with masses $(M_\nu/3, M_\nu/3, M_\nu/3)$, that we will call DEG; the case $(M_\nu/2, M_\nu/2, 0)$ that we will call 2M and the case $(M_\nu, 0, 0)$ that we will call 1M. These three unrealistic cases are potentially interesting to use as a fitting model in a forecast, because the total mass can be varied down to zero: thus, on top of estimating the value of $M_\nu$, one can assess the significance of the neutrino mass detection by comparing the probability of $M_\nu=0$ to that of the mean or best-fit value. Any of the DEG, 1M, or 2M models can achieve this purpose, however, we can already discard 1M and 2M, as a detailed inspection of the small difference between the matter power spectrum of these three models for fixed $M_\nu$ shows that the spectrum of the DEG model is much closer to that of the two realistic models (NH, IH) than the spectrum of 1M or 2M~(see e.g. Figure 16 in \cite{neutrinoreview}). Even current data starts to be slightly sensitive to the difference between 1M and (NH, IH)~\cite{Giusarma:2016phn}. Hence we only need to address the question: can we fit future data with the DEG model, even if the true underlying model is probably either NH or IH, or does this lead to an incorrect parameter reconstruction?

\begin{table}[!h]
\begin{center}
\footnotesize
\scalebox{0.95}{\begin{tabular}{| c || c | c | l |}
\hline
	Run & Fiducial model & Fitted model & posterior curve in Figure~\ref{fig:mass_splitting}\\
	\hline
	1. & NH with $M_\nu = 0.06$~eV & DEG & top panels, green\\
	2. & NH with $M_\nu = 0.06$~eV & NH & top panels, grey\\
	\hline
	3. & NH with $M_\nu = 0.10$~eV & DEG & bottom panels, solid green\\
	4. & NH with $M_\nu = 0.10$~eV & NH & bottom panels, solid grey\\
	5. & NH with $M_\nu = 0.10$~eV & IH & bottom panels, solid red\\
	\hline
	6. & IH with $M_\nu = 0.10$~eV & DEG & bottom panels, dashed green \\
	7. & IH with $M_\nu = 0.10$~eV & NH & bottom panels, dashed grey \\
	8. & IH with $M_\nu = 0.10$~eV & IH & bottom panels, dashed red \\
\hline
\end{tabular}
}
\end{center}
\caption{List of fiducial and fitted model used to check for possible parameter reconstruction bias when using the wrong assumptions on neutrino mass splitting. \label{tab:splitting_models}}
\end{table}

\begin{figure}
	\centering
		\includegraphics[width=5cm]{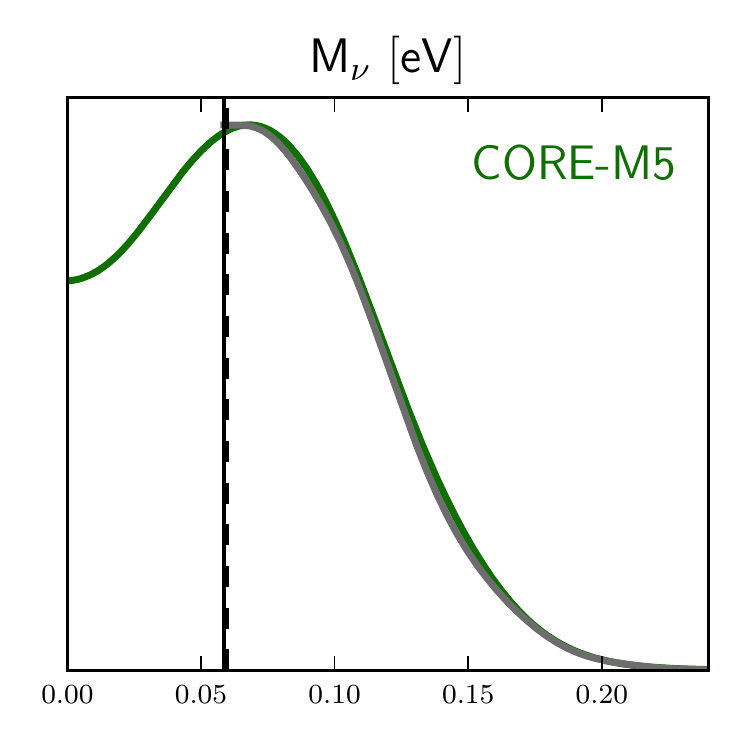}\includegraphics[width=5cm]{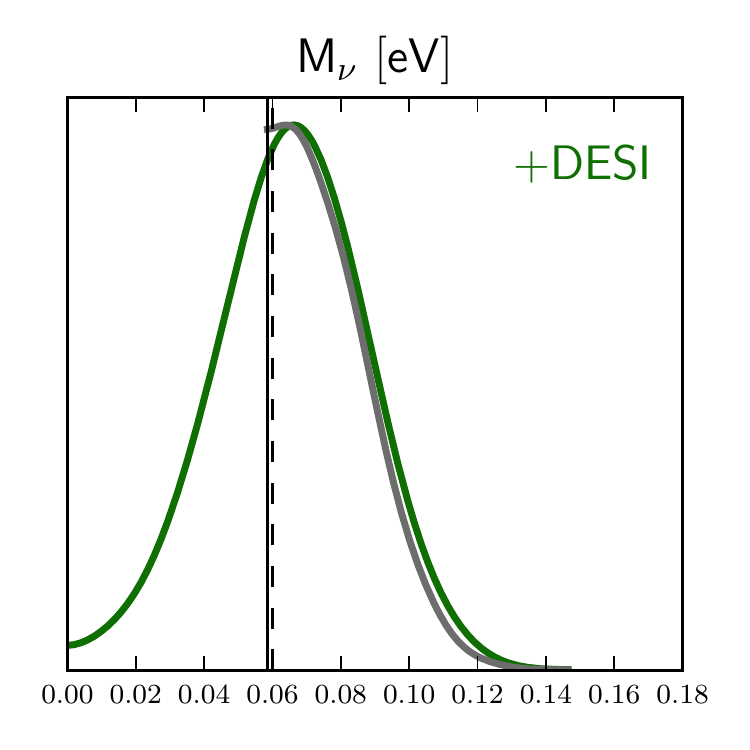}\includegraphics[width=5cm]{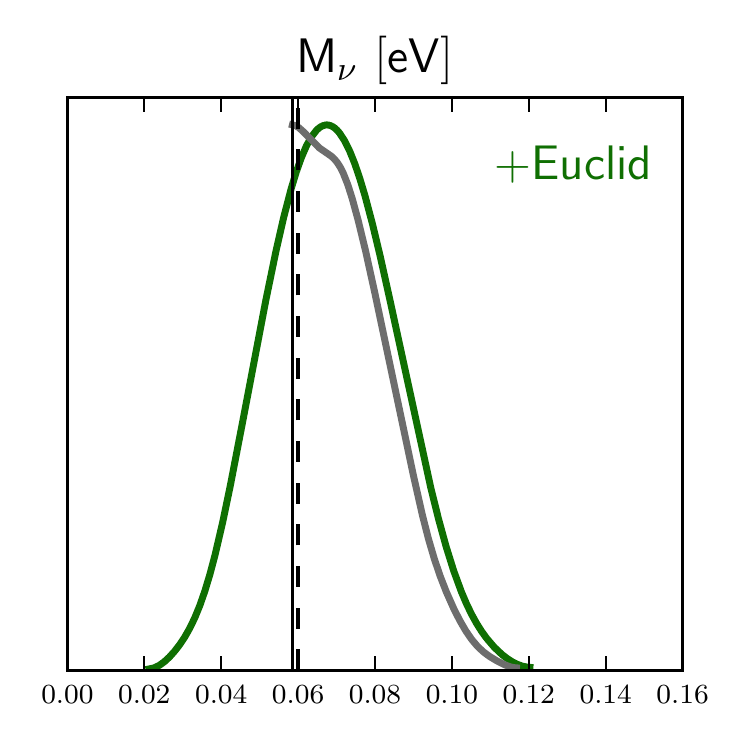}\\
		\includegraphics[width=5cm]{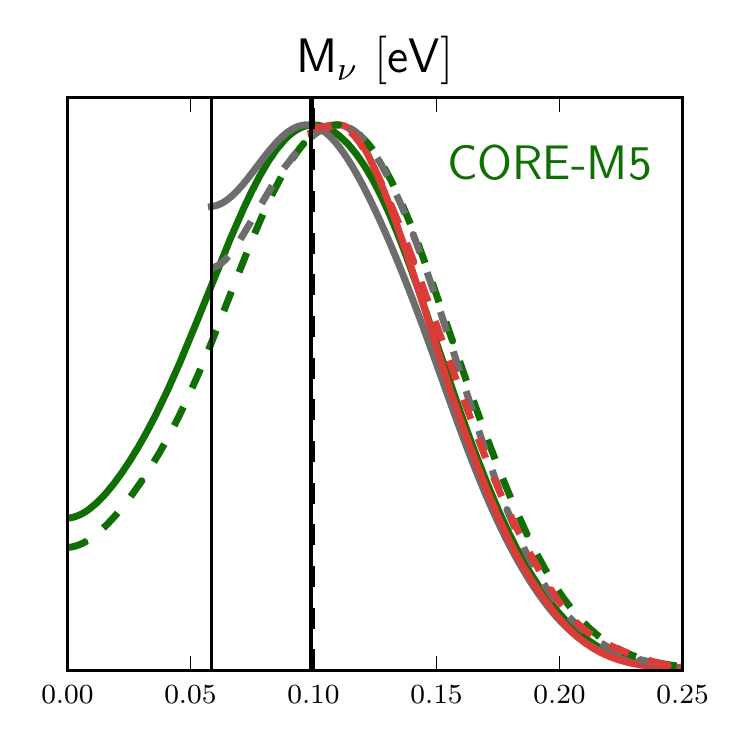}\includegraphics[width=5cm]{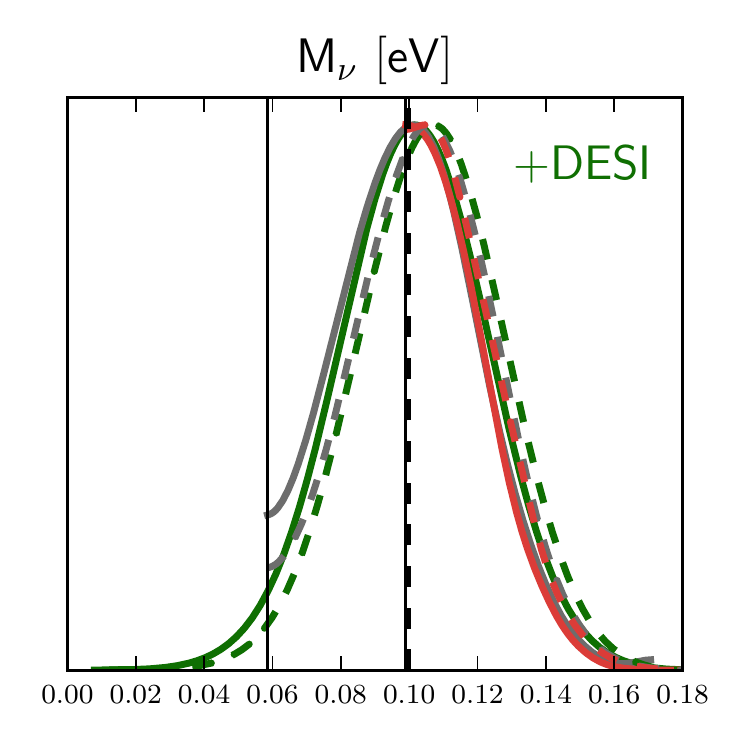}\includegraphics[width=5cm]{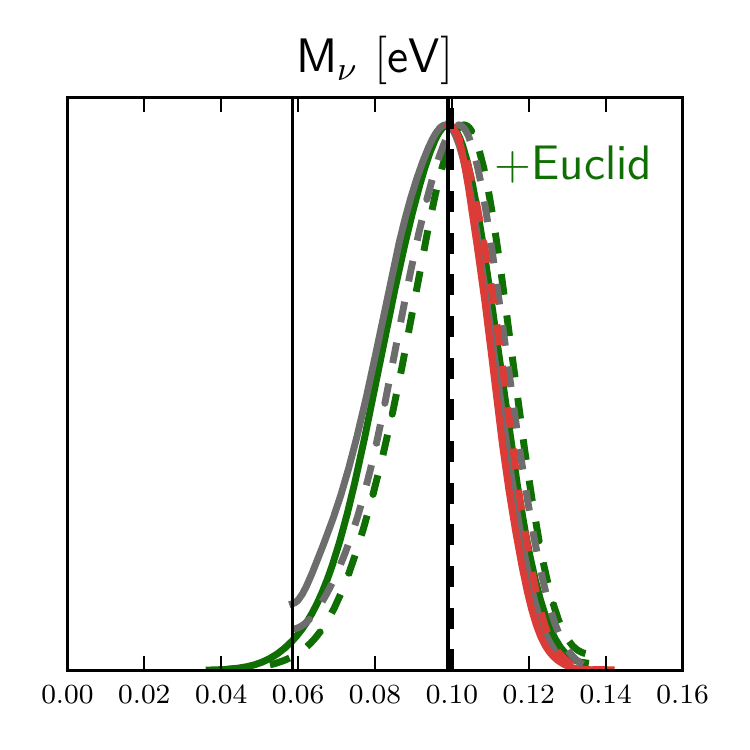}	
	\caption{Reconstruction of the total neutrino mass using various schemes for the mass splitting (NH, IH, degenerate), not always matching the assumed fiducial model. Table~\ref{tab:splitting_models} gives the explicit correspondence between the different curves and the assumptions made on the fiducial and fitting models. Vertical solid lines show lower prior edges in the NH and IH cases, while dashed lines show the fiducial values. The MCMC runs extracts the fiducial mass up to some reconstruction bias never exceeding 0.5$\sigma$.}
\label{fig:mass_splitting}
\end{figure}

We first consider a fiducial model with a total mass $M_\nu=60$~meV, thus necessarily given by NH. We generate mock data using the precise mass splitting of NH for such a value, with $\Delta m^2_{atm} = 2.45\times 10^{-3}$~(eV)$^2$ and $\Delta m^2_{sol} = 7.50\times 10^{-5}$~(eV)$^2$. We then compare the results of forecasts that assume either DEG or NH as a fitting model (still with fixed square mass differences). In both cases, the free parameters are the usual 6-parameter $\Lambda$CDM (with fiducial values given in footnote \ref{foot:dec_fid}) and $M_\nu$.
These two forecasts correspond to the first two lines in Table~\ref{tab:splitting_models}. The results for the CORE-M5 satellite\footnote{The next section will include a discussion of other possible sensitivity choices.}, alone or in combination with DESI BAOs and Euclid cosmic shear, are shown in the top three panels of Figure~\ref{fig:mass_splitting}. This is the most pessimistic case for measuring the neutrino mass, since it corresponds to the minimal total mass allowed by oscillation data. When looking at the results, one should keep in mind that we are fitting directly the fiducial spectrum, hence the posterior would peak at the fiducial value in absence of reconstruction bias; while with real scattered data the best fit would be shifted randomly, typically by one sigma. By looking at the results of the DEG fit (green curves in Figure~\ref{fig:mass_splitting} and numbers in Table~\ref{table:mnu}), we see that CORE-M5 alone would not detect $M_\nu=60$~meV with high significance, but it would typically achieve a 3$\sigma$ detection in combination with DESI BAOs, or a 4$\sigma$ detection when adding also Euclid cosmic shear data. There is a small offset between the mean value of $M_\nu$ found in the DEG fit and the fiducial value, corresponding respectively to 0.2$\sigma$, 0.2$\sigma$, 0.5$\sigma$ in the CORE, CORE+DESI, and CORE+DESI+Euclid cases. This can be attributed to bias reconstruction from assuming the wrong fitting model. However, in this situation, the conclusion of fitting real data with DEG would be that the preferred scenario is NH, since $M_\nu=100$~meV would be disfavoured typically at the 2$\sigma$ level by CORE+DESI+Euclid, and one would then perform a second fit assuming NH in order to eliminate this reconstruction bias. 
More detailed discussions on the discrimination power of future data between NH and IH can be found e.g. in \cite{neutrinocosmo3,neutrinocosmo4}.

Next, we considered a fiducial total mass $M_\nu=100$~meV, which could be achieved either within the NH or IH model. We are not interested in the possibility of directly discriminating between these two models, because the sensitivity of CORE+DESI+Euclid is clearly too low for such an ambitious purpose. Instead we only want to check whether using the DEG model for the fits introduces significant parameter bias. For that purpose, we perform six forecasts for each data set, corresponding to the two possible fiducial models (NH or IH) fitted by each of the three models DEG, IH or NH. We see on the lower panels of Figure~\ref{fig:mass_splitting} that the fiducial mass is again correctly extracted by the DEG fits, up to a bias ranging from 0.1$\sigma$ to 0.3$\sigma$: this is smaller than with a fiducial mass of 60~meV because masses are now larger and relative differences between NH, IH, and DEG are reduced. The error bars are always the same up to less than 0.1$\sigma$ differences.

We have checked that regardless of the real mass splitting realised in nature, and with the experimental data sets discussed in this analysis, we can correctly reconstruct the mass simply by fitting the DEG model to the data. For the purpose of our forecasts, the most important things to check are that the error is stable under different assumptions, and that the reconstruction bias induced by fitting DEG to NH or DEG to IH is under control: this is found to be the case. So the next forecasts can be done using either NH or IH as a fiducial, and sticking to DEG as the fitted model. We can even do something simpler and use DEG as both fiducial and fitted model in the forecasts, since we know that if the fiducial model was NH or IH we would not have a large bias. This is exactly what we will do in the next sections\footnote{More precisely, in sections~\ref{sec:mnu_seven} and \ref{sec:mnu_eight} we choose to fit DEG to NH when mock Euclid data is not used, and DEG to DEG otherwise.}. However, we also see that in future analyses, we ought to be a little bit more careful, and compare the results of different fits using either NH or IH as a fitted model, to assess the impact of different assumptions on the posterior probability for $M_\nu$.

\subsection{Neutrino mass sensitivity in a minimal 7-parameter model\label{sec:mnu_seven}}

Choosing the same fiducial model as in footnote \ref{foot:dec_fid}, with a summed mass equal to $M_\nu=60$~meV, we fit the 7-parameter $\Lambda$CDM+$M_\nu$ model for different CORE settings, alone or in combination with mock DESI BAOs and Euclid cosmic shear data.

\begin{table}[h]
\begin{center}\footnotesize
\scalebox{0.81}{\begin{tabular}{|c||c|c|c|c|c|}
\hline
Parameter & Planck, TEP & LiteCORE-80, TEP & LiteCORE-120, TEP & CORE-M5, TEP & COrE+, TEP\\
\hline
$M_\nu$ (meV) & $<315$ (68\%CL)& $78^{+36}_{-59}$& $74^{+38}_{-53}$& $72^{+37}_{-51}$& $72^{+38}_{-49}$\\
$\Omega_b h^2$ & $0.02219\pm0.00017$& $0.022250\pm0.000059$& $0.022256\pm0.000042$& $0.022256\pm0.000039$& $0.022255\pm0.000034$\\
$\Omega_c h^2$ & $0.1198\pm0.0015$& $0.11987^{+0.00050}_{-0.00071}$& $0.11983^{+0.00048}_{-0.00067}$& $0.11980^{+0.00043}_{-0.00065}$& $0.11981^{+0.00044}_{-0.00066}$\\
$100\theta_s$ & $1.04069\pm0.00036$& $1.04080\pm0.00010$& $1.040800\pm0.000087$& $1.040800\pm0.000082$& $1.040800\pm0.000079$\\
$\tau$ & $0.074\pm0.017$& $0.0604\pm0.0021$& $0.0604\pm0.0021$& $0.0604\pm0.0020$& $0.0604\pm0.0021$\\
$n_s$ & $0.9637\pm0.0051$& $0.9643\pm0.0020$& $0.9643\pm0.0018$& $0.9644\pm0.0018$& $0.9644\pm0.0018$\\
$ln(10^{10} A_s)$ & $3.081\pm0.033$& $3.0951\pm0.0042$& $3.0950\pm0.0041$& $3.0949\pm0.0040$& $3.0949\pm0.0041$\\
\hline
$H_0$ (km/s/Mpc) & $65.6^{+2.5}_{-1.4}$& $66.79^{+0.82}_{-0.44}$& $66.84^{+0.77}_{-0.40}$& $66.88^{+0.73}_{-0.39}$& $66.87^{+0.73}_{-0.39}$\\
$\sigma_8$ & $0.783^{+0.040}_{-0.021}$& $0.828^{+0.011}_{-0.006}$& $0.8288^{+0.0098}_{-0.0059}$& $0.8293^{+0.0095}_{-0.0057}$& $0.8291^{+0.0093}_{-0.0058}$\\

\hline
\end{tabular}}
\scalebox{0.81}{\begin{tabular}{|c||c|c|c|c|c|}
\hline
Parameter & Planck, TEP & LiteCORE-80, TEP & LiteCORE-120, TEP & CORE-M5, TEP & COrE+, TEP\\
 & + DESI & + DESI & + DESI & + DESI & + DESI\\ 
\hline
$M_\nu$ (meV) & $72^{+36}_{-48}$& $65\pm22$& $65^{+23}_{-20}$& $65^{+23}_{-19}$& $66^{+22}_{-20}$\\
$\Omega_b h^2$ & $0.02226\pm0.00012$& $0.022260\pm0.000053$& $0.022259\pm0.000041$& $0.022257\pm0.000039$& $0.022258\pm0.000033$\\
$\Omega_c h^2$ & $0.11966\pm0.00071$& $0.11970\pm0.00031$& $0.11971\pm0.00030$& $0.11971\pm0.00028$& $0.11971\pm0.00028$\\
$100\theta_s$ & $1.04080\pm0.00030$& $1.04080\pm0.00010$& $1.040800\pm0.000083$& $1.040800\pm0.000078$& $1.040800\pm0.000074$\\
$\tau$ & $0.0608\pm0.0044$& $0.0603\pm0.0021$& $0.0603\pm0.0021$& $0.0602\pm0.0021$& $0.0603\pm0.0020$\\
$n_s$ & $0.9647\pm0.0027$& $0.9646\pm0.0017$& $0.9646\pm0.0015$& $0.9646\pm0.0015$& $0.9645\pm0.0015$\\
$ln(10^{10} A_s)$ & $3.0954\pm0.0086$& $3.0944\pm0.0039$& $3.0944\pm0.0039$& $3.0944\pm0.0039$& $3.0944\pm0.0039$\\
\hline
$H_0$ (km/s/Mpc) & $66.94\pm0.28$& $66.98\pm0.27$& $66.98\pm0.26$& $66.98\pm0.26$& $66.97\pm0.26$\\
$\sigma_8$ & $0.829^{+0.012}_{-0.009}$& $0.8309\pm0.0043$& $0.8305\pm0.0039$& $0.8305\pm0.0038$& $0.8304\pm0.0037$\\

\hline
\end{tabular}}
\scalebox{0.81}{\begin{tabular}{|c||c|c|c|c|c|}
\hline
Parameter & Planck, TEP & LiteCORE-80, TEP & LiteCORE-120, TEP & CORE-M5, TEP & COrE+, TEP\\
 & + DESI + Euclid & + DESI + Euclid & + DESI + Euclid & + DESI + Euclid  & + DESI + Euclid \\ 
\hline
$M_\nu$ (meV) & $65\pm23$& $63^{+16}_{-18}$& $63^{+15}_{-17}$& $62\pm16$& $62^{+15}_{-17}$\\
$\Omega_b h^2$ & $0.02226\pm0.00011$& $0.022256\pm0.000052$& $0.022256\pm0.000040$& $0.022256\pm0.000037$& $0.022255\pm0.000032$\\
$\Omega_c h^2$ & $0.11977\pm0.00032$& $0.11978\pm0.00018$& $0.11977\pm0.00018$& $0.11977\pm0.00017$& $0.11979\pm0.00018$\\
$100\theta_s$ & $1.04080\pm0.00031$& $1.040800\pm0.000095$& $1.040800\pm0.000084$& $1.040800\pm0.000077$& $1.040800\pm0.000072$\\
$\tau$ & $0.0606\pm0.0046$& $0.0603\pm0.0020$& $0.0603\pm0.0021$& $0.0603\pm0.0020$& $0.0595\pm0.0020$\\
$n_s$ & $0.9644\pm0.0025$& $0.9645\pm0.0016$& $0.9645\pm0.0014$& $0.9645\pm0.0014$& $0.9645\pm0.0014$\\
$ln(10^{10} A_s)$ & $3.0951\pm0.0086$& $3.0945\pm0.0039$& $3.0947\pm0.0039$& $3.0947\pm0.0038$& $3.0944\pm0.0038$\\
\hline
$H_0$ (km/s/Mpc) & $66.96^{+0.23}_{-0.17}$& $66.97^{+0.21}_{-0.14}$& $66.97^{+0.21}_{-0.13}$& $66.98^{+0.20}_{-0.13}$& $66.98^{+0.19}_{-0.13}$\\
$\sigma_8$ & $0.8314^{+0.0039}_{-0.0030}$& $0.8316^{+0.0034}_{-0.0027}$& $0.8314^{+0.0033}_{-0.0026}$& $0.8315\pm0.0028$& $0.8316^{+0.0031}_{-0.0024}$\\

\hline
\end{tabular}}
\end{center}
\caption{$68\%$~CL constraints on cosmological parameters in the $\Lambda$CDM+$M_\nu$ model (accounting for the summed mass of standard neutrinos) from the different CORE experimental specifications and with or without external data sets (DESI BAOs, Euclid cosmic shear).
For Planck alone, we quote the results from the 2015 data release, while for combinations of Planck with future surveys, we fit mock data with a fake Planck likelihood mimicking the sensitivity of the real experiment (although a bit more constraining).}
\label{table:mnu}
\end{table}

\begin{figure}
	\centering
	\hspace{0cm}\includegraphics[width=6cm]{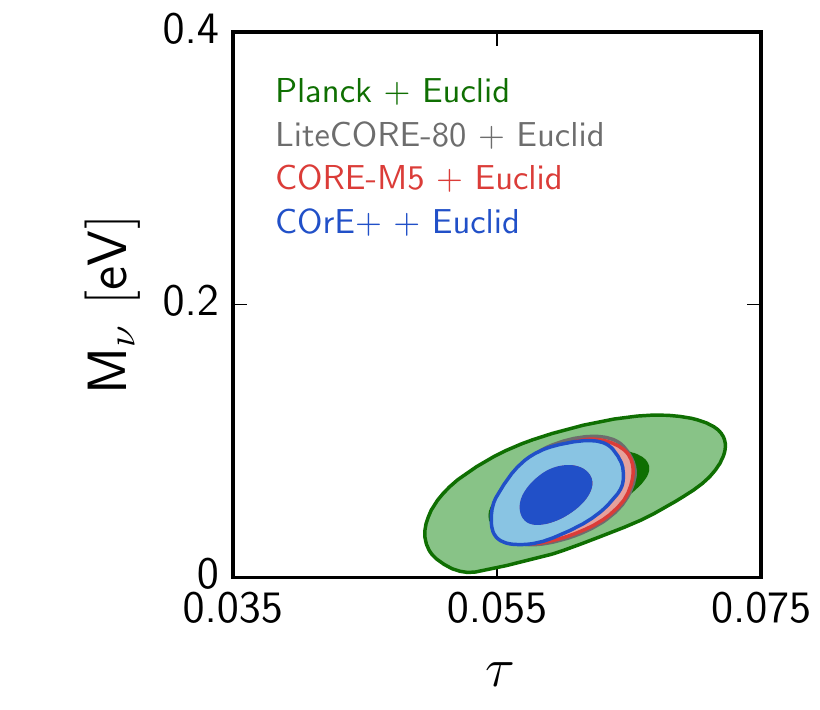}
	\hspace{-11.0cm}\includegraphics[width=6cm]{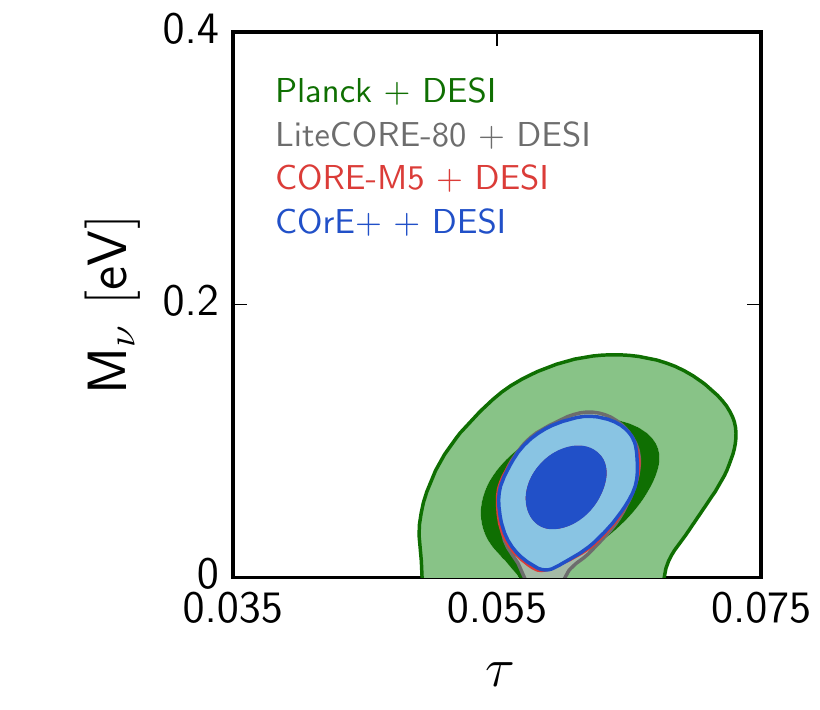}
	\hspace{-11.0cm}\includegraphics[width=6cm]{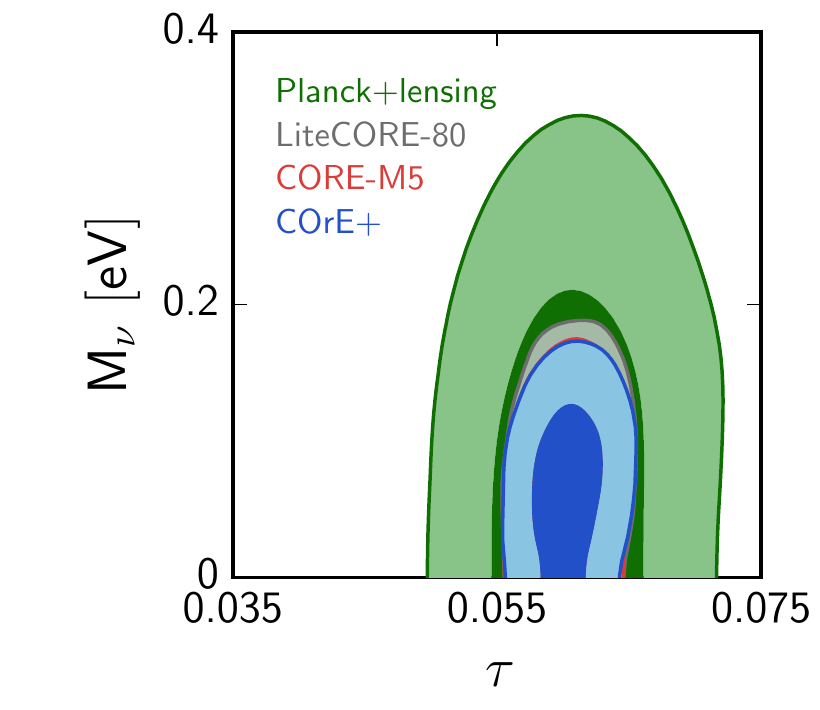}
	\caption{Results for the minimal model with massive neutrinos (discussed in section~\ref{sec:mnu_seven} and Table~\ref{table:mnu}).}
\label{fig:mnu-tau}
\end{figure}

\begin{figure}
	\centering
	\hspace{0cm}\includegraphics[width=6cm]{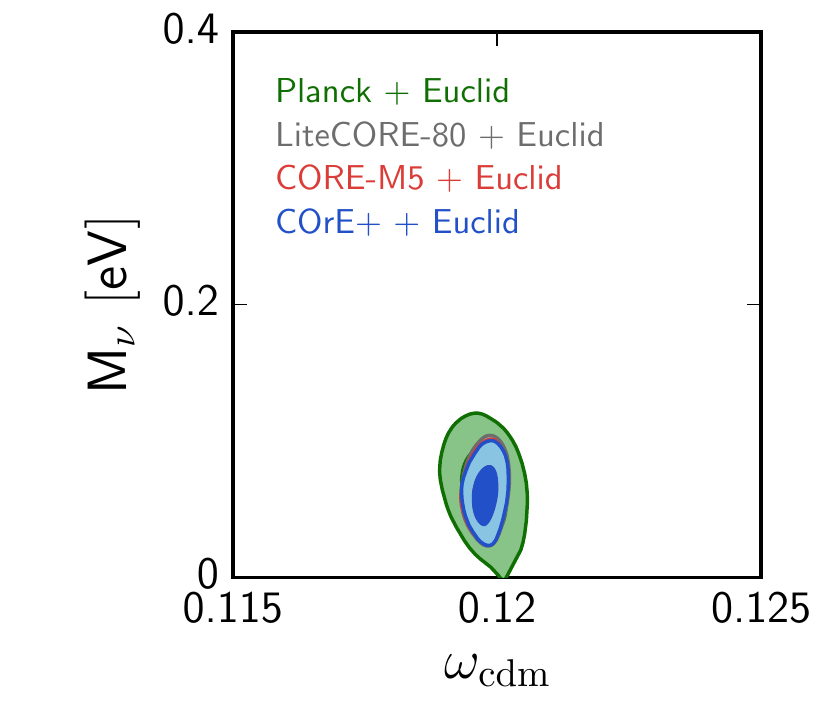}
	\hspace{-11.0cm}\includegraphics[width=6cm]{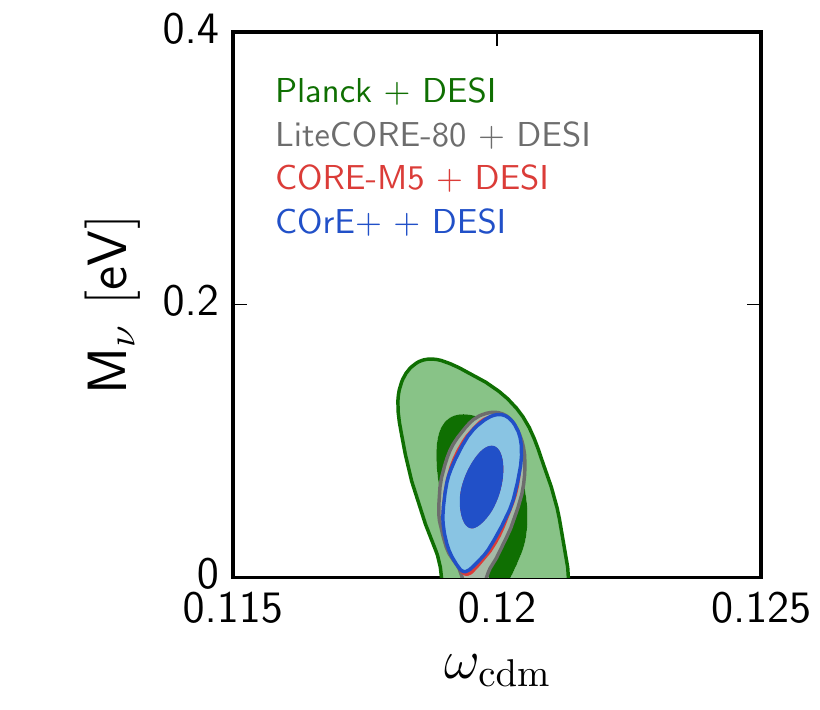}
	\hspace{-11.0cm}\includegraphics[width=6cm]{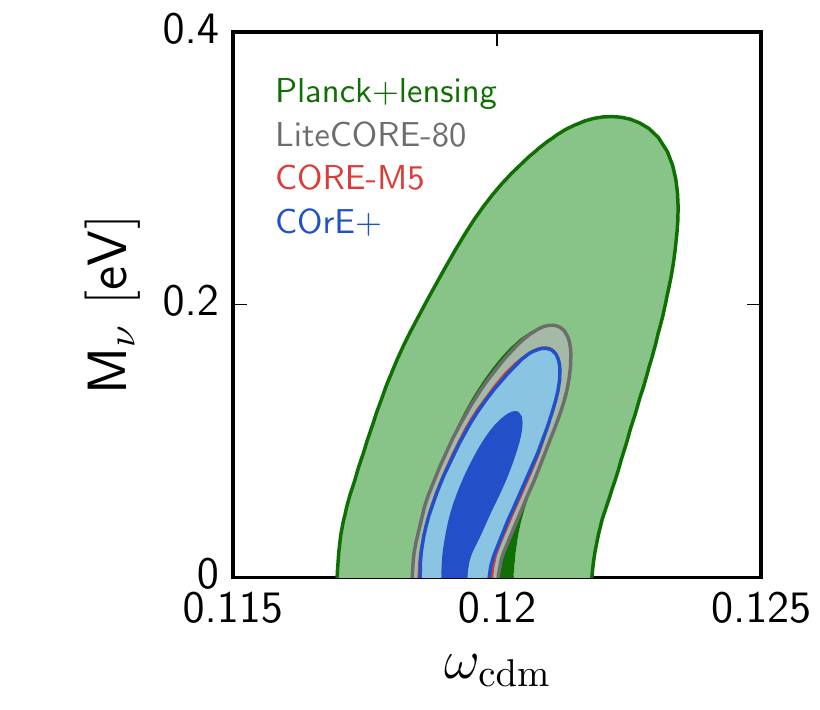}
		\caption{Results for the minimal model with massive neutrinos (discussed in section~\ref{sec:mnu_seven} and Table~\ref{table:mnu}).}
\label{fig:mnu-ocdm}
\end{figure}

Since we are looking at very small individual masses (mainly in the range $m_\nu<100$~meV), we expect the sensitivity of the CMB to $M_\nu$ to be dominated by CMB lensing effects. The different CORE settings considered here lead to different sensitivities to the CMB lensing potential. However, we only observe marginal differences between the forecasted mass sensitivities shown in Table~\ref{table:mnu}, with a symmetrized error ranging from 48~meV for LiteCORE-80 to 44~meV for CORE-M5 and COrE+. The reason is that the neutrino mass effect on the CMB lensing potential does not peak at the highest multipoles: rather it consists of a nearly constant suppression for a wide range of angular scales with $l>100$. Hence, in order to achieve a good detection of $M_\nu$, it is sufficient to have data in the region where the signal-to-noise ratio (S/N) is the largest, which is roughly from $\ell=200$ to 700 for CMB lensing. Lensing extraction on smaller angular scale will always have a smaller S/N and would bring little additional information. In the range $200<\ell<700$, LiteCORE-80 has a slightly worse sensitivity to the CMB lensing spectrum than other settings considered here, and hence a larger $\sigma(M_\nu)$; the other settings mainly differ for $\ell>700$. We conclude that the determination of $M_\nu$ cannot drive the choice between different possible CORE settings, unlike the determination of other parameters (e.g. tensor-to-scalar ratio, $N_\mathrm{eff}$) that critically depend on the sensitivity and/or resolution of the instrument.

However, a next-generation CMB satellite is essential for getting such tight bounds on the summed neutrino mass, because of its potential to measure small-scale polarisation and to constrain the optical depth to reionization $\tau$ (this is true for all CORE configurations). Indeed, the suppression induced by neutrino masses in the CMB lensing potential could be nearly cancelled by an increase in the primordial spectrum amplitude $A_s$. Since the product $e^{-2 \tau} A_s$ is fixed by the global amplitude of the CMB temperature/polarisation spectra, increasing $A_s$ requires increasing $\tau$. Future ground-based CMB experiments would only marginally improve on the $\tau$ determination from Planck, due to their limited sky coverage and large sampling variance for small multipoles. Hence, they would be affected by an ($M_\nu$, $\tau$) degeneracy for the reasons discussed above. To prove the importance of this effect we repeated the forecast for CORE-M5, cutting however all polarisation information for $\ell < 30$, and replacing it by a gaussian prior on $\tau$ with the sensitivity of Planck, $\sigma(\tau) \simeq 0.01$. We did find a degeneracy between $M_\nu$ and $\tau$ and the error bar on the summed mass degraded by a factor 2. Instead, we can clearly see in the left panel of Figure~\ref{fig:mnu-tau} that there is no such degeneracy, neither in the Planck-alone contours, caused by the too weak sensitivity to the CMB lensing spectrum, nor in CORE-alone contours because they break this degeneracy by measuring $\tau$ with good enough precision.

We can check how the combination of CMB data with other probes can achieve better constraints with CORE than with Planck. We find that CORE+DESI BAOs is about two times more constraining than Planck+DESI. This is related again to the better CMB lensing spectrum extraction {\it and} optical depth measurement by CORE. There are actually two ways to compensate the CMB lensing spectrum suppression induced by neutrino masses: by increasing $A_s$ and $\tau$, or by increasing $\omega_\mathrm{cdm}$~\cite{Archidiacono:2016lnv}. This leads to a strong ($M_\nu$, $\omega_\mathrm{cdm}$) degeneracy when using only CMB data (Figure~\ref{fig:mnu-ocdm}, left plot). However, future BAO data will fix $\omega_\mathrm{cdm}$ with very good accuracy. In the Planck+BAO case, the ($M_\nu$, $\tau$) degeneracy would then still remain (Figure~\ref{fig:mnu-tau}, middle plot). In the CORE+BAO case, with $\omega_\mathrm{cdm}$ fixed by BAOs and $\tau$ nearly fixed by polarisation measurements, very little degeneracies remain: in Figure~\ref{fig:mnu-tau}, middle plot, we just see a small positive correlation controlled by the error bar on $\tau$. Hence CORE will powerfully exploit the synergy between CMB and BAO measurements for measuring the neutrino mass. The combination with Euclid will further reduce degeneracies and errors by independently measuring the lensing spectrum at smaller redshifts than CORE. Even with very conservative assumptions on Euclid (i.e. including only cosmic shear data for $k\leq0.5h$/Mpc) we find that CORE+DESI+Euclid would have a sensitivity of $\sigma(M_\nu)=16$~meV, almost guaranteeing at least a 4$\sigma$ detection.

This claim relies on a 7-parameter forecast only, so we should still check its robustness against non-minimal assumptions on the cosmological model. 

\subsection{Degeneracy between neutrino mass and other parameters in extended 8-parameter models\label{sec:mnu_eight}}

In the previous section we found a sensitivity of about $\sigma(M_\nu)=44$~meV for CORE-M5, using any configuration, or 21~meV in combination with future BAOs, and 16~meV with future cosmic shear data. We explained why the sensitivity to $M_\nu$ has a very weak dependence on the assumed instrumental settings for CORE. To check how much these predictions depend on the assumed cosmological model, we do several extended forecasts with 8 free parameters instead of 7. 

The new parameters studied here are the primordial helium fraction, the tensor-to-scalar ratio, the constant Dark Energy equation of state parameter, the primordial scalar tilt running, and the effective density fraction of spatial curvature. Since our focus here is on neutrino masses, we do not investigate the sensitivity to these parameters in as much detail as in the sections devoted to them. For instance, we use here a (weak energy principle) prior $w>-1$, while in the Dark Energy section we will also consider phantom Dark Energy or a time-varying $w$. Also, as in the rest of this paper, we stick to a mock CORE likelihood including only temperature, E-polarisation and lensing data, and not using B-mode information: hence we obtain much worse constraints on $r$ than in the companion ECO paper on inflation~\cite{ECOinflation}, in which B modes play an essential role; but at least the present forecast allows to conservatively prove the absence of parameter correlation between $M_\nu$ and $r$ at the level of precision of CORE combined with DESI and Euclid. 

\begin{table}[h]
\begin{center}\footnotesize
\scalebox{0.82}{\begin{tabular}{|c||c|c|c|c|c|}
\hline
Parameter & Planck, TEP & LiteCORE-80, TEP & LiteCORE-120, TEP & CORE-M5, TEP & COrE+, TEP\\
\hline
$Y_\mathrm{He}$ & $0.247\pm0.014$& $0.2475\pm0.0042$& $0.2476\pm0.0030$& $0.2476\pm0.0028$& $0.2476\pm0.0025$\\
$M_\nu$ (meV) & fixed& $77^{+36}_{-60}$& $74^{+37}_{-53}$& $74^{+37}_{-52}$& $73^{+37}_{-51}$\\

\hline
$r$ & $<0.055$ (68\%CL)& $<0.020$ (68\%CL)& $<0.019$ (68\%CL)& $<0.020$ (68\%CL)& $<0.020$ (68\%CL)\\
$M_\nu$ (meV) & fixed& $<95$ (68\%CL)& $<97$ (68\%CL)& $68^{+33}_{-52}$& $66^{+31}_{-51}$\\

\hline
$w \,\,\, (>-1)$ & $<-0.74$ (68\%CL)& $<-0.886$ (68\%CL)& $<-0.895$ (68\%CL)& $<-0.900$ (68\%CL)& $<-0.904$ (68\%CL)\\
$M_\nu$ (meV) & fixed& $<73$ (68\%CL)& $<72$ (68\%CL)& $<72$ (68\%CL)& $<71$ (68\%CL)\\

\hline
$10^2 \, d\ln n_s/d \ln k$ & $-0.20\pm0.67$& $-0.01\pm0.30$& $0.00\pm0.25$& $-0.00\pm0.25$& $0.00\pm0.23$\\
$M_\nu$ (meV) & fixed& $<110$ (68\%CL)& $74^{+37}_{-53}$& $74^{+37}_{-52}$& $72^{+37}_{-50}$\\

\hline
$10^2 \, \Omega_k$ & $-0.37^{+0.83}_{-0.69}$& $-0.37^{+0.60}_{-0.33}$& $-0.33^{+0.56}_{-0.31}$& $-0.30^{+0.51}_{-0.29}$& $-0.28^{+0.48}_{-0.30}$\\
$M_\nu$ (meV) & fixed& $<200$ (68\%CL)& $<197$ (68\%CL)& $<193$ (68\%CL)& $<188$ (68\%CL)\\

\hline
\end{tabular}}
\scalebox{0.82}{\begin{tabular}{|c||c|c|c|c|c|}
\hline
Parameter & Planck, TEP & LiteCORE-80, TEP & LiteCORE-120, TEP & CORE-M5, TEP & COrE+, TEP\\
 & + DESI & + DESI & + DESI & + DESI & + DESI\\ 
\hline
$Y_\mathrm{He}$ & $0.248\pm0.010$& $0.2477\pm0.0041$& $0.2477\pm0.0030$& $0.2477\pm0.0027$& $0.2477\pm0.0024$\\
$M_\nu$ (meV) & $70^{+33}_{-51}$& $65^{+25}_{-21}$& $65^{+23}_{-20}$& $65\pm21$& $66^{+23}_{-19}$\\

\hline
$r$ & $<0.062$ (68\%CL)& $<0.020$ (68\%CL)& $<0.019$ (68\%CL)& $<0.020$ (68\%CL)& $<0.020$ (68\%CL)\\
$M_\nu$ (meV) & $72^{+37}_{-47}$& $62^{+26}_{-21}$& $62^{+23}_{-21}$& $63\pm22$& $63^{+23}_{-19}$\\

\hline
$w \,\,\, (>-1)$ & $<-0.961$ (68\%CL)& $<-0.965$ (68\%CL)& $<-0.9653$ (68\%CL)& $<-0.965$ (68\%CL)& $<-0.965$ (68\%)\\
$M_\nu$ (meV) & $<66$ (68\%CL)& $<56$ (68\%CL)& $<57$ (68\%CL)& $<60$ (68\%CL)& $<57$ (68\%CL)\\

\hline
$10^2 \, d\ln n_s/d \ln k$ & $-0.01\pm0.58$& $0.01\pm0.31$& $0.01\pm0.25$& $0.01\pm0.24$& $-0.00\pm0.23$\\
$M_\nu$ (meV) & $71^{+35}_{-48}$& $66^{+25}_{-22}$& $66^{+24}_{-21}$& $66^{+24}_{-20}$& $61^{+22}_{-19}$\\

\hline
$10^2 \, \Omega_k$ & $0.07\pm0.19$& $0.01\pm0.11$& $0.00\pm0.10$& $-0.00\pm0.10$& $0.00\pm0.10$\\
$M_\nu$ (meV) & $<122$ (68\%CL)& $66\pm31$& $65^{+31}_{-27}$& $65^{+29}_{-26}$& $66\pm28$\\

\hline
\end{tabular}}
\scalebox{0.82}{\begin{tabular}{|c||c|c|c|c|c|}
\hline
Parameter & Planck, TEP & LiteCORE-80, TEP & LiteCORE-120, TEP & CORE-M5, TEP & COrE+, TEP\\
 & + DESI + Euclid & + DESI + Euclid & + DESI + Euclid & + DESI + Euclid  & + DESI + Euclid \\ 
\hline
$Y_\mathrm{He}$ & $0.2470\pm0.0091$& $0.2477\pm0.0040$& $0.2477\pm0.0029$& $0.2477\pm0.0027$& $0.2477\pm0.0024$\\
$M_\nu$ (meV) & $63\pm26$& $63\pm17$& $63\pm16$& $62\pm16$& $62^{+15}_{-17}$\\

\hline
$r$ & $<0.063$ (68\%CL)& $<0.020$ (68\%CL)& $<0.020$ (68\%CL)& $<0.020$ (68\%CL)& $<0.020$ (68\%CL)\\
$M_\nu$ (meV) & $62\pm23$& $60^{+16}_{-18}$& $60^{+15}_{-17}$& $60^{+15}_{-17}$& $60^{+15}_{-17}$\\

\hline
$w \,\,\, (>-1)$ & $<-0.9914$ (68\%CL)& $<-0.9920$ (68\%CL)& $<-0.9921$ (68\%CL)& $<-0.9923$ (68\%)& $<-0.9926$ (68\%)\\
$M_\nu$ (meV) & $44^{+23}_{-29}$& $45^{+23}_{-19}$& $46^{+22}_{-18}$& $48^{+22}_{-17}$& $48^{+21}_{-17}$\\

\hline
$10^2 \, d\ln n_s/d \ln k$ & $0.01\pm0.52$& $0.01\pm0.29$& $0.01\pm0.24$& $0.00\pm0.23$& $0.01\pm0.22$\\
$M_\nu$ (meV) & $63\pm25$& $63\pm17$& $63\pm16$& $63\pm16$& $63\pm16$\\

\hline
$10^2 \, \Omega_k$ & $0.03\pm0.14$& $0.012\pm0.100$& $-0.002\pm0.094$& $0.004\pm0.098$& $-0.003\pm0.096$\\
$M_\nu$ (meV) & $70^{+34}_{-30}$& $65\pm22$& $62\pm22$& $63\pm21$& $62\pm21$\\

\hline
\end{tabular}}
\end{center}
\caption{$68\%$~CL constraints on the additional parameters of several extended 8-parameter models,
 for the different CORE experimental specifications, and with or without external data sets (DESI BAOs, Euclid cosmic shear). For Planck alone, we quote the results from the 2015 data release, {\it obtained with a fixed mass} $M_\nu=60$~meV, while for combinations of Planck with future surveys, we fit mock data with a fake Planck likelihood mimicking the sensitivity of the real experiment (although a bit more constraining). 
 In the case with free tensor-to-scalar ratio $r$, we did not include B-modes in the likelihood, unlike in the companion ECO paper on inflation~\cite{ECOinflation}. In the case with free $w$ we used a (weak energy principle) prior $w>-1$, that will be relaxed in the Dark Energy section of this paper.}
\label{tab:mnu_ext}
\end{table}

\begin{figure}
	\centering
		\includegraphics[width=5cm]{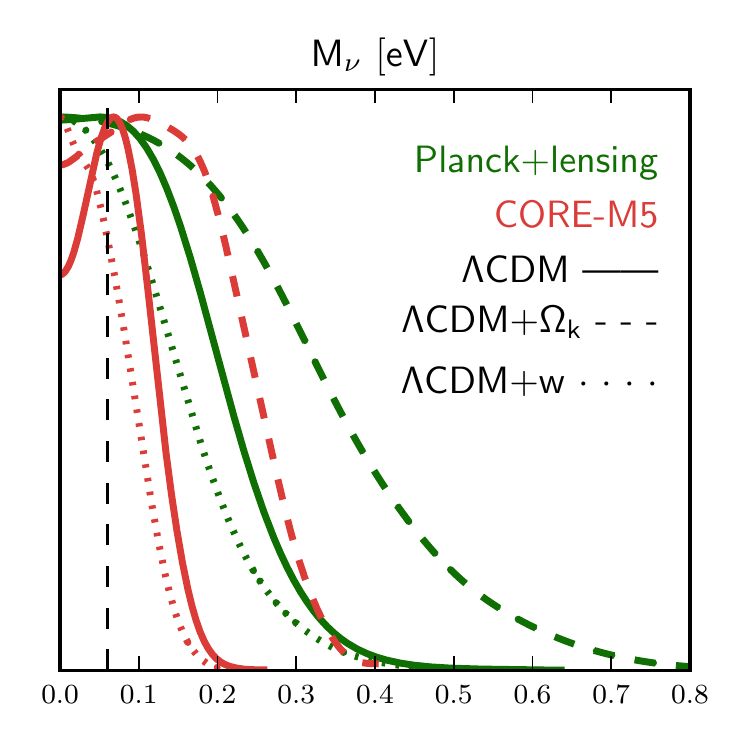} \hspace{-0.4cm}
		\includegraphics[width=5cm]{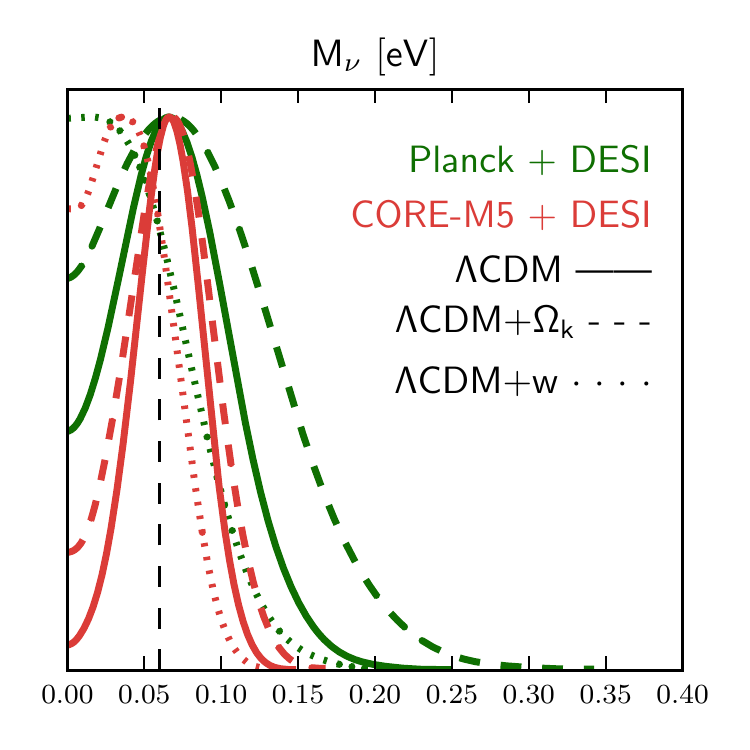} \hspace{-0.4cm}
		\includegraphics[width=5cm]{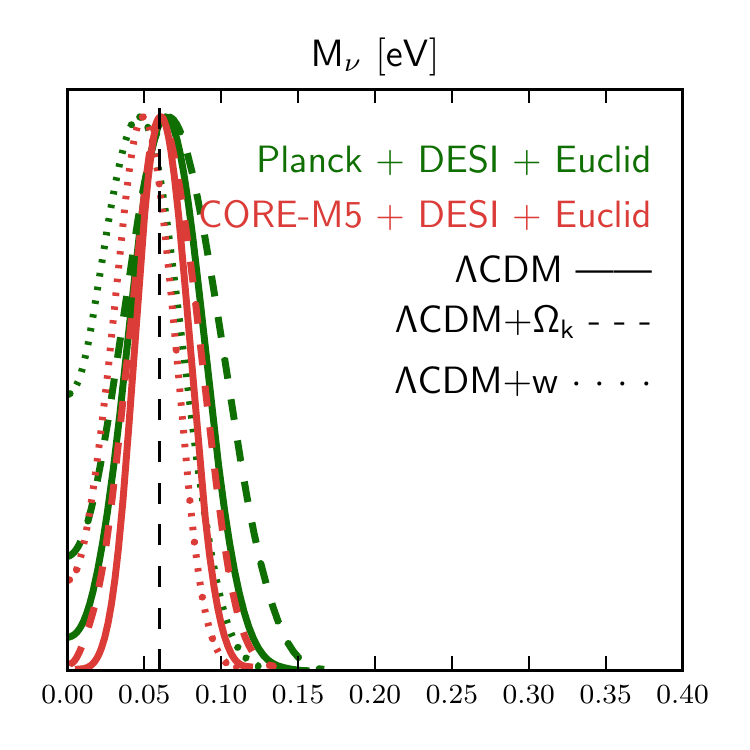}
	\caption{Posterior distribution of the summed neutrino mass in the extended models $\Lambda$CDM + $M_\nu$, $\Lambda$CDM + $M_\nu$ + $w$ and $\Lambda$CDM + $M_\nu$ + $\Omega_k$. The vertical dashed line shows the fiducial value.}
\label{fig:mnu-ok-w}
\end{figure}

\begin{figure}
	\centering
		\hspace{0cm}\includegraphics[width=6cm]{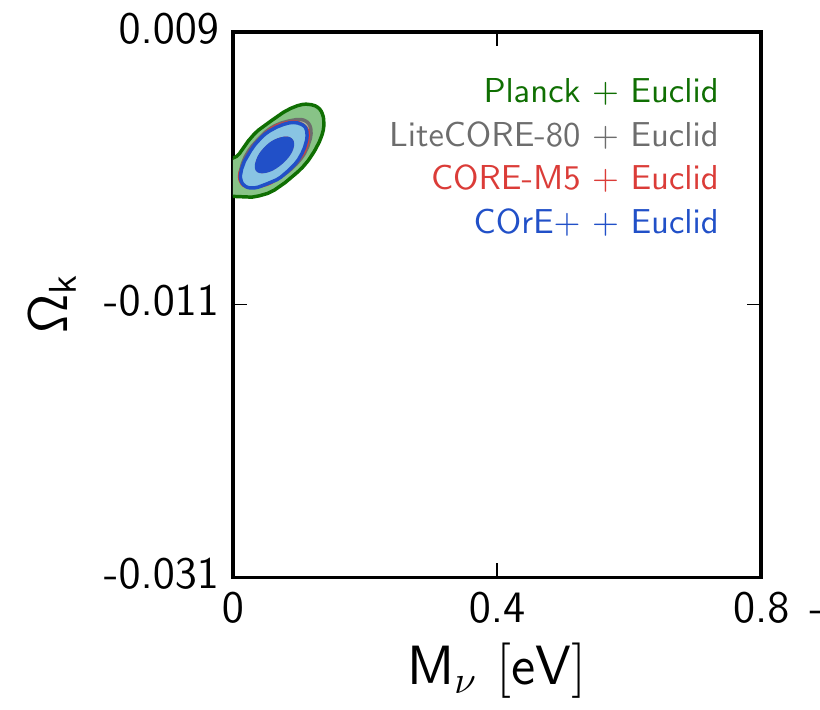}
		\hspace{-11.4cm}\includegraphics[width=6cm]{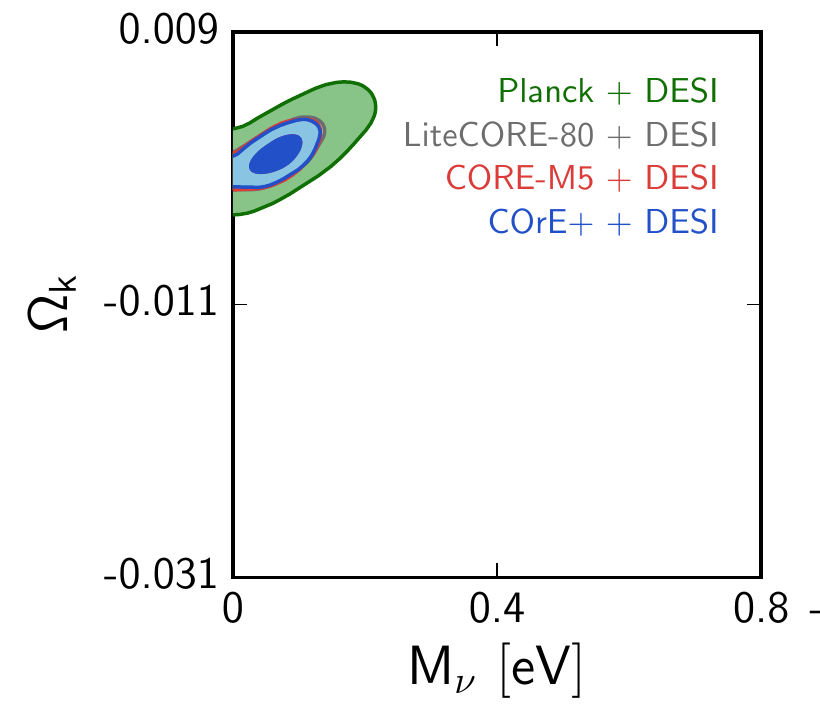}
		\hspace{-11.4cm}\includegraphics[width=6cm]{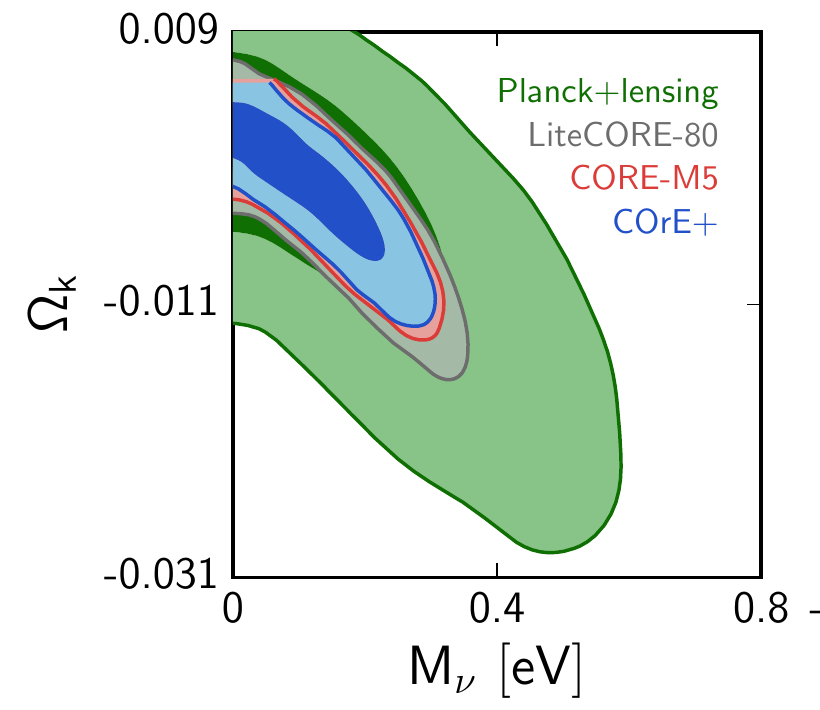}
	\caption{Results for the extended model $\Lambda$CDM + $M_\nu$ + $\Omega_k$. The $(M_\nu, \Omega_k)$ degeneracy is removed by adding BAO data..}
\label{fig:omegak}
\end{figure}

\begin{figure}
	\centering
		\hspace{0cm}\includegraphics[width=6cm]{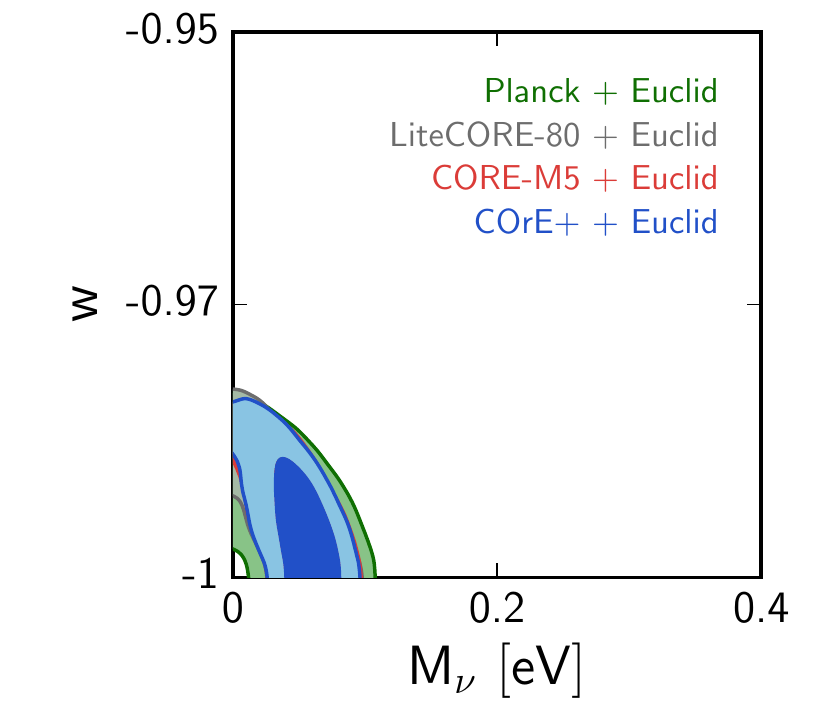}
		\hspace{-11.2cm}\includegraphics[width=6cm]{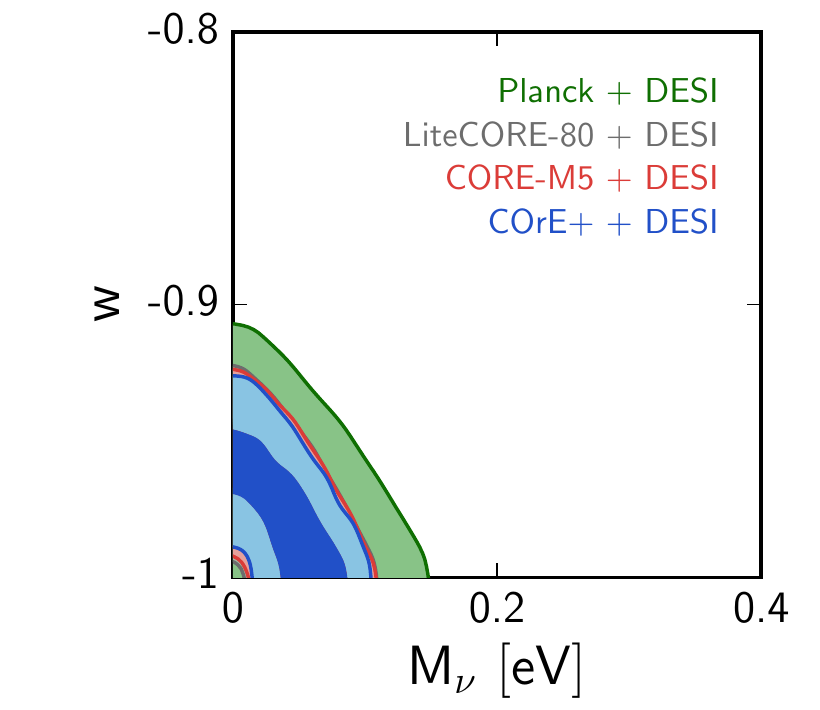}
		\hspace{-11.2cm}\includegraphics[width=6cm]{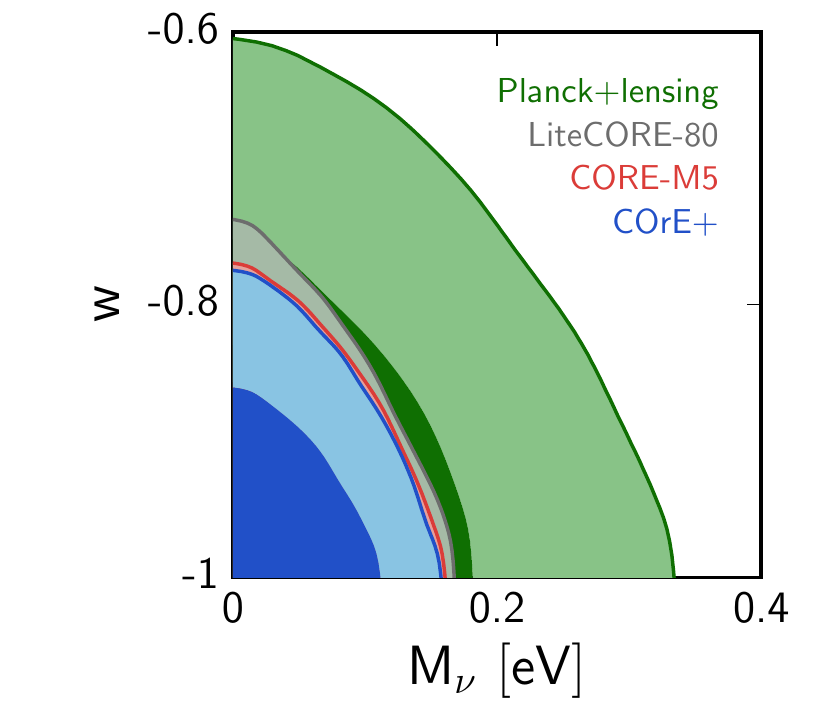}	
	\caption{Results for the extended model $\Lambda$CDM + $M_\nu$ + $w$ (with a prior $w>-1$). The $w$ axis scale changes between plots because of the huge difference of sensitivity between data sets. The $(M_\nu, w)$ degeneracy gets partially resolved by adding Euclid cosmic shear data.}
\label{fig:mnu-w}
\end{figure}

Our extended forecast results are summarised in Table~\ref{tab:mnu_ext}. When varying the helium fraction, the tensor-to-scalar ratio\footnote{With a free $r$ and in the forecasts based on CMB data alone, the error bar $\sigma(M_\nu)$ can be slightly smaller in the extended model than in the 7-parameter model, which may sound odd. In fact, it comes from a volume effect in Bayesian parameter extraction: models with $r\neq0$ are less discrepant with the data when $M_\nu$ is small, so after marginalising over $r$ the posterior for $M_\nu$ is shifted to lower values.}, or the tilt running, we find essentially  the same sensitivity to $M_\nu$ as in the 7-parameter model. Nonetheless, the cases with free $w$ or $\Omega_k$ make the neutrino mass detection more difficult, due to clear parameter degeneracies with $M_\nu$ when using CMB data alone (see Figures \ref{fig:mnu-ok-w}, \ref{fig:omegak}, \ref{fig:mnu-w}). 

We see in  Figures~\ref{fig:mnu-ok-w}, \ref{fig:omegak} that the ($M_\nu$, $\Omega_k$) degeneracy (a particular case of the {\it geometrical degeneracy} described in \cite{Smith:2006nk,Howlett:2012mh}) gets broken by the inclusion of BAO data, bringing the error down to $\sigma(M_\nu)\simeq 28$~meV. With additional Euclid cosmic shear data, one would reach $\sigma(M_\nu)\simeq21$~meV, still guaranteeing a 3$\sigma$ detection, while Planck+DESI+Euclid could only achieve $\sigma(M_\nu)\simeq32$~meV for free $\Omega_k$. 

In the case with free $w$ (Figures~\ref{fig:mnu-ok-w}, \ref{fig:mnu-w}), the degeneracy remains problematic even with CORE+BAO data, but ultimately Euclid cosmic shear data could partly differentiate between the physical effects of $w$ and $M_\nu$ effects and lead to $\sigma(M_\nu)\simeq19$~meV under the prior $w>-1$, instead of 26~meV for Planck+DESI+BAO. The error bar would degrade by also allowing for phantom dark energy, but on the other hand, the inclusion of further Large Scale Structure data (e.g. the Euclid galaxy correlation function) would further help to break the degeneracy, since the effect of neutrino masses and $w$ have a different dependence on redshift and scales ~\cite{Hamann:2012fe}.

\subsection{Light sterile neutrinos}

Right-handed or sterile neutrinos are present in several well-motivated extensions of the standard model of particle physics~\cite{sterile0,Adhikari:2016bei}. If their mass is of the order of a few keV or bigger, they can play the role of warm or cold dark matter, and they are constrained mainly by X-ray and Lyman-alpha observations~\cite{Adhikari:2016bei}. If their mass is of the order of the meV or smaller, they will simply behave as extra relativistic relics contributing to $N_\mathrm{eff}$. There is another interesting range deserving a specific study: that of light sterile neutrinos with a mass in the meV to eV range. Such particles have been extensively discussed over the past years, for the reason that the oscillations between such sterile neutrinos and active neutrinos (or more precisely, between the mass eigenstates formed of active and sterile neutrinos) could explain a number of possible anomalies in short-baseline neutrino oscillation data (see e.g.~\cite{Giunti:2015wnd}).

Sterile neutrinos with large mixing angles would normally acquire a thermal distribution through oscillations with active neutrinos, and their mass would then be very constrained (essentially, as much as that of active neutrinos). However, the explanation of short baseline anomalies requires an ${\cal O}(1)$~eV mass in tension with cosmological data. To avoid these bounds, people have discussed several ways to prevent sterile neutrino thermalisation (see e.g.~\cite{sterile0,sterile1,Archidiacono:2016kkh}). In that case, the bounds on the sterile neutrino mass become model-dependent, but a wide category of models can be parametrised in good approximation with two numbers ($N_{\rm s}$, $m_{\rm s}^{\rm eff}$), related to the asymptotic density at early times, given by $\Delta N_{\rm eff} = N_{\rm s}$,  and the asymptotic density at late times, given by the effective mass $m_{\rm s}^{\rm eff} = 94.1 \omega_s$~eV \cite{planck2013,planck2015}, where $\omega_s$ is the sterile neutrino density. This covers both the case of light early-decoupled thermal relics, and that of Dodelson-Widrow (i.e. non-resonantly produced) sterile neutrinos. For the later case, the physical mass of the sterile neutrino is given by $m_s = m_{\rm s}^{\rm eff} / N_{\rm s}$. 

To investigate the sensitivity of CORE to a non-thermal sterile neutrino, we stick to the same fiducial model as in the last subsections (total mass $M_\nu=60$~meV and  $N_{\rm eff}=3.046$), but we now fit it with an extended model with 9 free parameters, including the summed mass of active neutrinos  $M_\nu^\mathrm{active}$, as well as $N_{\rm s}$ and $m_{\rm s}^{\rm eff}$.
We impose in our forecasts a top-hat prior $m_{\rm s}^{\rm eff} / N_{\rm s} < 5$~eV, designed to eliminate models such that the extra species has a large mass, a very small number density, and behaves like extra cold dark matter.

\begin{table}[h]
\begin{center}\footnotesize
\scalebox{0.87}{\begin{tabular}{|c||c|c|c|c|c|}
\hline
Parameter & Planck, TEP & LiteCORE-80, TEP & LiteCORE-120, TEP & CORE-M5, TEP & COrE+, TEP\\
\hline
$m_\mathrm{eff}^\mathrm{sterile}$ (meV) & $<332$ & $<56$ & $<41$ & $<37$ & $<34$ \\
$\Delta N_\mathrm{eff}^\mathrm{sterile}$ & $<0.20$ & $<0.080$ & $<0.058$ & $<0.053$ & $<0.048$ \\
$M_\nu^\mathrm{active}$ (meV) & fixed& $<79$ & $<76$ & $<77$ & $<82$ \\

\hline
\end{tabular}}
\scalebox{0.87}{\begin{tabular}{|c||c|c|c|c|c|}
\hline
Parameter & Planck, TEP & LiteCORE-80, TEP & LiteCORE-120, TEP & CORE-M5, TEP & COrE+, TEP\\
 & + DESI & + DESI & + DESI & + DESI & + DESI\\ 
\hline
$m_\mathrm{eff}^\mathrm{sterile}$ (meV) & $<123$ & $<52$ & $<42$ & $<39$ & $<35$ \\
$\Delta N_\mathrm{eff}^\mathrm{sterile}$ & $<0.168$ & $<0.077$ & $<0.059$ & $<0.054$ & $<0.049$ \\
$M_\nu^\mathrm{active}$ (meV) & $<65$ & $43^{+21}_{-33}$& $45^{+24}_{-28}$& $47\pm26$& $48\pm26$\\

\hline
\end{tabular}}
\scalebox{0.87}{\begin{tabular}{|c||c|c|c|c|c|}
\hline
Parameter & Planck, TEP & LiteCORE-80, TEP & LiteCORE-120, TEP & CORE-M5, TEP & COrE+, TEP\\
 & + DESI  & + DESI  & + DESI  & + DESI   & + DESI  \\ 
 &  + Euclid &  + Euclid &  + Euclid &  + Euclid  &  + Euclid \\ 
\hline
$m_\mathrm{eff}^\mathrm{sterile}$ (meV) & $<60$ & $<43$ & $<37$ & $<35$ & $<33$ \\
$\Delta N_\mathrm{eff}^\mathrm{sterile}$ & $<0.151$ & $<0.074$ & $<0.057$ & $<0.054$ & $<0.049$ \\
$M_\nu^\mathrm{active}$ (meV) & $<63$ & $43^{+23}_{-26}$& $44\pm23$& $44\pm22$& $44^{+22}_{-20}$\\

\hline
\end{tabular}}
\end{center}
\caption{$68\%$~CL constraints and upper bounds on cosmological parameters in the $\Lambda$CDM + $M_\nu^\mathrm{active}$ + $m_\mathrm{eff}^{sterile}$ + $\Delta N_\mathrm{eff}^{sterile}$ model (accounting for massive active neutrinos plus one light and non-thermalised sterile neutrino) from the different CORE experimental specifications and with or without external data sets (DESI BAOs, Euclid cosmic shear). For Planck alone, we quote the results from the 2015 data release, {\it obtained with a fixed active neutrino mass} $M_\nu^\mathrm{active}=60$~meV, while for combinations of Planck with future surveys, we fit mock data with a fake Planck likelihood mimicking the sensitivity of the real experiment (although a bit more constraining). For concision, we only show the bounds for the extended model parameters.}
\label{tab:mnu_sterile}
\end{table}

\begin{figure}
	\centering
		\hspace{0cm}\includegraphics[width=6cm]{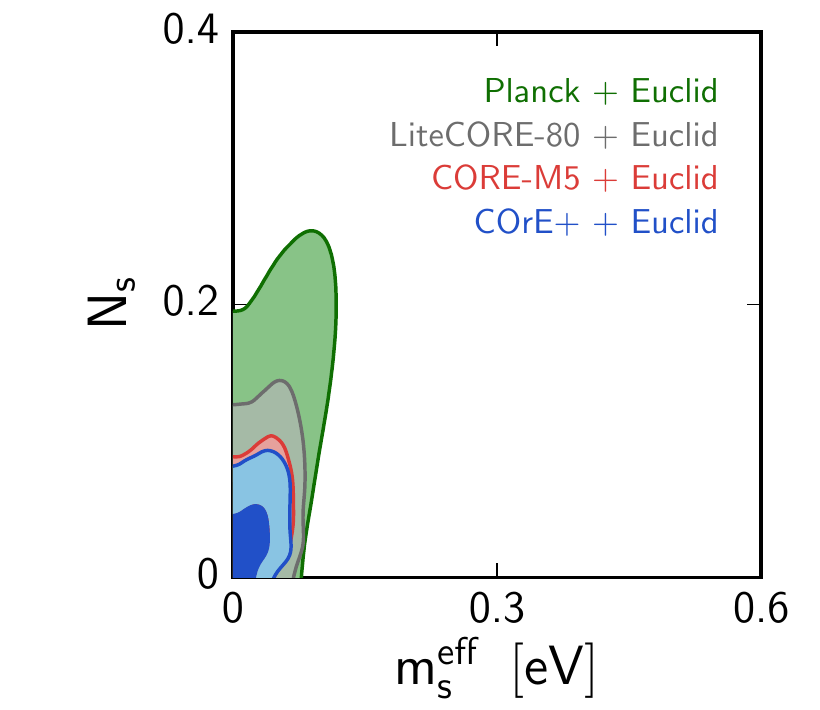}
		\hspace{-11.1cm}\includegraphics[width=6cm]{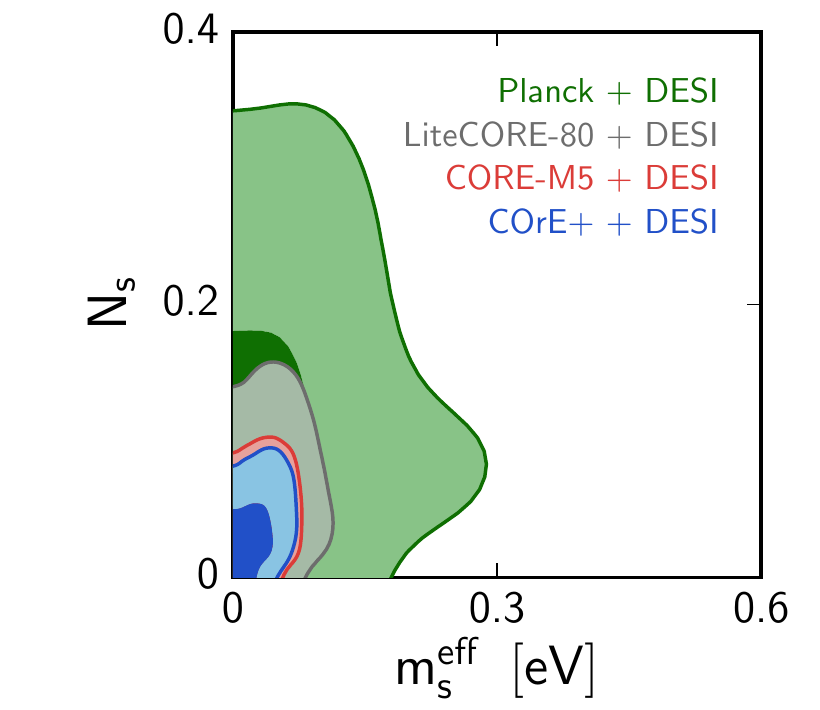}
		\hspace{-11.1cm}\includegraphics[width=6cm]{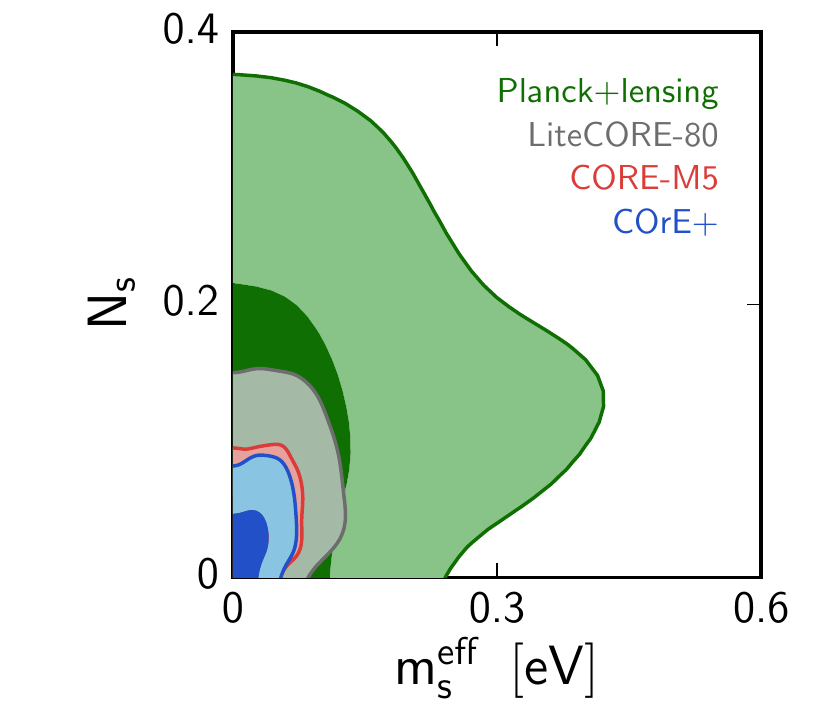}	
	\caption{Results for the extended model $\Lambda$CDM + $M_\nu$ + one light and non-thermalised sterile neutrino with effective mass $m_s^\mathrm{eff}$, contributing to the effective neutrino number as $N_s$.}
\label{fig:sterile}
\end{figure}

Our results for the parameters $(M_\nu^\mathrm{active}, N_{\rm s}, m_{\rm s}^{\rm eff} )$ are given in Table~\ref{tab:mnu_sterile}, and the probability contours for $(N_{\rm s}, m_{\rm s}^{\rm eff} )$ are shown in Figure~\ref{fig:sterile}. For CORE-M5, the bounds on the sterile sector are impressive: 
$$(m_{\rm s}^{\rm eff}, N_{\rm s}) < (37 \,{\rm meV}, 0.053), ~~~~\mathrm{(CORE-M5, \,\, 68\%CL)}$$ 
to be compared with 
$$(m_{\rm s}^{\rm eff}, N_{\rm s}) < (330 \,{\rm meV}, 0.2). ~~~~\mathrm{(PlanckTT+lowP+lensing+BAO, \,\, 68\%CL).}$$ 
The sensitivity to $(m_{\rm s}^{\rm eff}, N_{\rm s})$ depends heavily on the CORE settings. The error on $N_s$ varies by a factor two between LiteCORE-80 and COrE+. As discussed in section \ref{sec:relativistic}, this comes mainly from the ability to measure the temperature and polarisation damping tail up to high multipoles when the instrumental sensitivity and resolution are good enough. Besides, the measurement of the CMB lensing potential constrains the density of hot dark matter today, and hence roughly $M_\nu^\mathrm{active} + m_{\rm s}^{\rm eff}$.
If this were the only effect, all CORE configurations would lead essentially to $M_\nu^\mathrm{active} + m_{\rm s}^{\rm eff}=60\pm44$~meV at one sigma, and to the same constraints on $m_{\rm s}^{\rm eff}$. However, there is some extra sensitivity to $m_{\rm s}^{\rm eff}$ coming from the fact that for small $N_{\rm s}$, the physical mass associated to a given value of $m_{\rm s}^{\rm eff}$ can be large\footnote{More precisely, the velocity dispersion given by $\langle p \rangle / m_{\rm s}$ can be much smaller than for active neutrinos}, such that the sterile neutrinos have their non-relativistic transition before photon decoupling. In that case, there are additional effects on CMB primary anisotropies\footnote{Hot Dark Matter particles become non-relativistic before photon decoupling have a direct impact on CMB fluctuations at the level of primary anisotropies: they tend to suppress small-scale fluctuations.} that an experiment sensitive to smaller angular scales can constrain better. This explains the gain in sensitivity to $m_{\rm s}^{\rm eff}$ between LiteCORE-80 and COrE+. CORE-M5 appears as a good compromise, more constraining than LiteCORE-80 by 50\% for both $N_{\rm s}$ and $m_{\rm s}^{\rm eff}$. In summary, with a sensitivity to $m_{\rm s}^{\rm eff}$ ten times better than Planck, CORE-M5 appears as an ideal instrument for constraining light sterile neutrinos, and the CORE data release will play a key role in the discussion of anomalies in short baseline neutrino oscillations.

Note that with CORE data alone, we find no lower bound on the active neutrino mass $M_\nu$ in presence of a sterile neutrino, because the physical effect of the mass $M_\nu = 60$~meV in the fiducial model can be partially endorsed by the sterile neutrino mass. In other words, the data is not able to tell whether the fiducial mass of 60~meV belongs to active neutrinos, or to a mixture of sterile and active neutrinos. By removing degeneracies, BAO data from DESI makes the CMB lensing spectrum more sensitive to $M_\nu^\mathrm{active} + m_{\rm s}^{\rm eff}$, and given the upper bound on $m_{\rm s}^{\rm eff}$, one now finds a lower bound on $M_\nu^\mathrm{active}$. Cosmic shear data from Euclid directly probes the free-streaming effect associated with $M_\nu^\mathrm{active} + m_{\rm s}^{\rm eff}$, which results in a slightly better sensitivity to  $M_\nu^\mathrm{active}$, but the constraints on the sterile neutrino sector remain roughly the same as when considering CORE alone.


\subsection{Constraints on self-interacting neutrinos}

In the standard cosmological scenario, neutrinos decouple at $T\sim 1\,\mathrm{MeV}$, when the
rate for weak interactions becomes smaller than the expansion rate. After that moment, neutrinos behave as
free-streaming particles. This picture is a consequence of combining the standard model of particle physics with general relativity,
and can be tested already with present cosmological data   \cite{Cyr-Racine:2013jua,Archidiacono:2013dua,Smith:2011es,Gerbino:2013ova,Audren:2014lsa,Forastieri:2015paa,planck2015}; CORE will allow to test its validity even further. Moreover, the possibility of non-standard neutrino self-interactions 
that make the neutrino fluid collisional also at $T < 1\,\mathrm{MeV}$ is envisaged in some extensions of the standard model of particle physics
\cite{Chikashige:1980ui,Schechter:1981cv,Gelmini:1980re}.

Collisional neutrinos in a cosmological framework can be modelled in different ways. A popular approach
is to introduce effective viscosity and sound speed, following the parameterization introduced in Ref. \cite{Hu:1998kj}; this is the approach
followed in Refs. \cite{Trotta:2004ty,Smith:2011es,Archidiacono:2012gv,Diamanti:2012tg,Archidiacono:2013lva,Gerbino:2013ova,Audren:2014lsa}. This method has the advantage of being, to a good extent, model-independent;
however, the effective parameters are taken to be time-independent, a situation that is seldom realized in
physical models. Moreover, the interpretation of deviations from the free-streaming case is not immediate \cite{Cyr-Racine:2013jua,Oldengott:2014qra}. For these reasons
we choose not to use the effective parameterization. Alternative approaches consist in switching the behaviour of the neutrino fluid
from free streaming to highly collisional (or viceversa) at some redshift (like in Ref. \cite{Archidiacono:2013dua}), or to insert an (approximate) collision term modelling neutrino-neutrino scatterings 
directly in the Boltzmann equation (like in Refs. \cite{Cyr-Racine:2013jua,Forastieri:2015paa}); here we will stick to the latter method. In particular, we use the relaxation time approximation to
rewrite the Boltzmann hierarchy (in synchronous gauge) for massless neutrinos as:
\begin{subequations}
\begin{align}
&\dot{\delta} = -\frac{4}{3} \theta - \frac{2}{3} \dot{h} \, ,\\
&\dot{\theta} = k^2 \left(\frac{1}{4} \delta - \Pi \right) \, ,\\ 
&\dot{\sigma} = \frac{4}{15} \theta- \frac{3}{10} k F_{3} + \frac{2}{15} \dot{h} + \frac{4}{5}\dot{\eta} - a \Gamma_\mathrm{int} \sigma\, ,\\
&\dot{F_\ell} = \frac{k}{2\ell+1} \Big[ \ell F_{\ell-1} - (\ell+1) F_{\ell+1} \Big]  -a \Gamma_\mathrm{int} F_\ell \quad (\ell\ge 3)\, .
\end{align}
\label{eq:boltz_hierarchy_int}
\end{subequations}
where $\Gamma_\mathrm{int}$ is the scattering rate, and for the rest we follow the notation of Ma \& Bertschinger \cite{Ma:1995ey}.

The exact form of the collision term depends on the detail of the underlying particle physics model; however, two 
broad classes of models can be considered by means of an effective parameterization of the collision term. In models
in which the neutrino interaction is mediated by a scalar (like, e.g. in Majoron models), $\Gamma_\mathrm{int} \sim g^4 T_\nu$, 
($g$ being the typical value of the Yukawa couplings), so that $\Gamma_\mathrm{int}/H$ increases with time and neutrino become collisional again at some later time after decoupling.
In models in which the interaction is mediated by a vector, $\Gamma_\mathrm{int} \sim G_X^2 T_\nu^5$ ($G_X$ being
the ``Fermi constant'' of the new interaction) at low energies (below the mass of the mediator), so that neutrino possibly remain collisional for a longer time after weak decoupling.

\begin{figure}
	\centering
		\includegraphics[width=7cm]{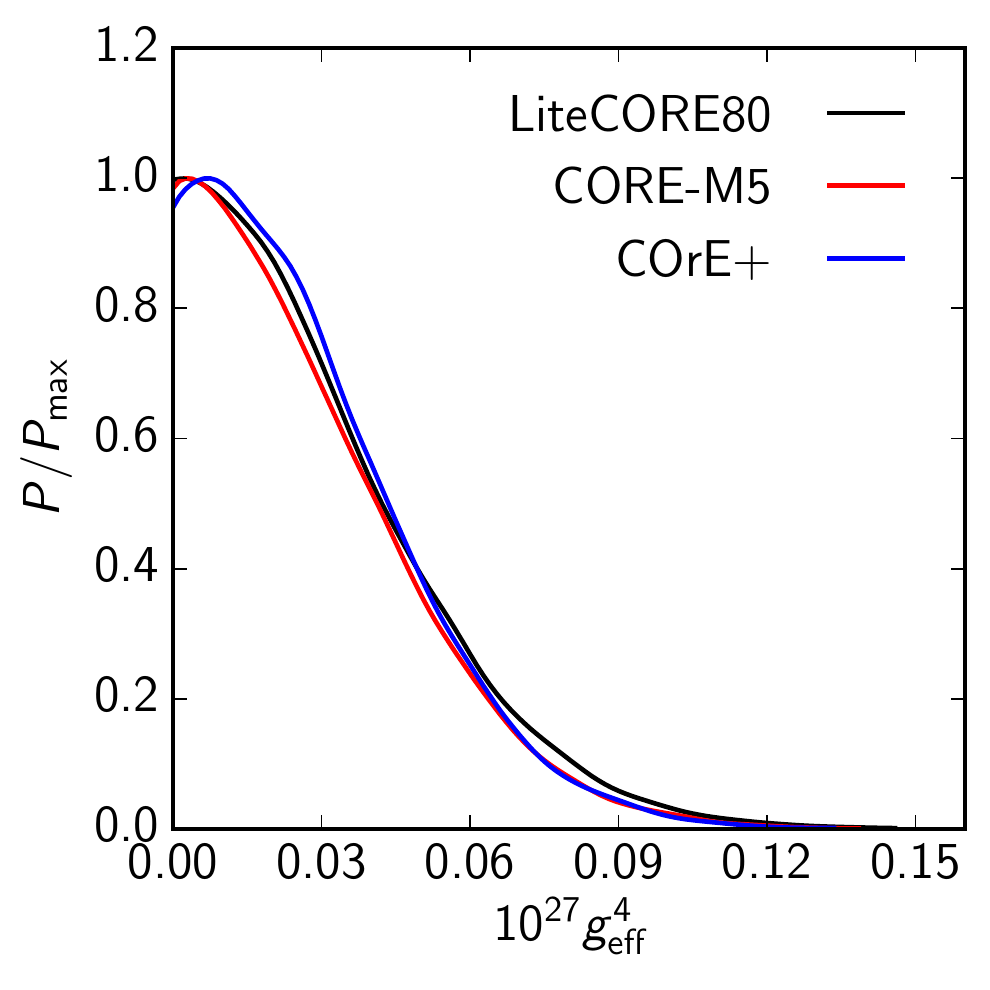}	
		\includegraphics[width=7cm]{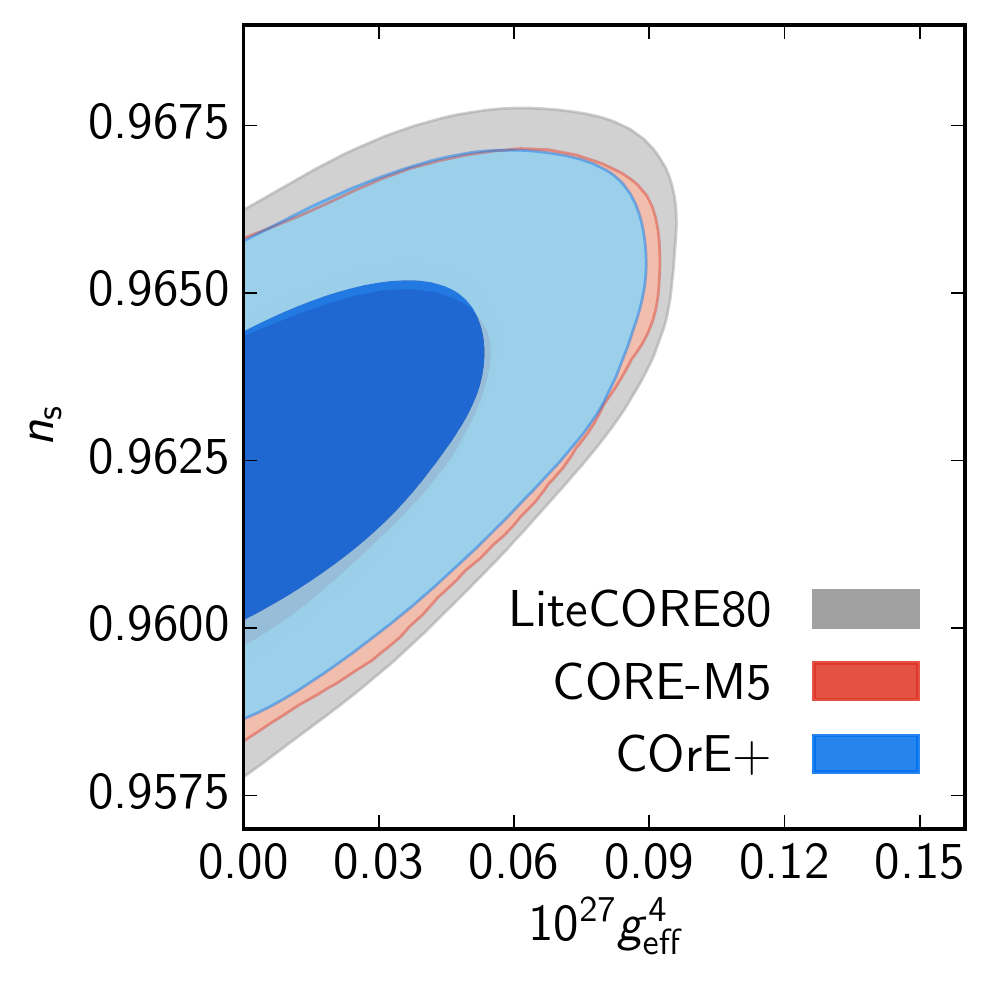}	
	\caption{Self-interacting neutrinos: (Left panel) one-dimensional posterior distribution for $g^4_{eff}$ for different CORE configurations;
	(Right) two-dimensional 68\% and 95\% credible regions in the ($n_s$, $g_{eff}^4$) plane, for the same configurations. }
\label{fig:intnu}
\end{figure}

Here we consider models of the first kind, i.e., scalar-mediated, and write the interaction rate as $\Gamma_\mathrm{int} = g_\mathrm{eff}^4 T_\nu$.
We then run a forecast for the model $\Lambda$CDM + $g_\mathrm{eff}$, with a flat prior on $g_\mathrm{eff}^4$, assuming massless neutrinos. The fiducial model
has $g_\mathrm{eff}^4 =0$  and $M_\nu = 0$, to be coherent with the assumption of massless neutrinos. We report our results in Tab. \ref{tab:intnu}
and show the one-dimensional posterior for $g_\mathrm{eff}^4$ for various CORE configurations in the left panel of Fig. \ref{fig:intnu}. 
We find that typical 95\% upper limits on $g_\mathrm{eff}^4$ are of
the order of $7\times 10^{-29}$ for all CORE configurations considered here, roughly a factor 8 improvement with respect to current limits from Planck
\cite{Forastieri:2015paa}. 
The marginal dependence of the sensitivity on the CORE settings is due to the fact that
the effect of the interaction mainly shows up on intermediate angular scales in the temperature spectrum, and even more clearly in the E-mode polarisation spectrum~\cite{Forastieri:2015paa}
(as would also be the case for the phenomenological model with effective viscosity and sound speed~\cite{Audren:2014lsa}). On those scales, all CORE configurations have a very good sensitivity to E-modes, close to cosmic variance.
Non-standard neutrino scalar interactions can also be probed by searches for neutrinoless double $\beta$ decay   \cite{Gando:2012pj,Albert:2014fya}
or observations of the neutrino signal from supernovae \cite{Choi:1989hi,Kachelriess:2000qc,Tomas:2001dh,Farzan:2002wx}.
A proper comparison between constraints from the various probes, including cosmology, is somehow model-dependent; however, for simple models,
cosmology gives the tightest limits on the couplings.

\begin{table}[h]
\begin{center} \footnotesize
\begin{tabular}{|c|c|c|c|c|}
\hline
Parameter & LiteCORE80, TEP & LiteCORE120, TEP & CORE-M5, TEP & COrE+, TEP \\ 
\hline
$10^{27}\, g^4_{\mathrm{eff}}$ & $ < 0.075\, (95\%\, CL) $ & $< 0.073\, (95\%\, CL)$ & $< 0.070\, (95\%\, CL) $ & $< 0.069\, (95\%\, CL) $ \\[0.25cm] 
 $\Omega_b h^2$ & $0.022175\pm0.000051$ & $0.022172\pm0.000041$ & $0.022176\pm0.000037$ & $0.022179\pm0.000034$ \\[0.25cm] 
 $\Omega_c h^2$ & $0.12051\pm0.00033$ & $0.12052\pm0.00030$ & $0.12051\pm0.00027$ & $0.12053\pm0.00026$ \\[0.25cm] 
 $100 \theta_{MC}$ & $1.04075\pm0.00011$ & $1.040736\pm0.000091$ & $1.040729^{+0.000086}_{-0.000084}$ & $1.040724\pm0.000078$ \\[0.25cm] 
 $\tau$ & $0.0600\pm0.0020$ & $0.0601^{+0.0020}_{-0.0022}$ & $0.0602^{+0.0020}_{-0.0021}$ & $0.0603^{+0.0019}_{-0.0021}$ \\[0.25cm] 
 $n_s$ & $0.9632^{+0.0018}_{-0.0020}$ & $0.9631^{+0.0017}_{-0.0018}$ & $0.9631\pm0.0017$ & $0.9633\pm0.0016$ \\[0.25cm] 
 $\ln(10^{10}A_s)$ & $3.0551\pm0.0037$ & $3.0553\pm0.0037$ & $3.0555\pm0.0036$ & $3.0556\pm0.0036$ \\[0.25cm] 
 \hline
 $H_0$ & $67.46^{+0.14}_{-0.13}$ & $67.45\pm0.12$ & $67.45^{+0.11}_{-0.11}$ & $67.45\pm0.10$ \\[0.25cm] 
 $\sigma_8$ & $0.8306\pm0.0013$ & $0.8307\pm0.0012$ & $0.8307\pm0.0011$ & $0.8308\pm0.0010$ \\[0.25cm] 
 \hline
\end{tabular}
\end{center}
\caption{Parameter constraints for $\Lambda$CDM + $g_\mathrm{eff}$ (68\% CL uncertainties, unless otherwise stated), for different CORE experimental configurations. \label{tab:intnu}}
\end{table}

Non-standard neutrino interactions also introduce additional parameter degeneracies. The extra pressure of the neutrino fluid induced by collisions
changes the height of the peaks in the CMB spectra, in a way that can the compensated by changing accordingly other parameters, 
most notably a combination of $\theta,\,\Omega_c h^2$ or $n_s$ \cite{Forastieri:2015paa}. In the right panel of Fig. \ref{fig:intnu} we show the two-dimensional posterior for $n_s$ and $g_\mathrm{eff}^4$, where the correlation
is particularly evident. We note however by comparing Tables \ref{tab:params_lcdm} and \ref{tab:intnu} that the precision
on the $\Lambda$CDM parameters is only slightly degraded in presence of interacting neutrinos.

\section{Constraints on the Dark Energy equation of state}

Since its discovery \cite{lambda1,lambda2}, one major goal of modern cosmology is to determine the nature of the dark energy component
responsible of the current accelerated expansion of the universe \cite{lambda3,lambda4,lambda5,lambda6,lambda7}. A crucial measurement in this
direction is the determination of the dark energy equation of state $w$, defined as the ratio between
the dark energy pressure and energy density: $w(a)=P_{de}/\rho_{de}$ (see, e.g. \cite{lambda8}).
In this section we forecast the constraints achievable by CORE on the dark energy equation of state parametrized 
either by a constant $w$ either by a Chevallier-Polarski-Linder (CPL) \cite{cpl1,cpl2} form where $w$ is a linear
function of the scale factor:

\begin{equation}
w(a)=w_0+(1-a)w_a
\end{equation}

\noindent with  $w_0$ and $w_a$ as free parameters, constants with redshift.

\subsection{Future constraints from CORE}

\begin{figure}
	\centering
	               \includegraphics[width=7.5cm]{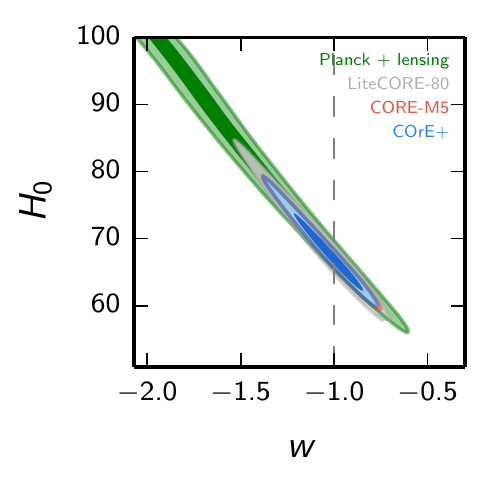}	
                \includegraphics[width=7.5cm]{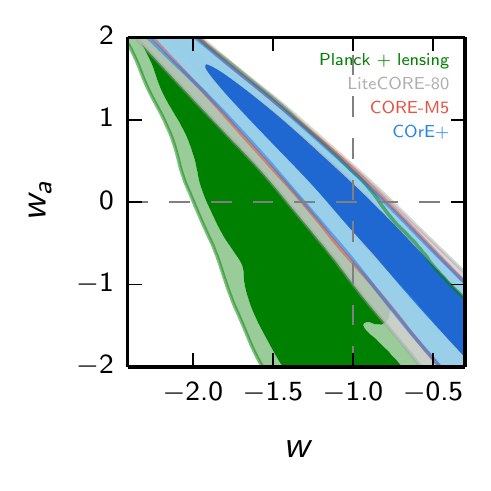}	
	\caption{Left Panel: Constraints on the $H_0$ vs $w$ plane from current Planck + CMB lensing data and from future CORE configurations. Notice the improvement on the bounds on $H_0$ from CORE with respect to Planck. Right Panel: Constraints on the $w_a$ vs $w_0$ plane from current Planck+Lensing datasets and from different future CORE configurations. In all simulated cases a cosmological constant is assumed.}
\label{fig:waw0}
\end{figure}

The recent Planck data alone, considering both temperature and polarization power spectra combined 
with lensing data, provide just a weak constraint on
the dark energy equation of state (assumed as constant with redshift) 
with $w=-1.42_{-0.47}^{+0.25}$ at $68 \%$ c.l. \cite{planck2015}. This is due
to the well known geometrical degeneracy between $w$ and $H_0$ since both
modify the angular diameter distance at recombination (see e.g. \cite{geometricdegeneracy1,geometricdegeneracy2,geometricdegeneracy3}). 
However, the improvement in the measurement of CMB lensing with CORE could provide more stringent constraints on
$w$ as we report in Table \ref{tab:params_w}. As we can see, a CORE-M5 configuration
alone could constrain the dark energy equation of state with a $\sim 10 \% $ accuracy almost
identical to the one provided by COrE+.
Weaker constraints, at the level of $\sim 15-20 \%$, could be reached by LiteCORE-120 and
LiteCORE-80 respectively. CORE-M5 could therefore improve current constraints on $w$ from Planck 
by a factor 2-3. This can be also seen  in the left panel of Figure \ref{fig:waw0} where we plot
the 2D posteriors in the $H_0$ vs $w$ plane from current Planck data and from future CORE
configurations. In particular, it is interesting to notice how CORE can now bound $H_0$ independently from
any external dataset.

\begin{table}[h]
\begin{center}\footnotesize
\scalebox{0.82}{\begin{tabular}{|c||c|c|c|c|c|}
\hline
Parameter         & Planck + lensing & LiteCORE 80, TEP& LiteCORE120, TEP& CORE-M5, TEP& COrE+, TEP  \\            
\hline
$w$ &      $-1.42\,_{-0.45}^{+0.26}$&      $-1.06\,_{-0.11}^{+0.19}$& $-1.05\,_{-0.10}^{+0.16}$& $ -1.04\,_{-0.09}^{+0.15}$&$ -1.04\,_{-0.09}^{+0.14}$ \\
$\Omega_bh^2$ &      $0.02225\pm0.00016$&      $0.022186\pm0.000056$& $0.022183\pm0.000041$& $ 0.022182\pm0.000039$&$ 0.022183\pm0.000034$ \\
$\Omega_ch^2$ &      $0.1192\pm0.0014$&      $0.12036\pm0.00061$& $0.12040\pm0.00055$& $ 0.12041\pm0.00051$&$ 0.12039\pm0.00047$ \\
$100\theta_{MC}$ &      $1.04087\pm0.00033$&      $1.04070\pm0.00010$& $1.040698\pm0.000088$& $ 1.040695\pm0.000083$&$ 1.040698\pm0.000078$ \\
$\tau$ &      $0.055\pm 0.015$&      $0.0597\,_{-0.0022}^{+0.0019}$& $0.0596\pm0.0020$& $ 0.0597\,_{-0.0021}^{+0.0019}$&$ 0.0597\,_{-0.0022}^{+0.0019}$ \\
$n_s$ &      $0.9653\pm0.0047$&      $0.9621\pm0.0020$& $0.9621\pm0.0018$& $ 0.9621\pm0.0018$&$ 0.9621\pm0.0017$ \\
$ln(10^{10}A_s)$ &      $3.042\pm0.028$&      $3.0558\pm0.0040$& $3.0557\,_{-0.0044}^{+0.0039}$& $ 3.0560\,_{-0.0042}^{+0.0037}$&$ 3.0558\,_{-0.0042}^{+0.0038}$ \\
\hline
$H_0$ [km/s/Mpc] & $>75.6$&      $69.0\,_{-6.1}^{+3.1}$& $68.5\,_{-5.1}^{+2.9}$& $ 68.1\,_{-4.6}^{+2.8}$&$ 68.3\,_{-4.4}^{+2.7}$ \\
$\sigma_8$ &      $0.93\,_{-0.07}^{+0.12}$&      $0.834\,_{-0.051}^{+0.028}$& $0.830\,_{-0.043}^{+0.026}$& $ 0.827\,_{-0.039}^{+0.025}$&$ 0.828\,_{-0.037}^{+0.024}$ \\
\hline
\end{tabular}}
\end{center}
\caption{$68\%$~CL constraints on cosmological parameters in the $\Lambda$CDM + $w$ model from the
Planck+Lensing real dataset (see \cite{planck2015}) and different CORE experimental specifications.}
\label{tab:params_w}
\end{table}

\begin{table}[h]
\begin{center}\footnotesize
\scalebox{0.82}{\begin{tabular}{|c||c|c|c|c|c|}
\hline 
Parameter         & Planck + lensing & LiteCORE 80, TEP& LiteCORE120, TEP& CORE-M5, TEP& COrE+, TEP  \\            
\hline
$w_0$ &       $-1.40\,_{-0.58}^{+0.41}$ &       $>-1.25$& $>-1.29$ &$ >-1.23$ &$ >-1.23$ \\
$w_a$ &      $<0.299$ &      $-0.1\,_{-1.3}^{+0.9}$& $-0.1\,_{-1.3}^{+0.9}$ & $ -0.1\,_{-1.2}^{+0.9}$ & $ -0.1\,_{-1.3}^{+0.8}$ \\
$\Omega_bh^2$ &      $0.02226\pm0.00016$ &      $0.022188\pm0.000055$& $0.022183\pm0.000041$ & $ 0.022183\pm0.000038$ & $ 0.022182\pm0.000034$ \\
$\Omega_ch^2$ & $0.1192\pm0.0014$ &      $0.12028\,_{-0.00067}^{+0.00059}$& $0.12033\pm0.00057$ & $ 0.12036\,_{-0.00056}^{+0.00049}$ & $ 0.12035\,_{-0.00054}^{+0.00046}$ \\
$100\theta_{MC}$ &      $1.04087\pm0.00032$ &      $1.04070\pm0.00011$& $1.040703\pm0.000090$ & $ 1.040699\pm0.000083$ & $ 1.040701\pm0.000078$ \\
$\tau$ &      $0.054\pm0.015$ &      $0.0597\pm0.0020$& $0.0598\pm0.0021$ & $ 0.0597\,^{+0.0019}_{-0.0021}$ & $ 0.0596\pm0.0020$ \\
$n_s$ &      $0.9653\pm0.0046$ &      $0.9624\pm0.0020$& $0.9623\pm0.0018$ & $ 0.9623\pm0.0018$ & $ 0.9622\pm0.0017$ \\
$ln(10^{10}A_s)$ &      $3.039\,_{-0.030}^{+0.026}$ &      $3.0555\,_{-0.0043}^{+0.0039}$& $3.0557\pm0.0041$ & $ 3.0557\,_{-0.0043}^{+0.0036}$ & $ 3.0555\,_{-0.0042}^{+0.0038}$ \\
\hline
$H_0$ [km/s/Mpc]&$>76.8$ &      $70\,_{-11}^{+5}$& $69\,_{-10}^{+4.5}$ & $ 68.8\,_{-9.6}^{+4.4}$ & $ 68.9\,_{-9.6}^{+4.3}$ \\
$\sigma_8$ &      $0.93\,_{-0.06}^{+0.12}$ &      $0.841\,_{-0.093}^{+0.045}$& $0.835\,_{-0.084}^{+0.041}$ & $ 0.831\,_{-0.081}^{+0.041}$ & $ 0.832\,_{-0.082}^{+0.039}$ \\
\hline
\end{tabular}}
\end{center}
\caption{$68\%$~CL constraints on cosmological parameters in the $\Lambda$CDM + $w_0+w_a$ model from the
Planck+Lensing real dataset (see \cite{planck2015}) and different CORE experimental specifications.}
\label{tab:params_w0_wa}
\end{table}

In Table \ref{tab:params_w0_wa} we present constraints on $w_0$ and $w_a$ using CORE data alone.
As we can see, the achievable constraints are rather weak, due to the intrinsic geometrical degeneracy
between these parameters that clearly affects CMB data. This is clearly visible in Figure
\ref{fig:waw0} (right panel) where we plot the constraints on the $w_a$ vs $w_a$ plane from current
Planck+CMB Lensing data and from future CORE configurations.
However is important to note that while $H_0$ is undetermined from current Planck constraints, CORE will
provide a $\sim 10\%$ determination of this parameter even in this very extended parameter space.

\subsection{Future constraints from CORE+DESI}

\begin{table}[h]
\begin{center}\footnotesize
\scalebox{0.82}{\begin{tabular}{|c||c|c|c|c|c|}
\hline 
Parameter         & Planck + lensing & LiteCORE 80, TEP& LiteCORE120, TEP& CORE-M5, TEP& COrE+, TEP  \\          
        &+DESI&+DESI&+DESI&+DESI&+DESI \\
\hline
$w$ &      $-1.002\pm0.043$&$-1.001\pm0.020$& $-1.000\pm0.020$& $ -1.001\pm0.020$&$ -1.000\pm0.019$ \\
$\Omega_bh^2$ &  $0.02218\pm0.00014$&    $0.022182\pm0.000052$& $0.022179\pm0.000041$& $ 0.022182\,\pm0.000039$&$ 0.022180\,_{-0.000033}^{+0.000037}$ \\
$\Omega_ch^2$ &   $0.1205\pm0.0012$&   $0.12049\pm0.00029$& $0.12051\pm0.00027$& $ 0.12049\,^{+0.00025}_{-0.00028}$&$ 0.12049\pm0.00025$ \\
$100\theta_{MC}$ &   $1.04069\pm0.00033$&   $1.04069\pm0.00010$& $1.040690\pm0.000081$& $ 1.040695\pm0.000092$&$ 1.040696\pm0.000072$ \\
$\tau$ &   $0.0605\,_{-0.0061}^{+0.0050}$&   $0.0598\,_{-0.0022}^{+0.0018}$& $0.0596\pm0.0021$& $ 0.0597\pm0.0022$&$ 0.0598\pm0.0023$ \\
$n_s$ &    $0.9620\pm0.0035$&  $0.9618\pm0.0017$& $0.9619\,_{-0.0014}^{+0.0016}$& $ 0.9619\,^{+0.0016}_{-0.0014}$&$ 0.9619\pm0.0014$ \\
$ln(10^{10}A_s)$ &   $3.058\,_{-0.012}^{+0.010}$&   $3.0564\,_{-0.0040}^{+0.0033}$& $3.0561\pm0.0038$& $ 3.0562\,\pm0.0043$&$ 3.0563\,_{-0.0040}^{+0.0035}$ \\
\hline
$H_0$  [km/s/Mpc]&     $66.98\pm0.85$& $66.97\pm0.56$& $66.95\pm0.57$& $ 66.98\pm0.57$&$ 66.96\pm0.56$ \\
$\sigma_8$ &   $0.818\pm0.016$&   $0.8175\pm0.0056$& $0.8173\pm0.0053$& $ 0.8175\pm0.0053$&$ 0.8174\pm0.0051$ \\
\hline
\end{tabular}}
\end{center}
\caption{$68\%$~CL constraints on cosmological parameters in the $\Lambda$CDM + w model from CORE+DESI. 
All the datasets (including Planck+lensing) consist of simulated datasets with a cosmological constant assumed as fiducial model.}
\label{tab:params_w_desi}
\end{table}

\begin{figure}
	\centering
		\includegraphics[width=7.5cm]{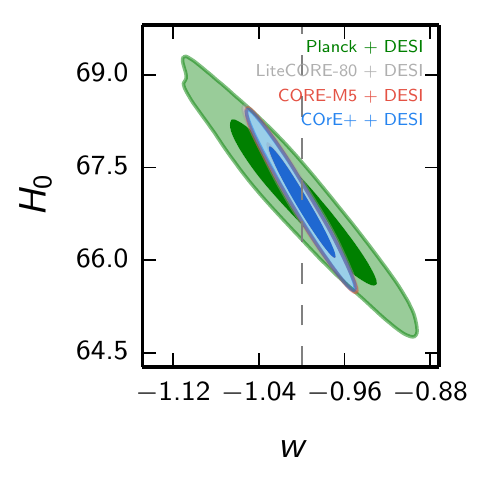}	
		\includegraphics[width=7.5cm]{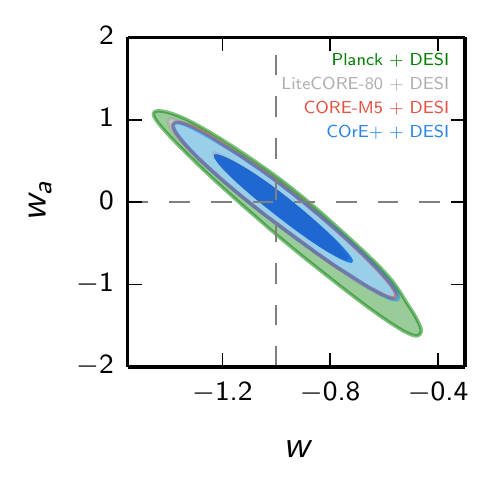}	
	\caption{Left Panel: Constraints on the $w$ vs $H_0$ plane from simulated Planck+DESI and CORE+DESI datasets. Right Panel:Constraints on the $w_a$ vs $w_0$ plane from simulated Planck+DESI and CORE+DESI datasets. }
\label{fig:waw0_desi}
\end{figure}

It is interesting to quantify the improvement on $w$ when future BAO datasets are included in the analysis.
In Table \ref{tab:params_w_desi} we present the constraints achievable by $CORE$ combined with a
future BAO survey as DESI. We compare the results with those coming from a satellite with experimental 
sensitivity as Planck again combined with DESI, derived assuming a cosmological constant as fiducial model.
As we can see, while there is no significant variation in the precision between the different $CORE$ configurations,
there is a relevant factor $\sim 2$ improvement with respect to Planck+DESI. In few words, the constraints on the
dark energy equation of state coming from a combination of cosmological data will significantly improve with
CORE. This is also evident from the left panel of Figure \ref{fig:waw0_desi}.

In Table \ref{tab:params_w0_wa_desi} we present costraints on $w_0$ and $w_a$ using CORE in combination with
 the simulated DESI dataset.  Again, the DESI data is able to significantly break the geometrical
 degeneracy that affects the CMB data. 
 This is clearly visible in Figure \ref{fig:waw0} where we plot the  constraints on the $w_a$ vs $w_a$ plane.
 When the DESI data is included there is no significant difference between the 
 constraints obtained from the CMB by the different configurations. CORE+DESI would improve the precision in the 
 constraints by $\sim 20 \%-30 \%$ with respect to Planck+DESI.
 This is also evident from the right panel of Figure \ref{fig:waw0_desi}. 
  
 Before concluding this section is important to note that the CORE-SZ cluster measurements at low redshift \cite{ECOclusters}
 will further increase the accuracy on the constraints presented here.
 
\begin{table}[h]
\begin{center}\footnotesize
\scalebox{0.82}{\begin{tabular}{|c||c|c|c|c|c|}
\hline
Parameter         & Planck + lensing & LiteCORE 80, TEP& LiteCORE120, TEP& CORE-M5, TEP& COrE+, TEP  \\          
        &+DESI&+DESI&+DESI&+DESI&+DESI \\      
\hline
$w_0$ &       $-0.96\,^{+0.18}_{-0.21}$&       $-0.97\pm0.16$& $-0.98\pm0.17$ &$-0.98\pm0.16$ &$-0.97\pm0.16$ \\
$w_a$ &       $-0.14\,_{-0.45}^{+0.60}$&       $-0.07\,_{-0.38}^{+0.42}$&$-0.05\pm0.39$ & $ -0.05\,_{-0.35}^{+0.41}$ & $ -0.08\,_{-0.35}^{+0.41}$ \\
$\Omega_bh^2$ &      $0.02217\pm0.00014$&      $0.022178\pm0.000052$& $0.022180\pm0.000041$ & $ 0.022182\pm0.000037$ & $ 0.022182\pm0.000033$ \\
$\Omega_ch^2$ &      $0.1208\pm0.0014$&      $0.12049\pm0.00032$& $0.12049\pm0.00029$ & $ 0.12048\pm0.00027$ & $ 0.12049\pm0.00026$ \\
$100\theta_{MC}$ &      $1.04067\pm0.00036$&      $1.040685\pm0.000097$& $1.040691\,_{-0.000080}^{+0.000092}$ & $ 1.040692\pm0.000077$ & $ 1.040691\,_{-0.000071}^{+0.000080}$ \\
$\tau$ &      $0.0600\,^{+0.0052}_{-0.0059}$&      $0.0597\pm0.0020$& $0.0598\pm0.0020$ & $ 0.0596\pm0.0020$ & $ 0.0595\,_{-0.0021}^{+0.0019}$ \\
$n_s$ &      $0.9613\pm0.0039$&      $0.9620\pm0.0017$& $0.9620\pm0.0015$ & $ 0.9619\,^{+0.0016}_{-0.0013}$ & $ 0.9619\pm0.0014$ \\
$ln(10^{10}A_s)$ &      $3.058\,^{+0.010}_{-0.011}$&      $3.0561\pm0.0038$& $3.0563\pm0.0037$ & $ 3.0561\,^{+0.0034}_{-0.0040}$ & $ 3.0559\pm0.0037$ \\
\hline
$H_0$ [km/s/Mpc]&      $66.7\pm1.8$&      $66.8\pm1.7$& $66.8\pm1.7$ & $ 66.8\pm1.6$ & $ 66.7\pm1.7$ \\
$\sigma_8$ &      $0.818\pm0.018$&      $0.816\pm0.015$& $0.816\pm0.015$ & $ 0.816\pm0.014$ & $ 0.815\pm0.015$ \\
\hline
\end{tabular}}
\end{center}
\caption{$68\%$~CL constraints on cosmological parameters in the $\Lambda$CDM + $w_0+w_a$ model when the DESI BAO
dataset is included. All the datasets (including Planck+lensing) consist of simulated datasets with a cosmological constant assumed as fiducial model.}
\label{tab:params_w0_wa_desi}
\end{table}

\section{Cosmological constraints from CORE-M5 in extended parameter spaces}

In the previous sections we have reported the constraints achievable from CORE-M5 on the $6$ parameters of the 
$\Lambda$CDM model and for one or two parameters extensions as, for example, the helium abundance and the neutrino effective number $Y_p+N_{\rm eff}$ (Section VI),
or the neutrino mass and effective number $M_{\nu} + N_{\rm eff}$ (Section VII). In this section, along the lines of recent analyses as \cite{DiValentino:2015ola},
we further extend the parameter space by considering $3$ or $4$ more parameters with respect to $\Lambda$CDM. 
The reason of this kind of analysis is clear: we need to assess the stability of the constraints under the assumption of $\Lambda$CDM. Moreover, if we extend the
parameter space the CORE constraints will be clearly relaxed since degeneracies are present between the parameters.
It is therefore useful to quantify how much future datasets as DESI will help in breaking these degeneracies and what is the
gain of CORE-M5 in this case with respect to current results from Planck.

\subsection{CORE-M5 constraints in a $\Lambda$CDM+$N_{\rm eff}$+$Y_p$+$M_{\nu}$ model.}

\begin{table}[h]
\begin{center}\footnotesize
\scalebox{0.87}{\begin{tabular}{|c||c|c|c|c|}
\hline
Parameter         & Planck & Planck + DESI& CORE-M5, TEP& CORE-M5, TEP + DESI  \\            
\hline
$M_{\nu}$ [eV]&              $<0.378$&$0.106\,^{+0.036}_{-0.098}$&$0.081\,^{+0.042}_{-0.052}$&$0.072\,^{+0.027}_{-0.024}$ \\
$\neff$ &         $3.11\,^{+0.36}_{-0.41}$ &$3.19\,^{+0.32}_{-0.39}$& $3.05\,\pm 0.10$ &$3.043\,\pm 0.099$  \\
$Y_{P}^{BBN}$  &  $0.238\,^{+0.022}_{-0.020}$&$0.239\,^{+0.021}_{-0.018}$& $0.2463\,\pm0.0057$&$0.2466\,\pm0.0057$\\
$\Omega_bh^2$  &  $0.02203\,\pm0.00029$& $0.02222\pm0.00019$  & $0.022178\pm0.000061$& $0.022181\pm0.000060$ \\
$\Omega_ch^2$  &  $0.1222\,^{+0.0055}_{-0.0063}$&  $0.1224\,^{+0.0052}_{-0.0063}$&  $0.1206\,^{+0.0015}_{-0.0017}$&  $0.1204\,\pm0.0015$\\
$100\theta_{MC}$  &   $1.0401\pm0.0014$& $1.0403\pm0.0013$&   $1.04066\pm0.00034$& $1.04069\,\pm0.00034$\\
$\tau$ &   $0.0598\,^{+0.0053}_{-0.0062}$& $0.0608\,^{+0.0052}_{-0.0062}$&  $0.0596\,\pm0.0020$& $0.0596\,\pm0.0021$ \\
$n_s$ &  $0.957\,\pm0.011$&$0.9642\,\pm0.0072$  & $0.9616\,\pm0.0029$&$0.9618\,\pm0.0026$  \\
$ln(10^{10}A_s)$ & $3.057\,\pm0.016$& $3.061\,\pm0.015$& $3.0563\,\pm0.0048$& $3.0558\,\pm0.0048$\\
\hline
$H_0$  [km/s/Mpc] &$64.7\,^{+3.7}_{-3.3}$ &$67.5\,^{+1.5}_{-1.7}$&$66.81\,\pm0.75$ &$66.91\,\pm0.49$  \\
$\sigma_8$ &$0.768\,^{+0.065}_{-0.034}$& $0.813\,^{+0.021}_{-0.017}$  & $0.8146\,_{-0.0074}^{+0.0089}$& $0.8162\,\pm0.0037$ \\
\hline
\end{tabular}}
\end{center}
\caption{$68\%$~CL constraints on cosmological parameters in the $\Lambda$CDM + $M_{\nu}$ + $\neff$ + $Y_{P}$ model.
All datasets (including Planck) are simulated assuming as fiducial model a $\lcdm$ model with $M_{\nu}=0.06$ eV.}
\label{tab:params_9p_yhe}
\end{table}

\begin{figure}
	\centering
		\includegraphics[width=7.0cm]{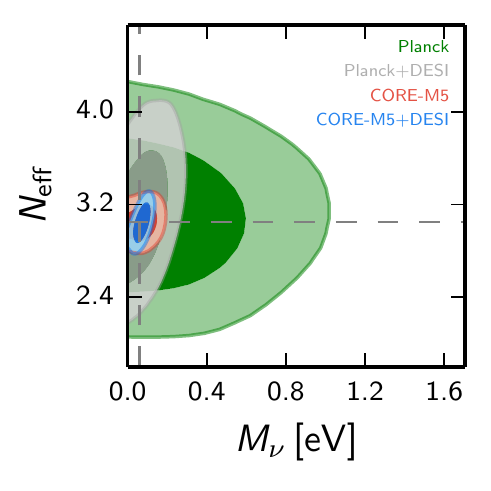}	
		\includegraphics[width=7.0cm]{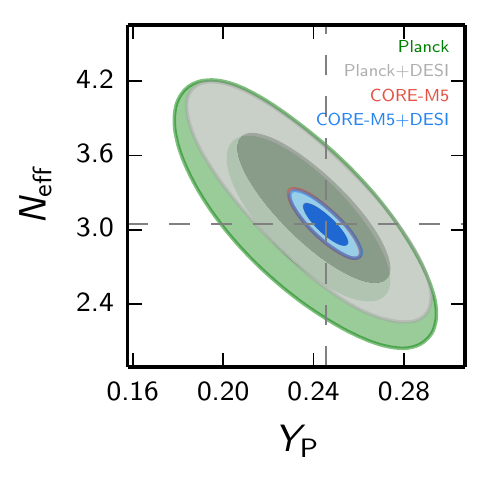}	
	\caption{Constraints on the $N_{\rm eff}$ vs $M_{\nu}$  plane (Left Panel) and  $N_{\rm eff}$ vs $Y_p$ plane (Right Panel) from Planck, Planck+DESI,
	CORE-M5, and CORE-M5+DESI simulated datasets. The dashed line corresponds to the $\Lambda$CDM prediction. $\Lambda$CDM with $M_{\nu}=0.06$ eV
	is assumed as fiducial model.}
\label{fig:yhemnunnu}
\end{figure}

We first consider an extension to $\Lambda$CDM varying at the same time three parameters already well
discussed in the previous section: the number of relativistic degrees of freedom at recombination, $N_{\rm eff}$,
the primordial Helium abundance, $Y_p$, and the total neutrino mass,  $M_{\nu}$.
The forecasted constraints are reported in Table~\ref{tab:params_9p_yhe} and in Figure~\ref{fig:yhemnunnu}. 
As we can see, in this extended parameter space
CORE-M5 will provide a significant improvement on the determination of the cosmological parameters
with respect to Planck (about a factor $\sim 7-4$) and, also, with respect to Planck+DESI (about a factor $2-3$).
CORE-M5+DESI will strongly improve the constraints respect to Planck+DESI by a factor $3-5$.
Is important for example to notice that a safe detection for a neutrino mass at the level of two standard deviations 
will be impossible from Planck+DESI, while it will still be achievable by CORE-M5+DESI.
The inclusion of the DESI dataset will not substantially improve the CORE-M5 constraints on $\neff$ and
$M_{\nu}$.

\subsection{CORE-M5 constraints in a $\Lambda$CDM+$N_{\rm eff}$+$Y_p$+$M_{\nu}$+$w$ model.} 

\begin{table}[h]
\begin{center}\footnotesize
\scalebox{0.87}{\begin{tabular}{|c||c|c|c|c|}
\hline
Parameter         & Planck & Planck + DESI& CORE-M5, TEP& CORE-M5, TEP + DESI  \\            
\hline
$M_{\nu}$ [eV]&              $<0.479$&$0.13\,^{+0.04}_{-0.12}$&$0.110\,\pm0.060$&$0.084\,^{+0.041}_{-0.060}$ \\
$\neff$ &         $3.08\,^{+0.36}_{-0.45}$ &$3.17\,^{+0.34}_{-0.40}$& $3.07\,\pm 0.11$ &$3.05\,\pm 0.10$  \\
$Y_{P}^{BBN}$  &  $0.238\,^{+0.024}_{-0.019}$&$0.241\,^{+0.021}_{-0.019}$& $0.2457\,\pm0.0057$&$0.2462\,^{+0.0066}_{-0.0054}$\\
$w$  &  $-1.46\,^{+0.60}_{-0.47}$& $-1.023\,^{+0.070}_{-0.056}$ &  $-1.15\,^{+0.26}_{-0.13}$& $-1.014\,^{+0.055}_{-0.036}$ \\
$\Omega_bh^2$  &  $0.02198\,\pm0.00030$& $0.02221\pm0.00022$  & $0.022182\pm0.000061$& $0.022178\pm0.000059$ \\
$\Omega_ch^2$  &  $0.1219\,^{+0.0056}_{-0.0068}$&  $0.1222\,^{+0.0054}_{-0.0063}$&  $0.1209\,\pm0.0017$&  $0.1206\,\pm0.0016$\\
$100\theta_{MC}$  &   $1.0401\pm0.0015$& $1.0403\pm0.0013$&   $1.04060\pm0.00036$& $1.04065\,\pm0.00035$\\
$\tau$ &   $0.0596\,^{+0.0055}_{-0.0064}$& $0.0608\,^{+0.0054}_{-0.0063}$&  $0.0598\,\pm0.0020$& $0.0597\,\pm0.0020$ \\
$n_s$ &  $0.956\,\pm0.011$&$0.9636\,\pm0.0081$  & $0.9622\,\pm0.0031$&$0.9616\,\pm0.0027$  \\
$ln(10^{10}A_s)$ & $3.055\,\pm0.016$& $3.060\,\pm0.015$& $3.0571\,\pm0.0050$& $3.0565\,\pm0.0049$\\
\hline
$H_0$ [km/s/Mpc] &$76\,^{+10}_{-20}$ &$67.8\,^{+1.5}_{-1.9}$&$71.2\,^{+3.7}_{-7.6}$ &$67.2\,^{+0.9}_{-1.1}$  \\
$\sigma_8$ &$0.85\,\pm0.11$& $0.812\,\pm0.020$  & $0.848\,_{-0.059}^{+0.031}$& $0.8175\,^{+0.0057}_{-0.0064}$ \\
\hline
\end{tabular}}
\end{center}
\caption{$68\%$~CL constraints on cosmological parameters in the $\Lambda$CDM+$w$+$M_{\nu}$+$\neff$+$Y_{P}$ model.
All datasets (including Planck) are simulated assuming as fiducial model a $\lcdm$ model with $M_{\nu}=0.06$ eV.}
\label{tab:wyhemnunnu}
\end{table}

\begin{figure}
	\centering
		\includegraphics[width=7.0cm]{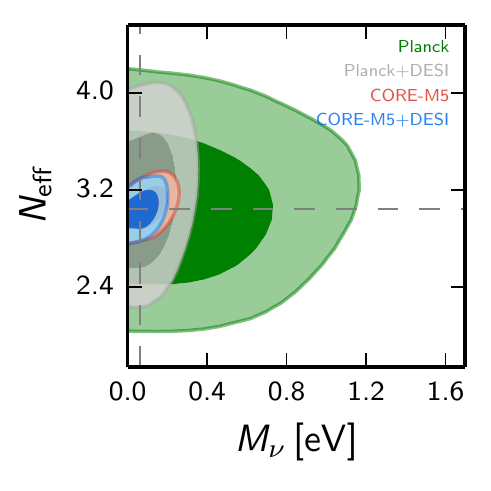}	
		\includegraphics[width=7.0cm]{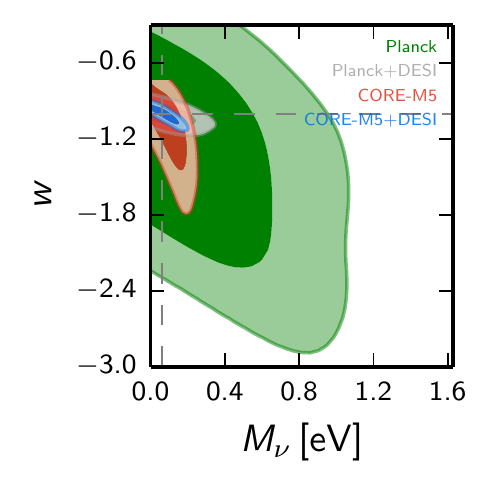}	
	\caption{Constraints on the $N_{\rm eff}$ vs $M_{\nu}$  plane (Left Panel) and  $w$ vs $M_{\nu}$ plane (Right Panel) from Planck, Planck+DESI,
	CORE-M5, and CORE-M5+DESI simulated datasets. The dashed line corresponds to the $\Lambda$CDM prediction. $\Lambda$CDM with $M_{\nu}=0.06$ eV
	is assumed as fiducial model.}
\label{fig:yhemnunnu}
\end{figure}

We now examine a further extension of the parameter space by considering also variations in the dark energy equation
of state $w$ (assumed as constant with redshift). In this case we then vary $10$ parameters at the same time:
the $6$ parameters of the standard $\Lambda$CDM model, plus $w$, $M_{\nu}$, $\neff$, and $Y_{P}$.
The forecasted constraints are reported in Table~\ref{tab:wyhemnunnu}, while in Figure~\ref{fig:yhemnunnu}
we report the 2D posteriors on the $N_{\rm eff}$ vs $M_{\nu}$  plane (Left Panel) and  $w$ vs $M_{\nu}$ 
plane (Right Panel). As we can see, also in this case the improvement in the parameter constraints from CORE-M5 with respect to Planck is
extremely significant.

\subsection{Figure of Merit} 

It is interesting to quantify the improvement of CORE-M5 with respect to current and future datasets by comparing the Figure of Merit (hereafter, FoM) for several cases.
Given a covariance matrix of parameter uncertainties for an experimental configuration, we can define the FoM, for $\Lambda$CDM, as:
\begin{equation}
FoM_{\Lambda\mathrm{CDM}} = (\det[\mathrm{cov}\{\Omega_{\rm b}h^2,\, \Omega_{\rm c}h^2,\, \theta,\, \tau,\, A_{\rm s},\, n_{\rm s}\}])^{-1/2}
\end{equation}

\noindent that is roughly inversely proportional to the volume of the constrained parameters space (see for example \cite{Bennett:2012zja}). 
Clearly, we can also consider the FoM for an extended parameter space, simply defined as
\begin{equation}
FoM_{ext} = (\det[\mathrm{cov}\{\Omega_{\rm b}h^2,\, \Omega_{\rm c}h^2,\, \theta,\, \tau,\, A_{\rm s},\, n_{\rm s},\,p_i\}])^{-1/2}
\end{equation}

\noindent where $p_i$ with $i=(1,...N)$ are the $N$ additional parameters one can consider.

\begin{table}[h]
\begin{center}\footnotesize
\scalebox{0.95}{\begin{tabular}{|c||c|c|c|}
\hline
    Model     &  Planck+DESI & CORE-M5 & CORE-M5+DESI \\                   
\hline
$\Lambda$CDM & $3.3$ & $2.3\times 10^3$& $2.3\times 10^3$ \\
$\Lambda{\rm CDM} + M_{\nu}$&  $11$ & $8.9\times10^{3}$& $2.0\times10^{4}$ \\
$\Lambda{\rm CDM}+ w$&  $24$ & $5.4\times10^{3}$& $2.2\times10^{4}$ \\
$\Lambda{\rm CDM} + M_{\nu} + \neff$  & $15$ & $4.7\times10^{4}$& $1.0\times10^{5}$ \\
$\Lambda{\rm CDM} + w_0 + w_a$&  $42$ & $4.7\times10^{3}$& $1.3\times10^{5}$ \\
$\Lambda{\rm CDM} + Y_{\rm P} + M_{\nu} + \neff$ & $13$ & $2.5\times10^{5}$& $5.0\times10^{5}$ \\
$\Lambda{\rm CDM} + w + Y_{\rm P} + M_{\nu} + \neff$ & $140$ & $5.2\times10^{5}$& $9.1\times10^{6}$ \\
\hline
\end{tabular}}
\end{center}
\caption{\small  Improvement with respect to simulated Planck data of the global figure of merit in the different cosmological scenarios specified in the first column for various data combinations involving CORE-M5 and future BAO measurements from the DESI survey.}
\label{tab:FOMs}
\end{table}
%

It is therefore interesting to compare the constraining power of different experimental configurations by considering the ratios of the FoM, given a cosmological model.
In Table~\ref{tab:FOMs} we indeed report these ratios, using the Planck+Lensing 2015 FoM as a baseline, for Planck+DESI, CORE-M5 and CORE-M5+DESI, for
$\Lambda$CDM and several extensions. As we can see, the improvement in adding DESI to Planck will be important ($>10$) only for extensions of $\Lambda$CDM. 
Indeed, as also discussed in the third section of this paper, adding DESI to Planck or CORE will not improve the constraints on the $\Lambda$CDM parameters significantly.
However, CORE-M5 can reduce the volume of the $\Lambda$CDM parameter space by almost $3$ orders of magnitude with respect to the current constraints from Planck.
As we can see from the results reported in the Table, CORE-M5 will reduce the currently viable parameter space by almost four orders of magnitude in case of
single parameter extensions and up to more than five orders of magnitude in case of a $4$ parameters extensions.

\section{Recombination physics\label{sec:recombination}}

The cosmological recombination epoch marks an important era in the thermal history of our Universe \cite{1968ZhETF..55..278Z, 1968ApJ...153....1P}. It determines the transition of the fully-ionized primordial plasma ($z\gtrsim 8000$), consisting mainly of free electrons, protons, and $\alpha$-particles all immersed in the bath of CMB photons, to the quasi-neutral phase\footnote{After recombination some tiny fraction $\simeq 2\times 10^{-4}$ of the hydrogen atoms remain ionized even before reionization at $z\lesssim 10$.}, with hydrogen and helium atoms at $z\lesssim 500$. The fine details of the evolution of doubly-ionized helium ($5000\lesssim z\lesssim 7000$) and neutral helium\footnote{In earlier recombination calculations, the recombination of helium extended into the recombination era of hydrogen; however, detailed recombination treatments have shown that helium recombination finished at $z\simeq 1700$, significantly diminishing its direct impact on the CMB anisotropies \cite{2007MNRAS.378L..39K, 2008PhRvD..77h3006S, 2008A&A...485..377R, 2012MNRAS.423.3227C}.} ($1700 \lesssim z\lesssim 3000$) only have a tiny direct impact on the CMB anisotropies, because they occur too deep inside the scattering medium to affect them strongly. Anisotropies in the medium mainly become visible during the recombination of hydrogen around $z\simeq 1100$, when photons have the largest probability of last scattering off of free electrons \cite{1970Ap&SS...7....3S, 1970ApJ...162..815P}.

Today's measurements of the CMB anisotropies are so precise that tiny variations of the free electron fraction at the $\gtrsim 0.1\%-0.5\%$ level during hydrogen recombination can induce measurable differences and biases in the main cosmological parameters \citep{2010MNRAS.403..439R, 2011MNRAS.415.1343S}. Conversely, this means that measurements of the CMB anisotropies can be used to directly constrain recombination physics and alternative recombination scenarios \cite[e.g.,][]{2000ApJ...539L...1P, 2004PhRvD..70d3502C, 2009PhRvD..80b3505G, 2013ApJ...764..137F}. In this section we outline some of the possible future directions.

\subsection{Remaining uncertainties among recombination codes}


Already for Planck, significant improvements over the standard recombination code, {\sf recfast} \cite{2000ApJS..128..407S}, had to be included to achieve the necessary sub-percent accuracy in the ionization fraction. This lead to the development of the publicly available recombination codes {\sf CosmoRec} \cite{2011MNRAS.412..748C} and {\sf HyRec} \cite{2011PhRvD..83d3513A}, which agree at the $\lesssim0.1\%$ level around hydrogen recombination.  Both codes include much more detailed computations of radiative transfer and atomic physics than {\sf recfast}. 
However, it has been shown that the precise dynamics of hydrogen recombination could be captured with {\sf recfast} when using fitting functions calibrated on the detailed computations for a given reference cosmology \citep{2010MNRAS.403..439R, 2011MNRAS.415.1343S}. Thus, most analyses available in the literature, including the main papers of the Planck collaboration, use {\sf recfast} instead of the full - albeit slightly more time-consuming - computations with {\sf HyRec} or {\sf CosmoRec}. In this section we check that the accuracy of {\sf recfast} 1.5 \cite{2008MNRAS.386.1023W} is still sufficient for the analysis of COrE+ data\footnote{We restrict this test to the most sensitive configuration to make the point.}. We only compare {\sf recfast} 1.5 with {\sf HyRec} using {\sf CLASS}, which is sufficient given that {\sf HyRec} agrees very well with {\sf CosmoRec}. 

To determine possible biases in parameters caused by remaining differences in the modeling of the recombination process, we generate mock data using {\sf recfast}, and then analyse it with models computed either with {\sf recfast} or {\sf HyRec}, assuming COrE+ sensitivity. For most parameters we find negligible shifts of the recovered mean values in comparison with the standard deviations. The biggest shift is for the scalar spectral index $n_s$ (see Table~\ref{tab:Recfast_vs_HyRec}), which is found to be biased by $\Delta n_s = 0.00044 = 0.31\,\sigma(n_s)$ due to discrepancies between the recombination codes. The parameters $\theta_s$ and $\sigma_8$ also have non-negligible shifts, by 0.15$\sigma$ and 0.20$\sigma$ respectively. Overall, this shows that for next-generation experiments like COrE+ and Stage-IV CMB, the precision of {\sf recfast} 1.5 is marginally sufficient. For a high-precision interpretation of the real data, the full recombination models from {\sf CosmoRec} or {\sf HyRec} should be used.\footnote{Alternatively, the difference can be reduced by re-calibrating the fudge-functions of {\sf recfast} using {\sf HyRec} or {\sf CosmoRec} for the Planck cosmology; however, given that the performance of the full recombination codes does not cause any bottleneck for the parameter estimation, this seems moot.} Of course, for the purpose of parameter sensitivity forecasts it makes no difference to use {\sf recfast} 1.5 instead, and as such this is what is done throughout this work.

\begin{table}[!h]
\begin{center}\footnotesize
\scalebox{1.0}{\begin{tabular}{| c || c | c | c | c |} 
 \hline 
Parameter & Recfast & HyRec & shift & shift in $\sigma$ units \\ 
\hline 
$\Omega_b h^2$ & $0.02225618\pm0.000033$& $0.02225573\pm0.000032$& $4.5 \times 10^{-7}$& 0.014\\
$\Omega_c h^2$ & $0.119751\pm0.00026$& $0.119733\pm0.00026$& $1.8 \times 10^{-5}$& 0.068\\
$100\theta_s$ & $1.040802\pm0.000073$& $1.040791\pm0.000074$& $1.1 \times 10^{-5}$& 0.15\\
$\tau$ & $0.060292\pm0.0020$& $0.060377\pm0.0020$& $-8.5 \times 10^{-5}$& -0.042\\
$n_s$ & $0.96450\pm0.0014$& $0.96406\pm0.0014$& 0.00044& 0.31\\
$ln(10^{10} A_s)$ & $3.09452\pm0.0035$& $3.09465\pm0.0035$& -0.00013& -0.036\\
\hline
$H_0$ (km/s/Mpc) & $67.017\pm0.103$& $67.001\pm0.101$& 0.016& 0.16\\
$\sigma_8$ & $0.83004\pm0.00102$& $0.82983\pm0.00104$& 0.00021& 0.20\\

\hline
 \end{tabular}}
 \end{center}
\caption{Recovered $\Lambda$CDM parameters from mock COrE+ data, using either {\sf RecFast 1.5} or {\sf HyRec}.}
\label{tab:Recfast_vs_HyRec}
\end{table}


\subsection{Measuring $T_0$ at last scattering}

Our most precise determination of $T_0$ comes from the CMB energy spectrum measured by FIRAS, yielding $T_0 = 2.7255\pm0.0006$ K \cite{1994ApJ...420..439M, 1994ApJ...420..445F, 2009ApJ...707..916F}. However, the CMB power spectra can also help determine it \cite{2004PhRvD..70f3529O, planck2015, 2014MNRAS.443.1881C}, since different values of $T_0$ have peculiar effects on both CMB perturbations and recombination physics. If we were able to make a precise and unique measurement of $T_0$ at the redshift of CMB decoupling,
we would achieve a crucial test of the temperature-to-redshift relation $T(z)\propto (1+z)$, which can indeed be modified in exotic models \citep{1996PhRvD..54.2571L, 2000MNRAS.312..747L,Avgoustidis:2011aa}.

A change of $T_0$ has very strong effects on the CMB power spectra~\cite{2008A&A...478L..27C, Hamann:2007sk, planck2015},
although many of them are {\it exactly} degenerate with shifts in other parameters like $\omega_b$, $\omega_{\rm cdm}$, and $\Omega_\Lambda h^2$. Indeed, the CMB is only probing ratios between the density of different species, and a global rescaling of all densities is unconstrained (unless one uses external data, like direct measurements of $H_0$). In particular, if one artificially fixed $z_\mathrm{rec}$ while carrying out such a global rescaling, it would leave the angular scale of the sound horizon $\theta_s\equiv r_s(z_{\rm rec})/D_A(z_{\rm rec})$ unchanged. But a shift in $z_\mathrm{rec}$, which is affected by the absolute value of $T_0$, does change $\theta_s$, thus lifting the degeneracy. Therefore, it is possible to measure $T_0$ from CMB observations only, provided that we exquisitely measure the angular scale of acoustic oscillations $\theta_s$.
In the temperature spectrum, this measurement is slightly degraded by the presence of extra contributions, from the Doppler or early ISW effect.
The polarisation spectrum does not include such contributions, and offers an opportunity to make a clean, uncontaminated measurement of the acoustic scale at the recombination time.

Due to significant errors on the polarisation spectrum, Planck was not the ideal experiment to measure $T_0$, and could only give constraints on this number in combination with external data ($\sigma(T_0)=27$~mK for Planck 2015 TT+lowP+BAO \cite{planck2015}). Thanks to unprecedented polarisation sensitivity, one expects CORE to do much better.
To demonstrate this, we analysed mock data with a $\Lambda$CDM model adding $T_0$ as free parameter, with a flat prior on $T_0$ in the range $[2.5,3]$ K. We used the Nested Sampling algorithm {\sc Multinest}~\cite{Feroz:2008xx}, which proved to converge faster than the Metropolis-Hastings algorithm in this case.

\begin{table}[h]
\begin{center}\footnotesize
\scalebox{0.85}{\begin{tabular}{|c||c|c|c|c|c|}
\hline
Parameter & Planck + BAO & LiteCORE-80, TEP & LiteCORE-120, TEP & CORE-M5, TEP & COrE+, TEP\\
\hline 
$T_0$ [K] & $2.711\pm0.023$& $2.727\pm0.021$& $2.726\pm0.019$& $2.727\pm0.018$& $2.727\pm0.018$\\
$\Omega_b h^2$ & $0.02188\pm0.00064$& $0.02229\pm0.00049$& $0.02228\pm0.00044$& $0.02228\pm0.00043$& $0.02229\pm0.00044$\\
$\Omega_c h^2$ & $0.1182\pm0.0021$& $0.1200\pm0.0034$& $0.1199\pm0.0030$& $0.1199\pm0.0029$& $0.1200\pm0.0029$\\
$100\theta_s$ & $1.04190\pm0.00058$& $1.04080\pm0.00015$& $1.04080\pm0.00012$& $1.04080\pm0.00012$& $1.04080\pm0.00011$\\
$\tau$ & $0.069^{+0.018}_{-0.010}$& $0.0603\pm0.0023$& $0.0604\pm0.0022$& $0.0603\pm0.0022$& $0.0603\pm0.0023$\\
$n_s$ & $0.9648\pm0.0054$& $0.9644\pm0.0021$& $0.9644\pm0.0019$& $0.9644\pm0.0019$& $0.9644\pm0.0019$\\
$ln(10^{10} A_s)$ & $3.084\pm0.027$& $3.094\pm0.013$& $3.094\pm0.011$& $3.094\pm0.011$& $3.094\pm0.011$\\
\hline
$H_0$ (km/s/Mpc) & $67.67\pm0.59$& $66.97\pm0.89$& $66.98\pm0.78$& $66.97\pm0.75$& $66.97\pm0.75$\\
$\sigma_8$ & $0.834\pm0.021$& $0.830\pm0.012$& $0.829\pm0.010$& $0.829\pm0.010$& $0.8294\pm0.0100$\\

\hline
 \end{tabular}} 
\end{center}  
\caption{$68\%$~CL constraints on the parameters of the $\Lambda$CDM + $T_0$ model. The first column is for Planck (high-$\ell$ TT + lowP 2015 data) combined with current BAO results, and the next columns for the different CORE experimental specifications with {\it no} external data required. \label{tab:T0}}
\end{table}


Our results are shown in Table~\ref{tab:T0}.
We find that CORE, with whatever configuration, should be able to provide the first CMB-only measurement of $T_0$ at high-redshift, although not at the same precision level as FIRAS at $z=0$. 
Since the measurement is driven by the determination of the acoustic peak scale in the polarisation spectrum, all CORE settings perform well, because on intermediate angular scales they all measure the polarisation spectrum nearly up to cosmic variance; the error just starts to increase when the sensitivity is downgraded to the LiteCORE-80 level, with $\sigma(T_0) = 21 $~mK instead of $\sigma(T_0) = 18$~mK for CORE-M5 or COrE+.

These numbers can be compared to $\sigma(T_0)=0.6$~mK for the direct determination by FIRAS. FIRAS is of course much more accurate, but we should stress that the two measurements are complementary, given that FIRAS probes the temperature precisely today, while a fit to the CMB is sensitive to the temperature evolution around the time of recombination. An independent measurement of $T_0$ using the CMB anisotropies would place tight constraints on exotic changes in the temperature-redshift relation between recombination and today, which are in fact being actively searched for using SZ-clusters \cite{2009ApJ...705.1122L, 2014MNRAS.440.2610S, 2015JCAP...09..011L,deMartino:2015ema} and molecular line transitions \cite{2011A&A...526L...7N, 2016PhRvD..93d3521A}. Assuming $T(z)=T_0(1+z)^{1-\beta}$, the error $\sigma(T_0)=18\,{\rm mK}$ implies $\sigma(\beta)\simeq 0.001$. This is comparable to what was obtained using the Planck 2015 data release in combination with BAO measurements \cite{planck2015}; however, CORE would provide a CMB-only constraint, which directly complements the CORE-SZ cluster measurements at low redshift \cite{ECOclusters}.

\subsection{Measurement of the $A_{2s1s}$ transition rate}

The 2s$\to$1s two-photon decay rate in the vacuum is known to be a key parameter of recombination physics \cite{1968ZhETF..55..278Z, 1968ApJ...153....1P, 2006A&A...446...39C, 2012JCAP...06..040M}. Indeed, it is the dominant process through which a net number of excited hydrogen atoms can reach the ground state.\footnote{$57\%$ of all hydrogen atoms become neutral through this channel \citep{2006A&A...458L..29C}, the rest decay through the Lyman-$\alpha$ transition.} The bulk of the produced photons have too low energy to significantly re-excite or ionize another recombined hydrogen atom. In contrast, direct recombinations to the ground state are irrelevant because the released Lyman-continuum photons are efficiently reabsorbed. Similarly, 2p$\to$1s decay photons are trapped in the Lyman-$\alpha$ resonance, unless they have time to redshift away from the line center before their next interaction; a very inefficient process. 

For a CMB experiment as accurate as CORE, several strategies can be adopted: the rate can be fixed to the theoretical value calculated from first principles, or varied within the range allowed by experimental bounds, or treated as a free parameter determined only by fitting cosmological data. The most detailed theoretical calculation leads to $A_{\rm 2s1s} = 8.2206\, \mathrm{s}^{-1}$~\cite{Labzowski}. 
Laboratory measurements are extremely challenging and result in large uncertainties \citep[e.g.,][]{1996PhRvL..77..255C}, roughly $6$ times worse than the current (indirect) CMB measurement performed by Planck; $7.72 \pm 0.60$~s$^{-1}$ for Planck 2015 TT,TE,EE+lowP~\cite{planck2015}. Hence, COrE+ could provide the most precise measurement of this transition. This also serves as a consistency check \cite{planck2015}, since the theoretical prediction is expected to be very robust and model-independent. Thus,  if the measurement shifts significantly away from the expected value, it could hint towards tensions in the data, indicating that further work would have to be done on the interpretation and understanding of foregrounds/systematics (before eventually claiming a discovery of new physics if no other explanation remains). 

\begin{table}[h]
\begin{center}\footnotesize
\scalebox{0.82}{\begin{tabular}{|c||c|c|c|c|c|}
\hline
Parameter & Planck, TEP & LiteCORE-80, TEP & LiteCORE-120, TEP & CORE-M5, TEP & COrE+, TEP\\
\hline
$A_{2s1s}/8.2206$ & $0.959\pm0.074$& $1.000\pm0.022$& $1.000\pm0.018$& $1.000\pm0.016$& $1.000\pm0.015$\\
$\Omega_b h^2$ & $0.02231\pm0.00018$& $0.022255\pm0.000069$& $0.022255\pm0.000055$& $0.022256\pm0.000051$& $0.022257\pm0.000045$\\
$\Omega_c h^2$ & $0.1188\pm0.0017$& $0.11975\pm0.00036$& $0.11975\pm0.00031$& $0.11975\pm0.00029$& $0.11975\pm0.00028$\\
$100\theta_s$ & $1.04200\pm0.00044$& $1.04080\pm0.00012$& $1.040799\pm0.000095$& $1.040799\pm0.000090$& $1.040800\pm0.000083$\\
$\tau$ & $0.065\pm0.015$& $0.0603\pm0.0021$& $0.0603\pm0.0021$& $0.0603\pm0.0020$& $0.0603\pm0.0020$\\
$n_s$ & $0.9685\pm0.0076$& $0.9645\pm0.0024$& $0.9645\pm0.0023$& $0.9645\pm0.0021$& $0.9645\pm0.0021$\\
$ln(10^{10} A_s)$ & $3.063\pm0.028$& $3.0946\pm0.0039$& $3.0945\pm0.0037$& $3.0944\pm0.0036$& $3.0945\pm0.0036$\\
\hline
$H_0$ (km/s/Mpc) & $67.33\pm0.72$& $67.01\pm0.19$& $67.01\pm0.16$& $67.01\pm0.15$& $67.01\pm0.14$\\
$\sigma_8$ & $0.8145\pm0.0088$& $0.8302\pm0.0014$& $0.8300\pm0.0012$& $0.8300\pm0.0011$& $0.8300\pm0.0010$\\

\hline 
 \end{tabular}}
 \end{center}
\caption{$68\%$~CL constraints on the parameters of the $\Lambda$CDM + $A_{2s1s}$ model, for Planck and the different CORE experimental specifications.}
\label{tab:A2s1s}
\end{table}

\begin{figure}
	\centering
	\hspace{0cm}\includegraphics[width=6cm]{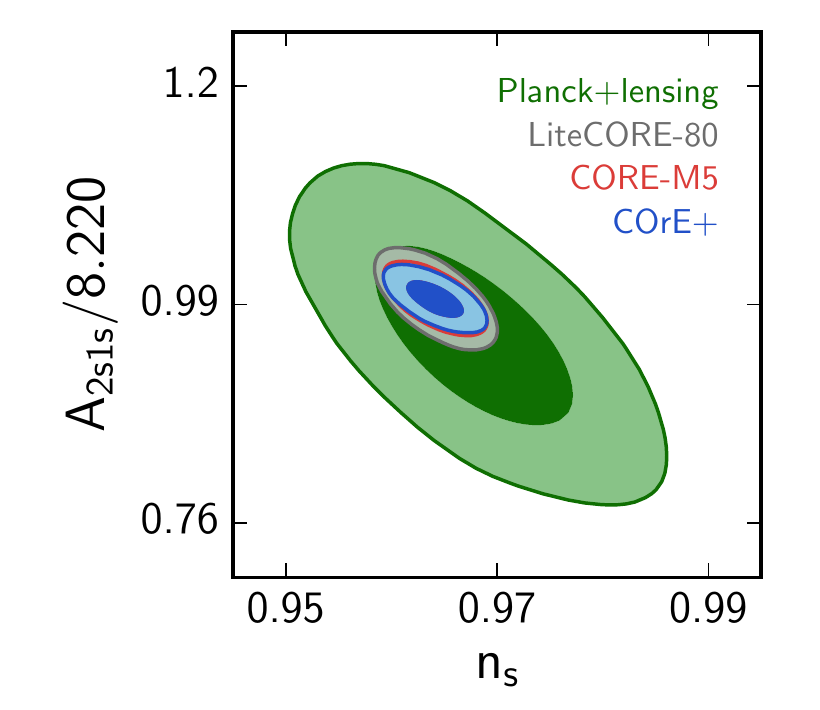}
	\hspace{-11.2cm}\includegraphics[width=6cm]{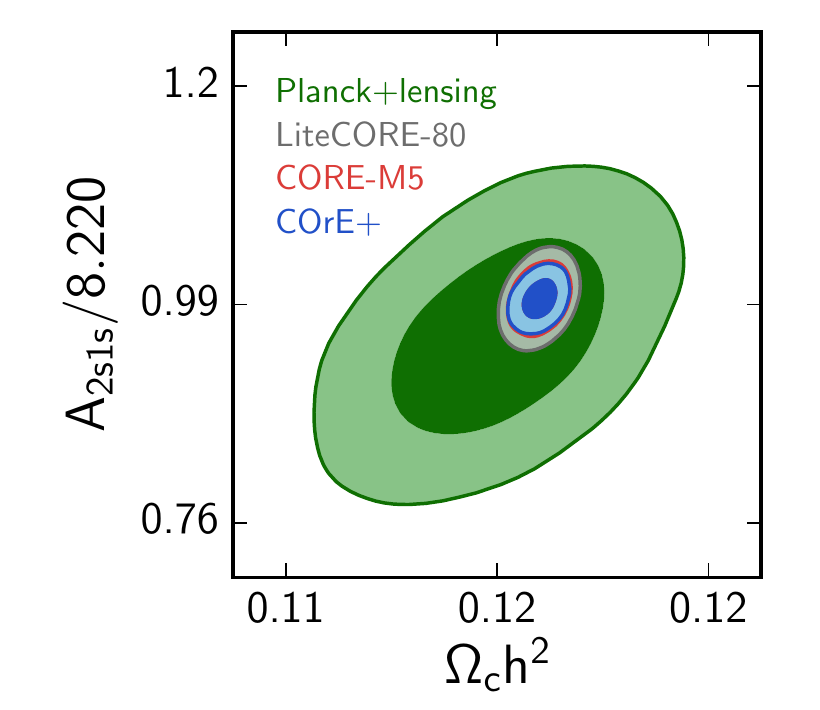}
	\hspace{-11.2cm}\includegraphics[width=6cm]{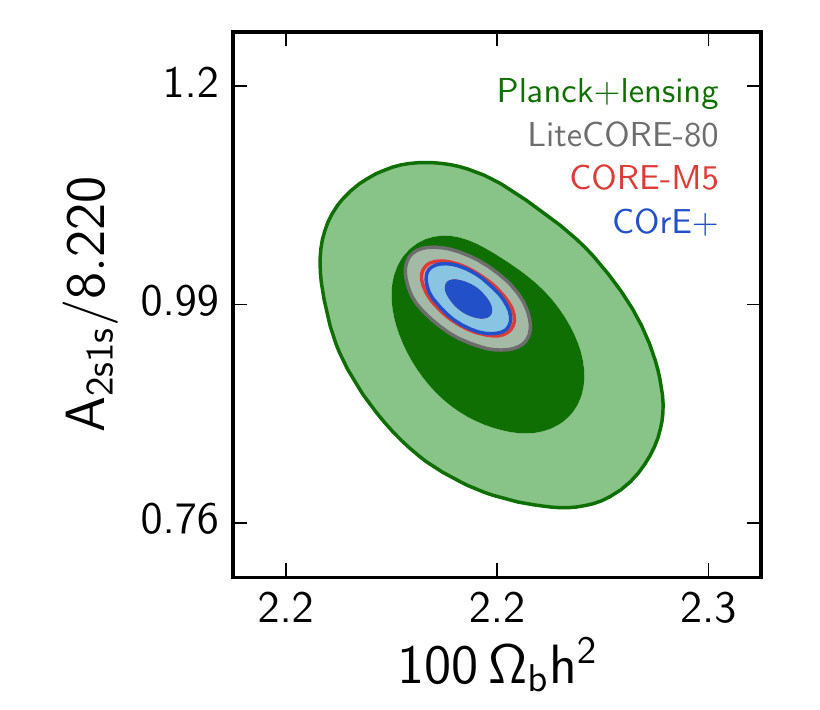}
	\caption{Results for the $\Lambda$CDM + $A_{2s1s}$ model, showing some of the parameters most correlated with $A_{2s1s}$ in the mock CORE data.}
\label{fig:A2s1s}
\end{figure}

Changes in the value of $A_{\rm 2s1s}$ affect the photon and baryon decoupling time. This has two effects in the CMB spectra: a shift in the position of acoustic peaks, and a change of amplitude in the envelope of the diffusion damping tail. The first effect should be probed equally well by all CORE configurations that measure the temperature and polarisation spectra up to cosmic variance around the scale of the first acoustic peaks, while the second effect should be better probed by the configurations most sensitive to high-$\ell$ polarisation. This is consistent with the results of our forecasts, shown in Table~\ref{tab:A2s1s}. We find that the 2s$\to$1s two-photon transition rate could be measured with a 2.2\% error by LiteCORE-80, 1.8\% error by LiteCORE-120, 1.6\% error by or CORE-M5 and 1.5\% error by COrE+, to be compared with 7.8\% only for Planck 2015 TT,TE,EE+lowP \cite{planck2015}
and the 46\% uncertainty from lab measurements \cite{1996PhRvL..77..255C}.
 

\section{Dark Matter properties\label{sec:DM}}
Although the existence of a Dark Matter (DM) component in the universe is by now well established, the nature of DM still lacks identification (see e.g. \cite{Bertone:2004pz} for a review).
In the WIMP paradigm, for instance, one can aim at detecting DM annihilation products. DM could also decay, provided its lifetime is much longer than the lifetime of the universe, as in R-parity breaking SUSY models (see e.g. \cite{Bouquet:1986mq}). We could then detect annihilation and decay products today, or probe their impact on the whole history of the Universe. If stable, DM could still be produced via the decay of a long-lived metastable heavier particle, releasing some electromagnetic energy (the so called the ``super-WIMP'' scenarios \cite{Feng:2004zu}). More generally, given our ignorance of the dark sector, there might be several components of DM, a fraction of which could be able to decay on a timescale smaller than the age of the Universe ($\Gamma > H_0$), leaving peculiar traces on cosmological observables. Most well-known candidates are e.g. unstable supersymmetric particles  \cite{Cabibbo:1981er, Salati:1984ii}, sterile neutrinos \cite{Hansen:2003yj}, and also scenarios in which DM is made of primordial black holes, either through matter accretion \cite{Ricotti:2007au} or Hawking radiation \cite{Carr:2016hva}. Cosmology, and especially the CMB, is a very sensitive and powerful probe of such models.
Typically, DM annihilation or decay via electromagnetic channels can alter the cosmological ionization history, either through modifications around the recombination epoch, or an early reionization of the Universe. This has been extensively studied in the literature and shown to have a strong impact on the CMB power spectra, especially that of polarisation \cite{2004PhRvD..70d3502C, 2005PhRvD..72b3508P, 2006PhRvD..74j3519Z, 2007PhRvD..76f1301Z, 2009PhRvD..80b3505G, dmannihilation4,2009A&A...505..999H, 2009PhRvD..80d3526S,2013MNRAS.436.2232C}. Already with WMAP and Planck, the CMB bounds on DM annihilation and decay are among the strongest in the literature, and have the major advantage of being almost free from theoretical and astrophysical uncertainties \cite{2009PhRvD..80b3505G, planck2015}. With very accurate CMB polarisation measurements, the CORE data could bring significant improvement on current bounds. For instance it could give the possibility of constraining scenarios of DM annihilation invoked to explain the so-called Fermi GeV galactic centre excess \cite{2015PhRvD..91f3003C}.

Moreover, the CMB has another remarkable property. It can probe scenarios in which DM can decay into non-electromagnetically interacting daughter particles  (like neutrinos or some kind of ``dark radiation'').  The modification of gravitational potential wells due to the decay leads to very peculiar signatures. Planck data alone can constrain the decay lifetime of such DM to be longer than 150 Gyr \cite{Poulin:2016nat}.  More accurate measurements of the temperature, polarisation, and CMB lensing spectra by CORE can greatly help towards constraining (or detecting) such models.

\subsection{Dark Matter annihilation}

We first study the 7-parameter model $\Lambda$CDM + $p_{\rm ann}$, where $$p_{\rm ann}\equiv f(z=600)\left\langle\sigma v\right\rangle/m_{\rm DM}$$ (reported here in units of cm${^3}$/s/GeV) parametrises the effect of Dark Matter ($s$-wave) annihilation\footnote{DM models for which a $p$-wave channel dominates the annihilation rate have been discussed in \cite{Diamanti:2013bia} and re-considered recently in \cite{Liu:2016cnk}. It has been shown that the limits coming from the CMB are much weaker than those coming from the upper limit on the intergalactic medium temperature $T_{IGM}$, typically $\simeq 10^4$~K at z = 4.8 \cite{Becker:2010cu}, as well as those coming from observations of the galactic diffuse gamma ray emission \cite{Essig:2013goa,Massari:2015xea,Boddy:2015efa,Albert:2014hwa}. For that reason, we only discuss here future bounds on s-wave annihilation.} on the ionization history \cite{2009PhRvD..80b3505G, 2012PhRvD..85d3522F, 2013PhRvD..88f3502G}. The efficiency factor, $f(z=600)$, accounts for the fraction of DM annihilation energy deposited into the medium, $\left\langle\sigma v\right\rangle$ is the thermal average of the cross section times velocity, and $m_{\rm DM}$ is the mass of the DM particles.
We choose a fiducial value $p_{\rm ann}=0$, and fit the corresponding mock data with a flat prior on $p_{\rm ann}\geq0$. 

\begin{table}[h]
\begin{center}\footnotesize
\scalebox{0.78}{\begin{tabular}{|c||c|c|c|c|c|}
\hline
Parameter & Planck, TEP & LiteCORE-80, TEP & LiteCORE-120, TEP & CORE-M5, TEP & COrE+, TEP\\
\hline
$10^{28} p_\mathrm{ann}$ [cm$^3$/s/GeV] & $<3.9$ (95\%CL)& $<1.44$ (95\%CL)& $<1.46$ (95\%CL)& $<1.37$ (95\%CL)& $<1.38$ (95\%CL)\\
$\Omega_b h^2$ & $0.02227\pm0.00016$& $0.022260\pm0.000054$& $0.022255\pm0.000042$& $0.022255\pm0.000038$& $0.022256\pm0.000033$\\
$\Omega_c h^2$ & $0.1196\pm0.0014$& $0.11970\pm0.00034$& $0.11969\pm0.00029$& $0.11970\pm0.00028$& $0.11970\pm0.00026$\\
$100\theta_s$ & $1.04178\pm0.00031$& $1.04079\pm0.00010$& $1.040794\pm0.000085$& $1.040798\pm0.000079$& $1.040798\pm0.000074$\\
$\tau$ & $0.059\pm0.013$& $0.0600\pm0.0021$& $0.0600\pm0.0020$& $0.0599\pm0.0020$& $0.0599\pm0.0020$\\
$n_s$ & $0.9654\pm0.0048$& $0.9651\pm0.0018$& $0.9651\pm0.0016$& $0.9651\pm0.0016$& $0.9651\pm0.0015$\\
$ln(10^{10} A_s)$ & $3.057\pm0.024$& $3.0956\pm0.0038$& $3.0955\pm0.0036$& $3.0953\pm0.0036$& $3.0954\pm0.0036$\\
\hline
$H_0$ (km/s/Mpc) & $67.41\pm0.64$& $67.03\pm0.14$& $67.03\pm0.12$& $67.03\pm0.11$& $67.03\pm0.10$\\
$\sigma_8$ & $0.8150\pm0.0085$& $0.8307\pm0.0014$& $0.8304\pm0.0012$& $0.8304\pm0.0011$& $0.8303\pm0.0011$\\

\hline 
 \end{tabular}}
 \end{center} 
\caption{$68\%$~CL constraints on the parameters of the $\Lambda$CDM + $p_{\rm ann}$ (annihilating Dark Matter) model,  for the different CORE experimental specifications. \label{tab:DM_ann}}
\end{table} 


\begin{figure}
	\centering
	\hspace{0cm}\includegraphics[width=6cm]{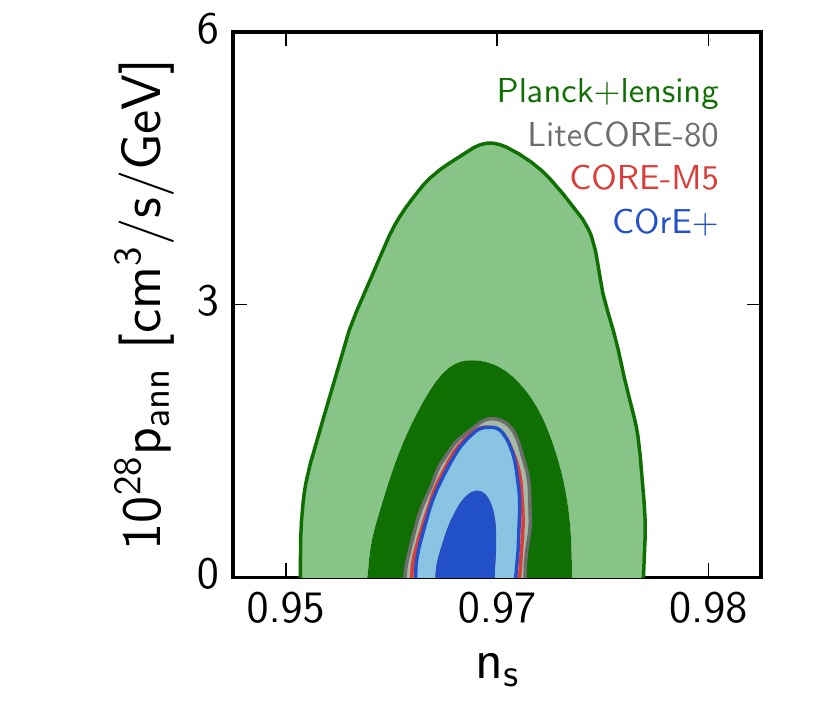}
	\hspace{-10.5cm}\includegraphics[width=6cm]{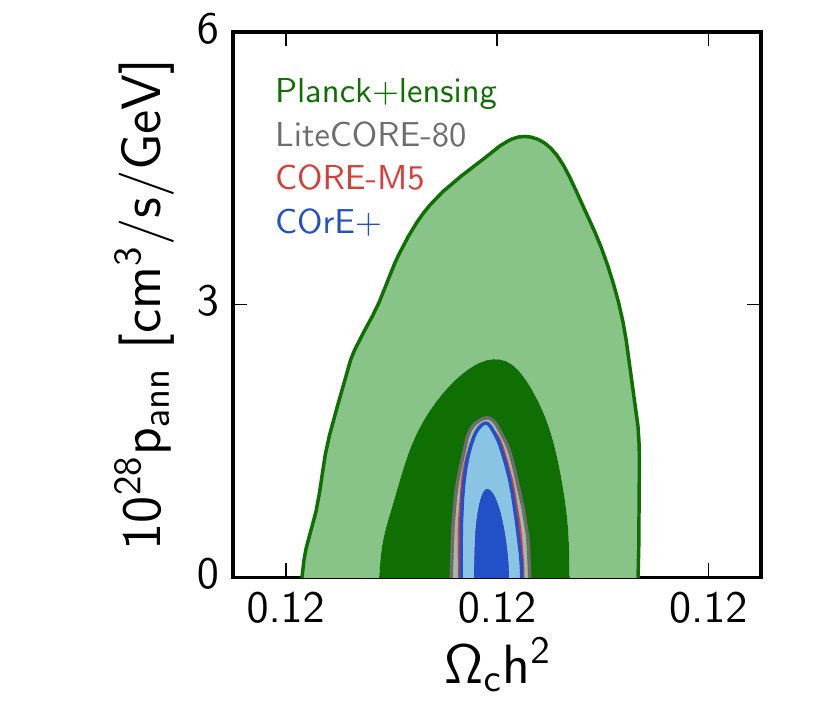}
	\hspace{-10.5cm}\includegraphics[width=6cm]{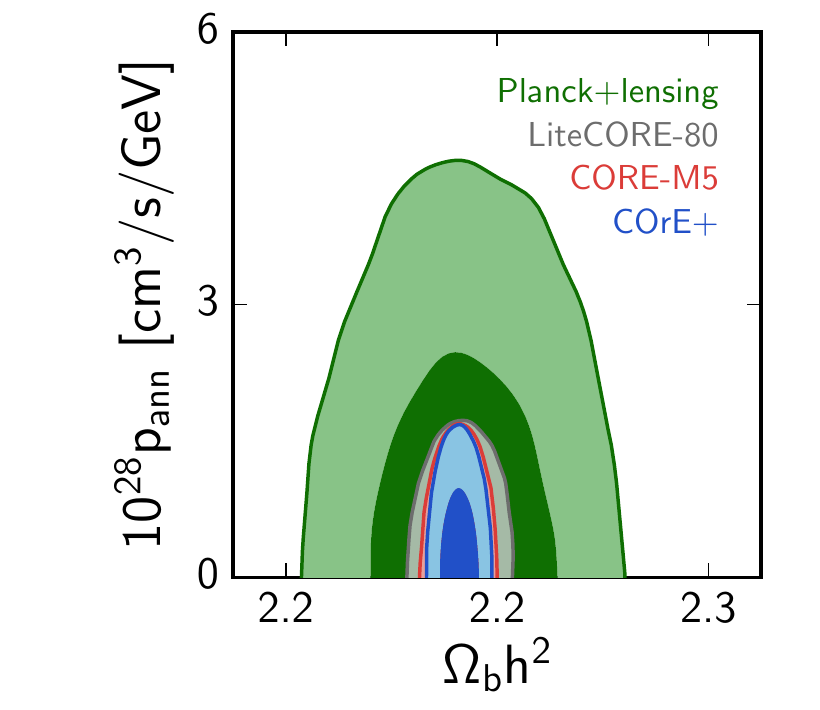}
	\caption{Results for the $\Lambda$CDM + $p_{ann}$ model, showing some of the parameters most correlated with $p_{ann}$ in the mock CORE data.}
\label{fig:pann}
\end{figure}

The effect of DM annihilation on the CMB is discussed e.g. in \cite{dmannihilation2, dmannihilation3,2012PhRvD..85d3522F}. The annihilation shifts the time of recombination and increases the free electron fraction after recombination. In the CMB temperature and polarisation spectra, the first effect can in principle affect the peak scale and the envelope of the diffusion damping tail at high-$\ell$, but only by a small amount. The clearest and most characteristic signature  of DM annihilation comes from the second effect, which changes the shape of the polarisation power spectrum on intermediate and large angular scales: this would be seen equally well by all CORE configurations. 

Our results for the different settings are presented in Table~\ref{tab:DM_ann} and Figure~\ref{fig:pann}. 
One could obtain a bound of $1.44\times10^{-28}$ cm$^3$/s/GeV with LiteCORE-80, $1.42\times10^{-28}$ cm$^3$/s/GeV with {LiteCORE}-120, $1.38\times10^{-28}$ cm$^3$/s/GeV with CORE-M5 and {COrE+}, to be compared with $4.1\times10^{-28}$ cm$^3$/s/GeV with Planck 2015 TT,TE,EE+low-P \cite{planck2015}. This is an improvement by roughly a factor of $3$, very close to the cosmic variance limit (see e.g. \cite{2012PhRvD..85d3522F}). Assuming a thermal annihilation cross-section, CORE-M5 could set a 95\%CL bound on the DM particles of mass $m \geq 100$~GeV for particles annihilating into $e^+e^-$ ($f(z=600) \simeq 0.45$), $m \geq 43$~GeV for annihilation into $\mu^+ \mu^-$ or $b\bar{b}$ ($f(z=600) \simeq 0.2$), and $m \geq 32$~GeV for annihilation into $\tau^+\tau^-$ ($f(z=600) \simeq 0.15$).

\subsection{Dark Matter decay}

Dark matter decay can also be tested using precise measurements of CMB anisotropies \cite{2004PhRvD..70d3502C, 2007PhRvD..76f1301Z, 2014JCAP...02..017D, Poulin:2016nat,dolgov_decay}. Here, we highlight constraints on decaying DM models that interact electromagnetically or purely gravitationally.

\subsubsection{Purely gravitational constraints}

We first focus on the constraints that CORE could place on the DM lifetime
through purely gravitational effects. Although these constraints are often not as strong as those that apply when electromagnetic decay channels are open, they can be the most stringent when the DM decays into neutrinos or some form of dark radiation (DR). Recently, these models have been reinvestigated in the light of tensions between low astronomical measurements of $H_0, \sigma_8$ and $\Omega_m$, and those inferred from CMB power spectra analyses. Indeed, DM decay can help in reconciling the discrepant datasets \cite{dolgov_decay}, although it does not totally solve the issue (see e.g. \cite{Poulin:2016nat} and references therein for a recent review).

DM decays affect the temperature power spectrum at small $\ell$'s through the late ISW effect, the polarisation spectrum at small $\ell$'s due to changes in the $\tau$ to $z$ relation around reionization, and all spectra through a different amount of CMB lensing (since a small fraction of the dark matter forming structures decays between recombination and today) \cite{Poulin:2016nat}. We expect CORE to improve upon Planck constraints, mainly through its better determination of the CMB lensing spectrum. We therefore analyse $\Lambda$CDM +  $\Gamma_{\rm dcdm}$ models, where $\Gamma_{\rm dcdm}$ is the decay rate of the DM particle. We also exchange $\Omega_{c} h^2$ with $\Omega_{\rm dcdm+dr}h^2$, the density parameter accounting for both decaying CDM and decay radiation, which would be equal to $\Omega_{c} h^2$ in the limit $\Gamma_{\rm dcdm} = 0$ (we refer to~\cite{Poulin:2016nat} for all relevant details on the parametrisation and computation of this model).
The fiducial model has $\Gamma_{\rm dcdm} = 0$ and we assume $\Gamma_{\rm dcdm} \geq 0$.

\begin{table}[h]
\begin{center}\footnotesize
\scalebox{0.80}{\begin{tabular}{|c||c|c|c|c|c|}
\hline
Parameter & Planck, TEP & LiteCORE-80, TEP & LiteCORE-120, TEP & CORE-M5, TEP & COrE+, TEP\\
\hline
$10^{20}\Gamma_\mathrm{dcdm}$ [s$^{-1}$] & $<20.9$ (95\%CL)& $<11.3$ (95\%CL)& $<9.9$ (95\%CL)& $<9.4$ (95\%CL)& $<8.9$ (95\%CL)\\
$\Omega_b h^2$ & $0.02224\pm0.00016$& $0.022241\pm0.000055$& $0.022249\pm0.000041$& $0.022250\pm0.000038$& $0.022253\pm0.000034$\\
$\Omega_\mathrm{dcdm+dr} h^2$ & $0.1177^{+0.0021}_{-0.0016}$& $0.11911^{+0.00067}_{-0.00045}$& $0.11919^{+0.00059}_{-0.00041}$& $0.11921^{+0.00059}_{-0.00037}$& $0.11924^{+0.00055}_{-0.00036}$\\
$100\theta_s$ & $1.04183\pm0.00031$& $1.04079\pm0.00010$& $1.040781\pm0.000085$& $1.040784\pm0.000082$& $1.040785\pm0.000076$\\
$\tau$ & $0.071\pm0.014$& $0.0606\pm0.0021$& $0.0606\pm0.0021$& $0.0607\pm0.0021$& $0.0607\pm0.0021$\\
$n_s$ & $0.9644\pm0.0048$& $0.9637\pm0.0018$& $0.9636\pm0.0016$& $0.9638\pm0.0016$& $0.9637\pm0.0016$\\
$ln(10^{10} A_s)$ & $3.075\pm0.026$& $3.0962\pm0.0040$& $3.0963\pm0.0039$& $3.0965\pm0.0038$& $3.0963\pm0.0038$\\
\hline
$H_0$ (km/s/Mpc) & $67.61\pm0.65$& $66.97\pm0.14$& $66.98\pm0.12$& $66.99\pm0.11$& $66.99\pm0.11$\\
$\sigma_8$ & $0.8189\pm0.0089$& $0.8302\pm0.0014$& $0.8301\pm0.0012$& $0.8302\pm0.0011$& $0.8301\pm0.0010$\\

\hline 
 \end{tabular}}
 \end{center}
\caption{$68\%$~CL constraints on the parameters of the $\Lambda$CDM + $\Gamma_{\rm dcdm}$ (decaying Cold Dark Matter) model for the different CORE experimental specifications. \label{tab:DM_dec}}
\end{table} 


We summarise our results in Table~\ref{tab:DM_dec}.
For Planck 2015 TT,TE,EE + lensing data we find $\Gamma_{\rm dcdm} < 21\times 10^{-20}$ s$^{-1}$ (95\%CL) in good agreement with Ref.~\cite{Poulin:2016nat}. LiteCORE-80 would already improve this constraint to $\Gamma_{\rm dcdm} < 11\times 10^{-20}$ s$^{-1}$ (equivalently to a lifetime $\tau_{\rm dcdm} > 280$ Gyr). However, the impact of $\Gamma_{\rm dcdm}$ on lensing appears to be slightly degenerate with that of $\Omega_{\rm dcdm+dr} h^2$ and $n_s$, leading to parameter correlations. By increasing the sensitivity and resolution of CORE, one can reconstruct the CMB lensing spectrum with larger leverage and reduce these degeneracies. We find that {CORE-M5} would give a bound $\Gamma_{\rm dcdm} <  9.4 \times 10^{-20}$~s$^{-1}$, nearly as strong as COrE+ which would obtain $\Gamma_{\rm dcdm} < 8.9 \times 10^{-20}$~s$^{-1}$  ($\tau_{\rm dcdm} > 360$ Gyr).

\subsubsection{Electromagnetic constraints}

We now run forecasts for the model $\Lambda$CDM + $\Gamma_{\rm eff}$, where $\Gamma_{\rm eff}$ is defined in a manner similar to the annihilation parameter as $\Gamma_{\rm eff}\equiv f_{\rm eff} \Gamma_{\rm DM}f_{\rm e.m.}$ in unit of s$^{-1}$. Here, $f_{\rm eff}$ is the typical efficiency with which the energy released by the decay of DM particles is deposited into the medium, $\Gamma_{\rm DM}$ is the DM decay rate, and $f_{\rm e.m.}=\Delta E/m_{\rm DM}c^2$ is the fraction of mass energy transferred to electromagnetic decay products. 
The effect of such a DM decay is typically to increase the free electron fraction in a way similar to reionization, but starting at much higher redshifts $z\geq 100$. 
As a consequence, one might expect constraints on $\Gamma_{\rm eff}$ to depend on the detailed way in which reionization itself is modelled. Recently, Ref.~\cite{Oldengott:2016yjc} has compared bounds on $\Gamma_{\rm eff}$ obtained in the nearly-instantaneous, or ``camb-like'', reionization\footnote{In the nearly-instantaneous reionization, the free electron fraction is given at low-$z$ by $x_e(z) = \frac{f}{2}\big[1+\tanh(\frac{y-y_{\rm re}}{\Delta y})\big]$ with $f = 1 + n_{\rm He}/n_{\rm H}$, $y = (1 + z)^{3/2}$ and $\Delta y = 3(1 + z)^{1/2}\Delta z$. The reionization is, therefore, redshift-symmetric, centred around the key parameter $z_{\rm re}$ with a width given by $\Delta z$. } to the more recent, redshift-asymmetric, parametrisation of \cite{Douspis:2015nca} given by\footnote{Following the authors of Ref.~\cite{Oldengott:2016yjc}, we replaced the argument of the exponent by $-\lambda \frac{(z-z_{\rm p})^3}{(z-z_{\rm p})^2+0.2}$ in order to improve the smoothness of the transition.}
\begin{equation} \label{eq:reio_param_2}
x_e(z) =f \times \left\{ \begin{array}{cl}
& \frac{1-Q_{\rm p}}{(1+z_{\rm p})^3-1}\big((1+z_{\rm p})^3-(1+z)^3\big)+Q_{\rm p}\:\:\;\textrm{for } z < z_{\rm p} \\
& Q_{\rm p}\exp\big(-\lambda (z-z_{\rm p})\big) \:\:\;\textrm{for } z \geq z_{\rm p}.
\end{array}\right.
\end{equation} 
Here, the parameters have been adjusted to match direct observations of the ionized hydrogen fraction $Q_{\rm HII} (z)$ (\cite{Bouwens:2015vha} and references therein) and are given by $z_{\rm p} = 6.1$, $Q_{\rm p}\equiv Q_{\rm HII} (z_{\rm p})  = 0.99986$ and $\lambda = 0.73$. The authors of~\cite{Oldengott:2016yjc} found that the bounds on $\Gamma_{\rm eff}$ obtained using nearly-instantaneous or redshift-asymmetric reionization differ by only 20\% when using Planck 2015 data, but we wish to check whether this will still be the case with very precise data from CORE. 
Similarly to Ref.~\cite{Oldengott:2016yjc}, when studying redshift-asymmetric reionization, we fix $z_{\rm p}$ and $Q_{\rm p}$ to their best-fit values, and let the evolution rate $\lambda$ vary in order to cover a large range of possible ionization histories. 
The fiducial model assumes $\Gamma_{\rm eff} = 0$ and we take a flat prior on this parameter, imposing only $\Gamma_{\rm eff} \geq 0$.

\begin{table}[h]
\begin{center}\footnotesize
\scalebox{0.82}{\begin{tabular}{|c||c|c|c|c|c|}
\hline
Parameter & Planck, TEP & LiteCORE-80, TEP & LiteCORE-120, TEP & CORE-M5, TEP & COrE+, TEP\\
\hline
$10^{27}\Gamma_\mathrm{eff}$ [s$^{-1}$] & $<69$ (95\%CL)& $<5.8$ (95\%CL)& $<5.7$ (95\%CL)& $<5.6$ (95\%CL)& $<5.7$ (95\%CL)\\
$\Omega_b h^2$ & $0.02229\pm0.00016$& $0.022256\pm0.000041$& $0.022256\pm0.000041$& $0.022256\pm0.000038$& $0.022255\pm0.000034$\\
$\Omega_c h^2$ & $0.1192\pm0.0013$& $0.11966\pm0.00030$& $0.11966\pm0.00031$& $0.11966\pm0.00028$& $0.11967\pm0.00027$\\
$100\theta_s$ & $1.04184\pm0.00030$& $1.040804\pm0.000082$& $1.040805\pm0.000084$& $1.040806\pm0.000079$& $1.040807\pm0.000074$\\
$\tau$ & $0.053^{+0.008}_{-0.017}$& $0.0598\pm0.0021$& $0.0598\pm0.0021$& $0.0598\pm0.0021$& $0.0597\pm0.0021$\\
$n_s$ & $0.9655\pm0.0046$& $0.9648\pm0.0015$& $0.9648\pm0.0015$& $0.9648\pm0.0015$& $0.9648\pm0.0014$\\
$ln(10^{10} A_s)$ & $3.066\pm0.023$& $3.0959\pm0.0037$& $3.0958\pm0.0037$& $3.0958\pm0.0037$& $3.0958\pm0.0036$\\
\hline
$H_0$ (km/s/Mpc) & $67.57\pm0.61$& $67.04\pm0.12$& $67.04\pm0.12$& $67.04\pm0.11$& $67.04\pm0.11$\\
$\sigma_8$ & $0.8173\pm0.0083$& $0.8303\pm0.0012$& $0.8303\pm0.0012$& $0.8303\pm0.0011$& $0.8303\pm0.0011$\\
$z_\mathrm{reio}$ & $7.3^{+1.1}_{-1.7}$& $8.25\pm0.21$& $8.24\pm0.21$& $8.25\pm0.20$& $8.24\pm0.21$\\
 
\hline
 \end{tabular}}
 \end{center}
\caption{$68\%$~CL constraints on the parameters of the $\Lambda$CDM + $\Gamma_{\rm eff}$ (decaying Dark Matter) model, assuming nearly-instantaneous reionization, for the different CORE experimental specifications. \label{tab:DM_dec_EM}}
\end{table} 

\begin{table}[h]
\begin{center}\footnotesize
\scalebox{0.90}{\begin{tabular}{|c||c|c|c|c|}
\hline
Parameter & LiteCORE-80, TEP & LiteCORE-120, TEP & CORE-M5, TEP & COrE+, TEP\\
\hline
$10^{27}\Gamma_\mathrm{eff}$ [s$^{-1}$] & $<7.3$ (95\%CL)& $<7.4$ (95\%CL)& $<7.3$ (95\%CL)& $<7.4$ (95\%CL)\\
$\lambda$ & $0.498^{+0.043}_{-0.062}$& $0.498^{+0.044}_{-0.060}$& $0.499^{+0.041}_{-0.060}$& $0.503^{+0.041}_{-0.061}$\\
$\Omega_b h^2$ & $0.022265\pm0.000054$& $0.022261\pm0.000041$& $0.022261\pm0.000038$& $0.022258\pm0.000033$\\
$\Omega_c h^2$ & $0.11956\pm0.00035$& $0.11955\pm0.00032$& $0.11956\pm0.00029$& $0.11957\pm0.00028$\\
$100\theta_s$ & $1.04080\pm0.00010$& $1.040809\pm0.000084$& $1.040808\pm0.000080$& $1.040809\pm0.000073$\\
$\tau$ & $0.0613\pm0.0023$& $0.0613\pm0.0023$& $0.0612\pm0.0022$& $0.0611\pm0.0022$\\
$n_s$ & $0.9648\pm0.0017$& $0.9649\pm0.0016$& $0.9649\pm0.0015$& $0.9649\pm0.0015$\\
$ln(10^{10} A_s)$ & $3.0979\pm0.0041$& $3.0979\pm0.0041$& $3.0977\pm0.0039$& $3.0974\pm0.0037$\\
\hline
$H_0$ (km/s/Mpc) & $67.09\pm0.15$& $67.09\pm0.13$& $67.08\pm0.12$& $67.08\pm0.11$\\
$\sigma_8$ & $0.8310\pm0.0014$& $0.8308\pm0.0013$& $0.8308\pm0.0012$& $0.8307\pm0.0011$\\

\hline
 \end{tabular}}
 \end{center}
\caption{$68\%$~CL constraints on the parameters of the $\Lambda$CDM + $\Gamma_{\rm eff}$ (decaying Dark Matter) model, assuming redshift-asymmetric reionization, for the different CORE experimental specifications. \label{tab:DM_dec_EM_II}}
\end{table} 
In the nearly-instantaneous reionization scenario, $\Gamma_{\rm eff}$ could be constrained to be smaller than $5.7\times 10^{-27}$ s$^{-1}$ at 95\%CL by essentially all CORE configurations (see Table~\ref{tab:DM_dec_EM}). This bound
represents a factor ten improvement with respect to the current limit from Planck 2015 TT,TE,EE + lensing data; $\Gamma_{\rm eff} < 69\times 10^{-27}$ s$^{-1}$. This comes from much better polarisation measurements.
However, all CORE configurations do equally well for this model, because EM decay effects only impact large angular scales, unlike gravitational decay effects which also modify CMB lensing. On large angular scales, all CORE settings provide cosmic-variance-limited measurements of both the temperature and polarisation. 

In the redshift-asymmetric reionization scenarios, the CORE limits are of $\mathcal{O}(30\%)$ looser than with nearly-instantaneous reionization (see Table~\ref{tab:DM_dec_EM_II}).
It is reassuring to find the same order of magnitude, since the reionization epoch is still poorly known. In the future, this type of uncertainty can be resolved by a better mapping of the reionization history coming from 21cm surveys~\cite{Pritchard:2011xb}.



\section{Constraints on the variation of the fine structure constant}

Fundamental constants of nature are numbers that characterize the theoretical framework we use to describe nature.  General relativity and the standard model of particles, on which the $\Lambda$CDM model is built on, are characterized by about 20 constants.
Measuring  the constancy, in space or time, of these numbers represents a very stringent test of the validity of such theories \citep{jpu-revue,jpu-llr,GarciaBerro:2007ir}. 

The fine structure constant characterizes the strength of the electromagnetic force. 
A wide variety of local experiments and astrophysical observations allows one to set constraints on the variation of $\alpha$ at very different redshifts, from the constraints set using atomic clocks ($z\sim 0$) \citep{resenband08} or the Oklo natural nuclear reactor ($z\sim 0.1$) \citep{damour96,Petrov:2005pu} to the ones from BBN ($z\sim 10^8$) \citep{Iocco:2008va}. The most stringent astrophysical bounds come from the observation of quasar spectra. Long-standing claims of a possible detection of the variation of $\alpha$ in these data at $z\sim 0.2-4$, at the level of $\sim \Delta\alpha/\alpha\sim 10^{-6}$, have further increased the interest for these measurements in the past decade \citep{webb01,dipole1,Evans:2014yva,Kotus:2016xxb}, although these claims are still the subject of controversy \citep{whitmore15}.

\begin{table}[h]
\begin{center}\footnotesize
\scalebox{0.80}{ \begin{tabular} { |c|c|c|c|c|c|}
 \hline
Parameter & Planck TTTEEE+lowP   & LiteCORE-80, TEP    & LiteCORE-120, TEP    & CORE-M5, TEP    & COrE+, TEP \\ 
\hline
{$\Omega_{\mathrm{b}} h^2$} & $0.02223\pm 0.00016        $ & $0.022179\pm 0.000053      $ & $0.022180\pm 0.000042      $ & $0.022181\pm 0.000039      $ & $0.022181\pm 0.000034      $\\

{$\Omega_{\mathrm{c}} h^2$} & $0.1190\pm 0.0019          $ & $0.12050\pm 0.00046        $ & $0.12049\pm 0.00038        $ & $0.12049\pm 0.00035        $ & $0.12049\pm 0.00032        $\\

{$100\theta_{\mathrm{MC}}$} & $67.1\pm 1.4               $ & $66.95\pm 0.42             $ & $66.95\pm 0.33             $ & $66.97\pm 0.30             $ & $66.96\pm 0.27             $\\

{$\tau           $} & $0.063\pm 0.015            $ & $0.0596\pm 0.0021          $ & $0.0597\pm 0.0020          $ & $0.0597^{+0.0020}_{-0.0022}$ & $0.0597\pm 0.0019          $\\

{$\alpha/\alpha_0$} & $0.9990\pm 0.0034          $ & $1.0000\pm 0.0010          $ & $0.999995\pm 0.00077       $ & $1.00004\pm 0.00070        $ & $1.00001\pm 0.00063        $\\

{$\ln(10^{10} A_\mathrm{s})$} & $3.059\pm 0.028            $ & $3.0561\pm 0.0039          $ & $3.0562\pm 0.0037          $ & $3.0562\pm 0.0037          $ & $3.0561\pm 0.0035          $\\

{$n_\mathrm{s}   $} & $0.9669\pm 0.0081          $ & $0.9619\pm 0.0030          $ & $0.9620\pm 0.0026          $ & $0.9619\pm 0.0025          $ & $0.9619\pm 0.0024          $\\
\hline
$H_0                       $ & $67.1\pm 1.4               $ & $66.95\pm 0.42             $ & $66.95\pm 0.33             $ & $66.97\pm 0.30             $ & $66.96\pm 0.27             $\\
$\sigma_8                  $ & $0.8135\pm 0.0095          $ & $0.8173\pm 0.0020          $ & $0.8173\pm 0.0015          $ & $0.8173\pm 0.0014          $ & $0.8173\pm 0.0013          $\\
\hline
\end{tabular}}
\caption{Constraints on the basic six-parameter $\Lambda$CDM model and the fine structure constant $\alpha/\alpha_0$ using different combination of datasets.}
\label{tab:alpha} 
\end{center}
\end{table}

The CMB is a very powerful probe of the value of the fine structure constant at redshift $z\sim 1000$ \citep{cmb-kap,cmb-han, Avelino:2001nr, Rocha:2003gc, Martins:2003pe}. A change in the value of $\alpha$ would in fact change the evolution of the recombination history of the universe, thus introducing a signature in the temperature and polarization power spectra.
Currently, the Planck experiment sets the strongest constraints on the fine structure constant from the CMB. The first release of the Planck data set a constraint on $\alpha/\alpha_0$, where $\alpha_0$ is the standard value, at the level of $0.4\%$, $\alpha/\alpha_0=0.9936 \pm 0.0043$ \citep{planckalpha15}. We calculate that the constraints from the second release of the Planck data \citep{planck2015}, combining both temperature and polarization and lensing reconstruction are at the level of  $0.34\%$ (Planck TTTEEE+lowTEB+lensing).

\begin{figure}
\centering
\includegraphics[width=0.85\textwidth]{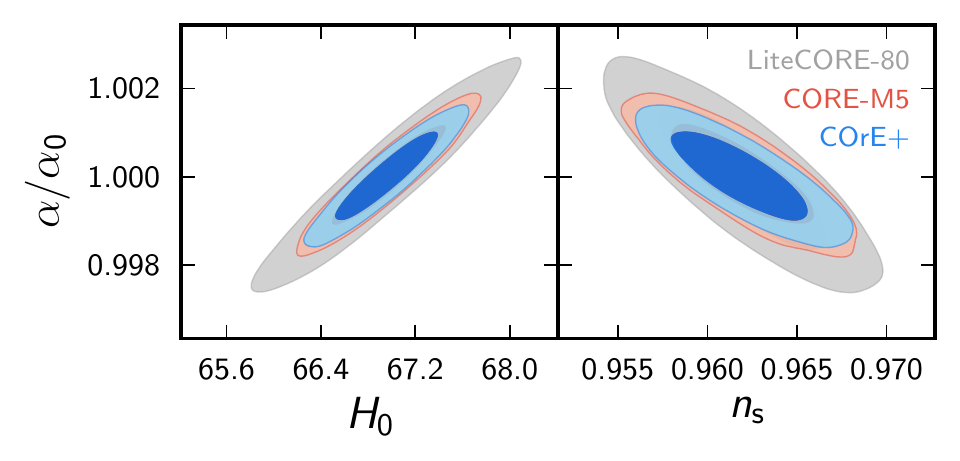}
\caption{Two-dimensional contour plots for $\alpha$ vs $H_0$ (Left Panel) and $n_s$ (Right Panel). }
\label{fig:alpha}
\end{figure}

Table \ref{tab:alpha} shows the improvement that a future satellite mission would bring to these constraints. We find that the LiteCoRE80, COrE-M5 and COrE configurations could improve the constraints by up to a factor of $5$ with respect to Planck, to $0.10\%$, $0.070\%$, and $0.063\%$  respectively. These constraints are essentially limited by the well known degeneracy between $\alpha$ and $H_0$ \citep{planckalpha15}, as also shown in Figure \ref{fig:alpha}. For this same reason, the constraint on $H_0$ is weakened by about a factor of $\sim 2$ when marginalizing over $\alpha$ with respect to the $\Lambda$CDM case. We find that adding the information from DESI would only marginally improve the results to $0.055\%$, $0.064\%$ and $0.077\%$. At face value, these constraints are still three orders of magnitude weaker compared to the latest quasar measurements (see e.g. \cite{Kotus:2016xxb}). However, the comparison is not straightforward since the CMB probes a very different range of redshifts compared to quasars. From an observational point of view, models where a dynamical degree of freedom yields a time variation of $\alpha$ can be divided into just two classes \cite{Martins:2014iaa}. If this degree of freedom is the one responsible for dark energy, then current low-redshift constraints imply that any $\alpha$ variations at $z\sim1100$ must be no larger than $10^{-5}$, and thus not directly detectable by the CMB. However, if the physical mechanism responsible for $\alpha$ variations is distinct from the one responsible for dark energy (or if the variations are environment-dependent rather than simply time-dependent) then no such extrapolation can be made, and variations at the level of $10^{-3}$ at $z\sim1100$ could easily be accommodated. Therefore improved high-redhift constraints which CORE can provide, when combined with the low-redshift spectroscopic ones, enable a key consistency test of the underlying theoretical paradigms.

\section{Constraints on the epoch of reionization}

The epoch of reionization (EoR) of the Universe is still largely unknown. The observation of the so-called  Gunn-Peterson trough \cite{Gunn:1965hd} in quasar spectra \cite{Becker:2001ee, Fan:2005eq,Venemans:2013npa,Becker:2015lua} indicates that hydrogen was almost fully reionized by $z\simeq 6$, possibly by the Lyman-$\alpha$ photons emitted by early star-forming galaxies. Quasars are then believed to be responsible for helium reionization between $z \simeq 6$ to $z \simeq 2$ (see e.g. \cite{Mesinger:2016} for a recent review).  

The CMB is a sensitive probe of the EoR, since the CMB photons can Compton scatter off free electrons generated by reionization. This leads to a suppression of the CMB anisotropies inside the Hubble horizon at the EoR, typically above $\ell \sim 10$, and to a regeneration of power below $\ell\sim 10$ in the TE and EE spectra (the so-called {\it reionization bump}) (see e.g. \cite{Kaplinghat:2002vt,Hu:2003gh,Lewis:2006ym,Mortonson:2007hq}). These two effects mostly depend on the column density of electrons along the line-of-sight\footnote{Note that potentially the CMB is sensitive to {\it inhomogeneous} (or patchy) reionization, which could also help in refining models. However, the non-gaussian signature of such a a process at small scales was shown to be very challenging to detect with a CORE-like experiment \cite{Su:2011ff,Natarajan:2012ie,Smith:2016lnt}.}, parametrized by the optical depth to reionization $\tau$. 

There are well known degeneracies between $\tau$ and other cosmological parameters, e.g., when using temperature data alone, with the amplitude of the primordial scalar perturbations $A_s$\footnote{The normalization of the $\ell>20$ part of the spectrum is mostly controlled by the product $A_s\exp^{-2\tau}$.} and the spectral index $n_s$. Moreover, in extensions of the $\Lambda$CDM model, there exists a degeneracy between $\tau$ and the sum of neutrino masses $M_{\nu}$, which gets strengthened by the addition of external datasets such as BAO measurements \cite{Liu:2015txa,Allison15}.
Thus, an accurate measurement of $\tau$ through the reionization bump at large scales is essential for the determination of other cosmological parameters as well.

Finally, the CMB, and in particular its polarization, could potentially provide more information about the evolution of the epoch of reionization than just the constraint on $\tau$\cite{Mortonson:2007hq}.

In this section, we thus quantify: i) how much the knowledge of the reionization epoch as observed by CORE would help constraining the other cosmological parameters; ii) how well CORE will be able to provide information about the evolution of the EoR, beyond an accurate measurement of $\tau$.

\begin{figure}[!t]
\includegraphics[width=15cm]{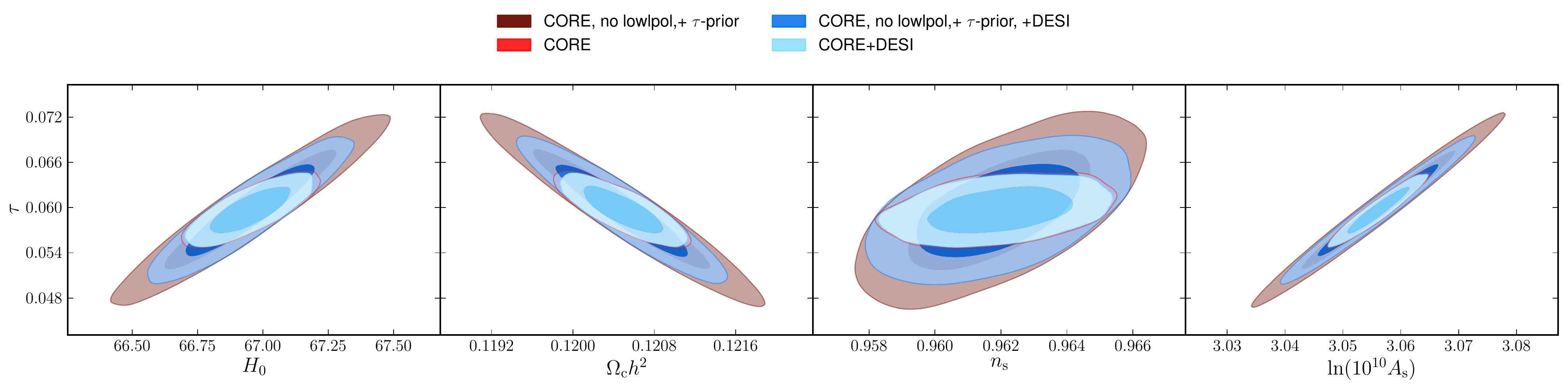}
\includegraphics[width=15cm]{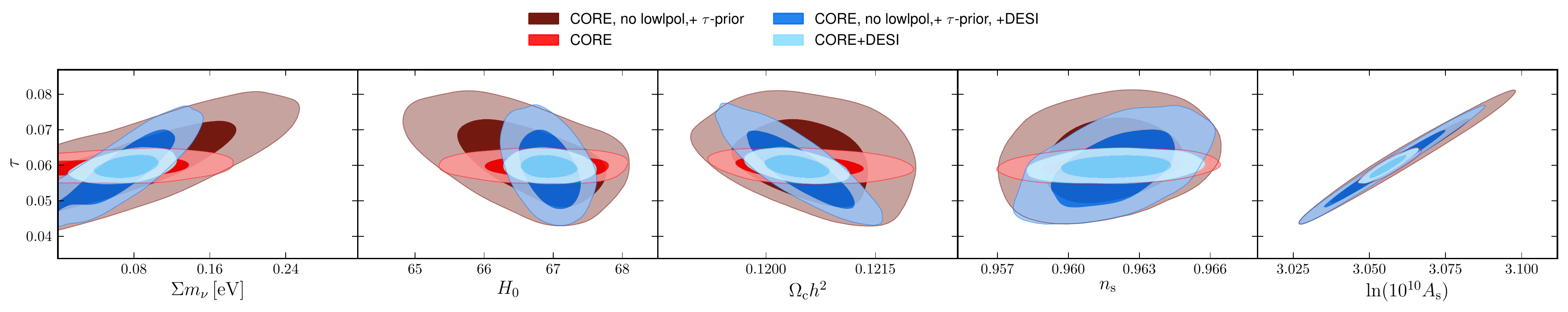}
\caption{Two-dimentional posterior distributions of cosmological parameters in the $\Lambda$CDM (top panel) and $\Lambda$CDM+$\sum m_\nu$ (bottom panel) cases estimated using CORE with the largest scales in polarization excised and using a prior on $\tau$ (dark  red) or CORE (light red). The same cases  with the addition of the DESI mock dataset are shown in dark blue and light blue respectively.  \label{fig:tau_mnu}}
\end{figure}

In order to tackle the first point, we forecast constraints on cosmological parameters excising the low-$\ell$ polarization spectra at $\ell<30$, and using a gaussian prior in $\tau$ with an uncertainty of  $\sigma_{\rm prior}(\tau)=\pm0.01$, consistent with the precision of the latest results from {\it Planck} \cite{Aghanim:2016yuo}. This is about $4$ times worse than the constraint that a CORE-like experiment could achieve using the full large scale polarisation information, as already shown in Section~\ref{sec:lcdm}.
We find that in the $\Lambda$CDM case, excising the large scales in polarization degrades the constraints in the case of CORE-M5 (CORE-M5 + DESI) by a factor of $\sim 2.5$ ($\sim 2$) on $\tau$ and $\log{A_{\rm s}}$, a factor $\sim 2$ ($\sim 1.6$) on $\Omega_{\mathrm c}$ and $ \theta_{*}$, and by $30\%$ ($14\%$) on  $n_{\mathrm s}$, leaving $\Omega_bh^2$  unaffected. Note that the recovered constraint on $\tau$ is {\it stronger} than the prior, at the level of $\sigma(\tau)=\pm0.005$. This is due to the fact that the degeneracy between $\tau$ and $A_{\rm s}$ is reduced by the information  provided by lensing on $A_{\rm s}$.
As the $\Lambda$CDM+$M_\nu$ case is concerned, we find that the upper limit on the sum of neutrino masses would be degraded in the case of CORE-M5 from $M_\nu< 152 \,\mathrm{meV}$ to $M_\nu< 201\, \mathrm{meV}$ (95 \% CL), and and that the constraint from CORE-M5 + DESI would worsen from $\sigma(M_\nu)=21\,\mathrm{meV}$ to $ \sigma(M_\nu)=34\,\mathrm{meV}$, while other cosmological parameters would be less affected than in the $\Lambda$CDM case, as also shown in Fig. \ref{fig:tau_mnu}.
This illustrates the importance of a precise $\tau$ measurement for a highly statistically significant  detection of the sum of neutrino masses.

We now turn to  quantify how much CORE could be sensitive to the evolution of the reionization history. In order to do so, we use the usual fiducial model generated assuming a redshift symmetric (nearly instantaneous) reionization parametrization, but we perform a cosmological parameter extraction using the redshift-asymmetric parametrization introduced in the Planck 2016 analysis of~\cite{Adam:2016hgk}:
\begin{equation} \label{eq:reio_param_asym}
x_e(z) =\left\{ \begin{array}{cl}
&\hspace{-0.2cm}f \hspace{3.12cm}\textrm{for } z < z_{\rm stop},\\
&\hspace{-0.2cm}f\times \bigg(\frac{z_{\rm early}-z}{z_{\rm early}-z_{\rm stop}}\bigg)^\alpha \:\:\;\textrm{for } z \geq z_{\rm stop},\quad\quad\text{with}\quad f = 1+f_{\rm He} = 1 + n_{\rm He}/n_{\rm H}. 
\end{array}\right.
\end{equation} 
We fix the redshift of formation of the first emitting sources to $z_{\rm early} = 20$, and vary two parameters: the redshift at which reionization ends $z_{\rm stop}$ (with flat prior in the range $[1,15]$), and the exponent $\alpha$ (with flat prior in the range $[2,50]$). 
This allows to quantify how sensitive CORE is to the {\it duration} of the reionization era,  $\Delta z_{\rm reio}=z_{\rm beg} - z_{\rm end}$, where $z_{\rm beg}\equiv z_{10\%}$ is defined by $x_e( z_{10\%})= 0.1\times f$ and $z_{\rm end}\equiv z_{99\%}$ by $x_e( z_{99\%})= 0.99\times f$. It also gives a hint of how accurately one could measure the redshift at which $x_e(z)= 0.5\times f$, usually called $z_{\rm reio} \equiv z_{50\%}$.

\begin{table}[h]
\begin{center}\footnotesize
\scalebox{0.87}{\begin{tabular}{|c||c|c|c|c|}
\hline
Parameter & LiteCORE-80, TEP & LiteCORE-120, TEP & CORE-M5, TEP & COrE+, TEP\\
\hline 
$z_\mathrm{stop}$ & $7.75^{+0.47}_{-0.23}$& $7.81^{+0.39}_{-0.24}$& $7.82^{+0.38}_{-0.24}$& $7.82^{+0.36}_{-0.25}$\\
$\Delta z_\mathrm{reio}$ & $<3.39$ (95\%CL)& $<2.65$ (95\%CL)& $<2.55$ (95\%CL)& $<2.37$ (95\%CL)\\
$\Omega_b h^2$ & $0.022258\pm0.000055$& $0.022257\pm0.000041$& $0.022256\pm0.000039$& $0.022256\pm0.000034$\\
$\Omega_c h^2$ & $0.11973\pm0.00034$& $0.11973\pm0.00030$& $0.11974\pm0.00028$& $0.11974\pm0.00028$\\
$100\theta_s$ & $1.04080\pm0.00010$& $1.040803\pm0.000084$& $1.040802\pm0.000081$& $1.040803\pm0.000074$\\
$\tau$ & $0.0605\pm0.0021$& $0.0604\pm0.0021$& $0.0604\pm0.0021$& $0.0604\pm0.0021$\\
$n_s$ & $0.9645\pm0.0017$& $0.9646\pm0.0016$& $0.9645\pm0.0015$& $0.9646\pm0.0014$\\
$ln(10^{10} A_s)$ & $3.0949\pm0.0038$& $3.0947\pm0.0037$& $3.0947\pm0.0037$& $3.0946\pm0.0037$\\
\hline
$H_0$ (km/s/Mpc) & $67.02\pm0.14$& $67.02\pm0.12$& $67.01\pm0.11$& $67.01\pm0.11$\\
$\sigma_8$ & $0.8303\pm0.0014$& $0.8300\pm0.0012$& $0.8300\pm0.0011$& $0.8301\pm0.0011$\\
$z_\mathrm{beg}$ & $9.05^{+0.19}_{-0.59}$& $8.97^{+0.23}_{-0.50}$& $8.95^{+0.23}_{-0.48}$& $8.94^{+0.22}_{-0.46}$\\
$z_\mathrm{reio}$ & $8.16\pm0.22$& $8.17\pm0.21$& $8.17\pm0.21$& $8.17\pm0.21$\\
$z_\mathrm{end}$ & $7.74^{+0.48}_{-0.23}$& $7.80^{+0.40}_{-0.24}$& $7.82^{+0.38}_{-0.24}$& $7.82^{+0.36}_{-0.24}$\\

\hline
 \end{tabular}} 
\end{center}  
\caption{$68\%$~CL constraints on the $\Lambda$CDM + $\alpha$ + $z_\mathrm{stop}$ model (asymmetric reionization), for the different CORE experimental specifications. Instead of lower bounds on $\alpha$, we report the more interesting upper bounds on the derived parameter $\Delta z_{\rm reio}$. We also show the results for the derived parameters ($z_\mathrm{beg}$, $z_\mathrm{reio}$, $z_\mathrm{end}$, $\Delta z_\mathrm{reio}$).\label{tab:reio_asym}}
\end{table}


Our results are displayed in Table~\ref{tab:reio_asym}. 
Within the prior range $[1,15]$, $\alpha$ is only bounded from below by the data, thus leading to an upper bound on the derived parameter $\Delta z_{\rm reio}=z_{\rm beg} - z_{\rm end}$. Since this parameter has a much more intuitive interpretation, we report bounds on $\Delta z_{\rm reio}$ instead of $\alpha$ in Table~\ref{tab:reio_asym}.
We also present the bounds on the derived parameters ($z_\mathrm{beg}$, $z_\mathrm{reio}$, $z_\mathrm{end}$).

Since most of the information on reionization comes from polarisation on large angular scales, on which CORE measurements are cosmic-variance-limited, we could expect all configurations to be equally sensitive to this model. In fact, the most sensitive configurations are able to extract the CMB lensing spectrum in a larger range of scales: thus they corner $A_s$ with better accuracy, and reionization results are less affected by the $\tau-A_s$ degeneracy.

We find that LiteCORE-80 could set a constraint on the duration of reionization given by $ \dzreio < 3.4$ (95 \% CL), while CORE-M5 (COrE+)  would improve the constraint to $\dzreio< 2.6 $ (2.4). This is about two times better than the constraints from Planck CMB anisotropies combined with Kinetic Sunyaev$-$Zel'dovich measurements, and a factor of order 4 better than Planck alone \cite{Adam:2016hgk}, without using any prior on $\zend$. Note also that CORE-M5 would be able to provide precise measurements of $\zbeg$, $\zend$ and $\zreio$, with $\sigma(\zbeg) \simeq 0.33$, $\sigma(\zend) \simeq 0.31$ and $\sigma(\zreio) \simeq 0.21$, to be compared to the recent Planck measurements,  $\sigma(\zbeg) \simeq 1.9$, $\zend \lesssim 10$ and $\sigma(\zreio) \simeq1.1$. We therefore conclude that a CORE-like experiment would be sensitive enough to constrain the end of the EoR from CMB data only, and would improve the determination of $\zreio$ and $\zbeg$ by a factor of 4 and 6 respectively.

\section{Constraints on Modified Gravity}

\subsection{Theoretical framework}

The current accelerated expansion of the universe could be also explained by introducing modifications
to general relativity and considering an energy content made just of dark matter and baryons and no
dark energy.
Several modified gravity scenarios have been proposed. One possible way to check for hints for modified
gravity in the data, without relying on a particular model, is to introduce additional parameters to
perturbation theory that can modify the evolution of the gravitational potentials $\Phi$ and $\Psi$
(see, for example \cite{Zhao:2010dz,Daniel:2010ky1,Daniel:2010ky2,Simpson:2012ra,DiValentino:2012yg,Giannantonio:2009gi,Macaulay:2013swa,
Hojjati:2013xqa,Marchini:2013oya,Hu:2013aqa,Munshi:2014tua,Boubekeur:2014uaa,Hu:2014sea, Cataneo:2014kaa,Dossett:2015nda,Johnson:2015aaa,planckmg,DiValentino:2015bja}).
For example, a now common approach, presented in the publicly available code \texttt{MGCAMB} \cite{Zhao:2008bn,Hojjati:2011ix}
and also recently applied in \cite{planckmg} and \cite{DiValentino:2015bja} to the Planck data, is to firstly modify the 
Poisson equation for $\Psi$:

\begin{equation}  
k^2\Psi=-4 \pi G a^2\mu(k,a)\rho_{dm}\Delta \,,
\end{equation}

\noindent introducing the scale-dependent function $\mu(k,a)$. In the above equation, $\rho_{dm}$ is the dark matter energy density and $\Delta$ is the comoving density perturbation. 
Secondly, one can also introduce the possibility of an additional anisotropic stress considering a second function $\eta(k,a)$,
such that:
\begin{equation}  
\frac{\Phi}{\Psi}=\eta(k,a) \,.
\end{equation}

\noindent A third function, $\Sigma(k ,a)$, which modifies the lensing/Weyl
potential $\Phi+\Psi$ can be introduced as:
\begin{equation}
 -k^2 (\Phi+\Psi) \equiv 8\pi G a^2 \Sigma(k,a) \rho  \Delta \,,
\end{equation}

\noindent This function is not independent from  $\mu(k,a)$ and $\eta(k,a)$ since: 
\begin{equation} 
 \Sigma=\frac{\mu}{2} (1+\eta ) \,.
\end{equation}

\noindent These functions can be used to study the effects of a possible modification to GR. If GR is valid
then $\mu = \eta = \Sigma=1$. 
Here we use the following parametrization:

\begin{eqnarray} \label{DErel}
\mu(k,a) &=& 1 + E_{\rm{11}} \Omega_{\rm{\Lambda}}(a)\,; \\
\eta(k,a) &=& 1 + E_{\rm{22}} \Omega_{\rm{\Lambda}}(a)\,.
\end{eqnarray}

\noindent where $E_{\rm{11}}$ and $E_{\rm{22}}$ are two parameters that are constant with
redshift and $\Omega_{\Lambda}$ is the energy density in the cosmological constant 
that we choose as a good approximation for the background evolution.

\noindent Computing the $\Sigma$ function today we have:

\begin{equation}
\Sigma_0=\frac{1+E_{\rm{11}}\Omega_{\rm{\Lambda}}}{2}(2 + E_{\rm{22}} \Omega_{\rm{\Lambda}})
\end{equation}

\noindent i.e. $\Sigma_0=1$ if GR is valid.

A detection of $\Sigma_0 -1\neq 0$ could therefore indicate a departure of the evolution of density
perturbations from GR. Interestingly, the recent Planck 2015 data suggest a value
of $\Sigma_0-1=0.23\pm0.13$ \cite{planckmg,DiValentino:2015bja} at $68$\% CL, i.e.
a presence for MG slightly above two standard deviations.
Of course, given the very low statistical significance, this indication can be just due to
a statistical fluctuation or to a small residual systematic.
However, it is clearly important to study what kind of constraint can be achieved by future CMB data
and at which level of confidence the current hint could be falsified.

\subsection{Future constraints from CORE}

\begin{figure}
	\centering
		\includegraphics[width=7.3cm]{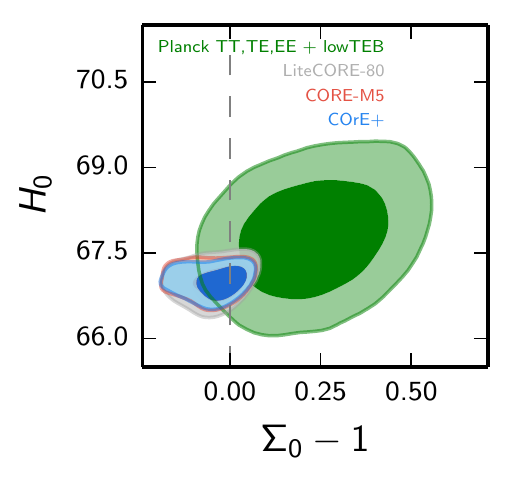}	
		\includegraphics[width=7.0cm]{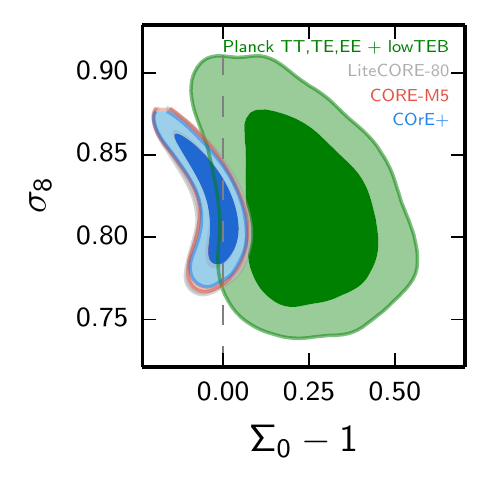}	
	\caption{Constraints on the $H_0$ vs $\Sigma_0-1$  plane (Left Panel) and  $\sigma_8$ vs $\Sigma_0-1$ plane (Right Panel) from different CORE configurations and from current Planck 2015 temperature and polarization data. The dashed line corresponds to the GR prediction.}
\label{fig:sigma0}
\end{figure}

\begin{table}[h]
\begin{center}\footnotesize
\scalebox{0.87}{\begin{tabular}{|c||c|c|c|c|}
\hline 
Parameter         & LiteCORE-80, TEP& LiteCORE-120, TEP& CORE-M5, TEP& COrE+, TEP  \\            
\hline
$E_{11}$ &  $0.03^{+0.35}_{-0.44}$ & $0.01^{+0.34}_{-0.45}$ & $0.04^{+0.34}_{-0.48}$ & $0.05^{+0.34}_{-0.43}$ \\
$E_{22}$&  $0.03^{+0.72}_{-1.0}$ &  $0.08^{+0.72}_{-1.0}$  &  $0.0^{+0.7}_{-1.0}$ &  $-0.03^{+0.68}_{-0.99}$ \\
$\Omega_bh^2$ &      $0.022186\pm0.000055$& $0.022183\pm0.000041$ & $ 0.022182\pm0.000038$ & $ 0.022181\pm0.000032$ \\
$\Omega_ch^2$ &      $0.12039\pm0.00056$& $0.12038\pm0.00050$ & $ 0.12037\,^{+0.00049}_{-0.00044}$ & $ 0.12038\pm0.00044$ \\
$100\theta_{MC}$ &      $1.04069\pm0.00010$& $1.040692\pm0.000087$ & $ 1.040697\pm0.000080$ & $ 1.040698\pm0.000077$ \\
$\tau$ &      $0.0597\,_{-0.0022}^{+0.0020}$& $0.0597\,^{+0.0019}_{-0.0022}$ & $ 0.0596\pm0.0020$ & $ 0.0597\pm0.0020$ \\
$n_s$ &      $0.9621\pm0.0019$& $0.9620\pm0.0017$ & $ 0.9621\pm0.0017$ & $ 0.9621\pm0.0016$ \\
$ln(10^{10}A_s)$ &      $3.0559\pm0.0042$& $3.0559\,^{+0.0039}_{-0.0044}$ & $ 3.0559\pm0.0040$ & $ 3.0560\,^{+0.0037}_{-0.0042}$ \\
\hline
$\mu_0 -1$& $0.02^{+0.24}_{-0.30}$ & $0.01^{+0.23}_{-0.31}$  & $0.03^{+0.23}_{-0.32}$ & $0.04^{+0.23}_{-0.29}$\\
$\eta_0 -1$&  $0.02^{+0.49}_{-0.70}$  & $0.06^{+0.49}_{-0.71}$ & $0.00^{+0.49}_{-0.71}$ & $-0.02^{+0.46}_{-0.67}$\\
$\Sigma_0 -1$& $-0.034 \,^{+0.062}_{-0.035}$ & $-0.033^{+0.060}_{-0.032}$  & $-0.036^{+0.063}_{-0.029}$ & $-0.034^{+0.061}_{-0.028}$\\
$H_0$ [km/s/Mpc]&      $66.99\pm0.23$& $66.99\pm0.20$ & $ 67.00\,^{+0.17}_{-0.19}$ & $ 66.99\pm0.17$ \\
$\sigma_8$ &      $0.819\,^{+0.025}_{-0.030}$& $0.818\,^{+0.025}_{-0.030}$ & $ 0.820\,^{+0.024}_{-0.032}$ & $ 0.821\,^{+0.024}_{-0.030}$ \\
\hline
\end{tabular}}
\end{center}
\caption{$68\%$~CL constraints on cosmological parameters from four different CORE configurations. The possibility of modified gravity is allowed.}
\label{tab:mg}
\end{table}

In Table \ref{tab:mg} we present the  constraints on modified gravity parameters using the three different experimental configurations
for CORE, under the assumption of GR (i.e. $\Sigma_0-1=0$). As we can see the current constraints on $\Sigma_0$ can be improved by nearly a factor
three with respect to current constraints from Planck 2015, quite independently from the choice of the experimental configuration.
Constraints on the $H_0$  vs $\Sigma_0-1$ and $\sigma_8$ vs $\Sigma_0-1$ planes are also reported in Figure \ref{fig:sigma0} from 
three CORE configurations and also from the current Planck 2015 data. The improvement of CORE with respect to Planck is clearly visible.
\noindent A future CMB experiment could therefore confirm or exclude at high significance (about four standard deviations) the 
current hints for MG from Planck.

\section{Cosmological Birefringence}
Cosmological birefringence is the in vacuo rotation of the photon polarization direction during propagation \cite{Carroll:1989vb}.
In general, such effect is unconstrained by the $TT$ spectrum, while results in a mixing between $Q$ and $U$ Stokes parameters that produces non-null CMB cross correlations between temperature and $B$-mode polarization, 
and between $E$- and $B$-mode polarization. 
Since these correlations are expected to vanish under parity conserving assumptions, cosmic birefringence
is a tracer of parity violating physics. 

Several theoretical models exhibit cosmological birefringence, such as coupling of the electromagnetic (EM) field with axion-like particles \cite{Finelli:2008jv} or a quintessence field \cite{Giovannini:2004pf}, quantum-gravity terms \cite{Gubitosi:2009eu} or Chern-Simons type interactions \cite{Carroll:1989vb} in the EM Lagrangian.
For the sake of simplicity, we restrict to the case of constant, isotropic $\alpha$, for which the effect can be parametrized as \cite{Lue:1998mq,Feng:2004mq,Gubitosi:2014cua}
\begin{eqnarray}\label{eq:aps}
C_{\ell}^{TE,obs}  &=&  C_{\ell}^{TE} \cos (2 \alpha) \, ,
\label{TEobs} \\
 C_{\ell}^{TB,obs}  &=&  C_{\ell}^{TE} \sin (2 \alpha) \, ,
\label{TBobs} \\
 C_{\ell}^{EE, obs} &=& C_{\ell}^{EE}  \cos^2 (2 \alpha) +  C_{\ell}^{BB} \sin^2 (2 \alpha) \, ,
\label{EEobs} \\
 C_{\ell}^{BB, obs}  &=&  C_{\ell}^{BB}  \cos^2 (2 \alpha) +  C_{\ell}^{EE} \sin^2 (2 \alpha) \, ,
\label{BBobs} \\
 C_{\ell}^{EB, obs} &=& {1 \over 2} \left(  C_{\ell}^{EE}  - C_{\ell}^{BB}  \right) \sin (4 \alpha) \, ,
\label{EBobs}
\end{eqnarray}
\noindent with $C_{\ell}^{XY,obs} $ and $C_{\ell}^{XY}$ being the observed and the unrotated power spectra for the $XY$ fields ($X$, $Y$ = $T$, $E$ or $B$),
i.e. the one that would arise in absence of birefringence. We set the primordial $TB$ and $EB$ spectra to zero, assuming a negligible role of parity violation effects up to CMB photon decoupling (this choice excludes e.g. chiral gravity theories). Recent constraints on this model employing Planck 2015 data are reported in \cite{Aghanim:2016fhp,Gruppuso:2015xza}.

We remind that the most relevant systematic effect affecting constraints on isotropic birefringence is the miscalibration of the detector polarization angle. An estimate of the error budget from current experiments is $\sim 1^{\circ}$, already dominant over the statistical error achievable on $\alpha$ \cite{Gruppuso:2015xza}. As a result, future experiments should require an exquisite control of systematic effects to really nail down constraints on isotropic birefringence.

\begin{figure}
\begin{center}
\includegraphics[width=0.8\textwidth]{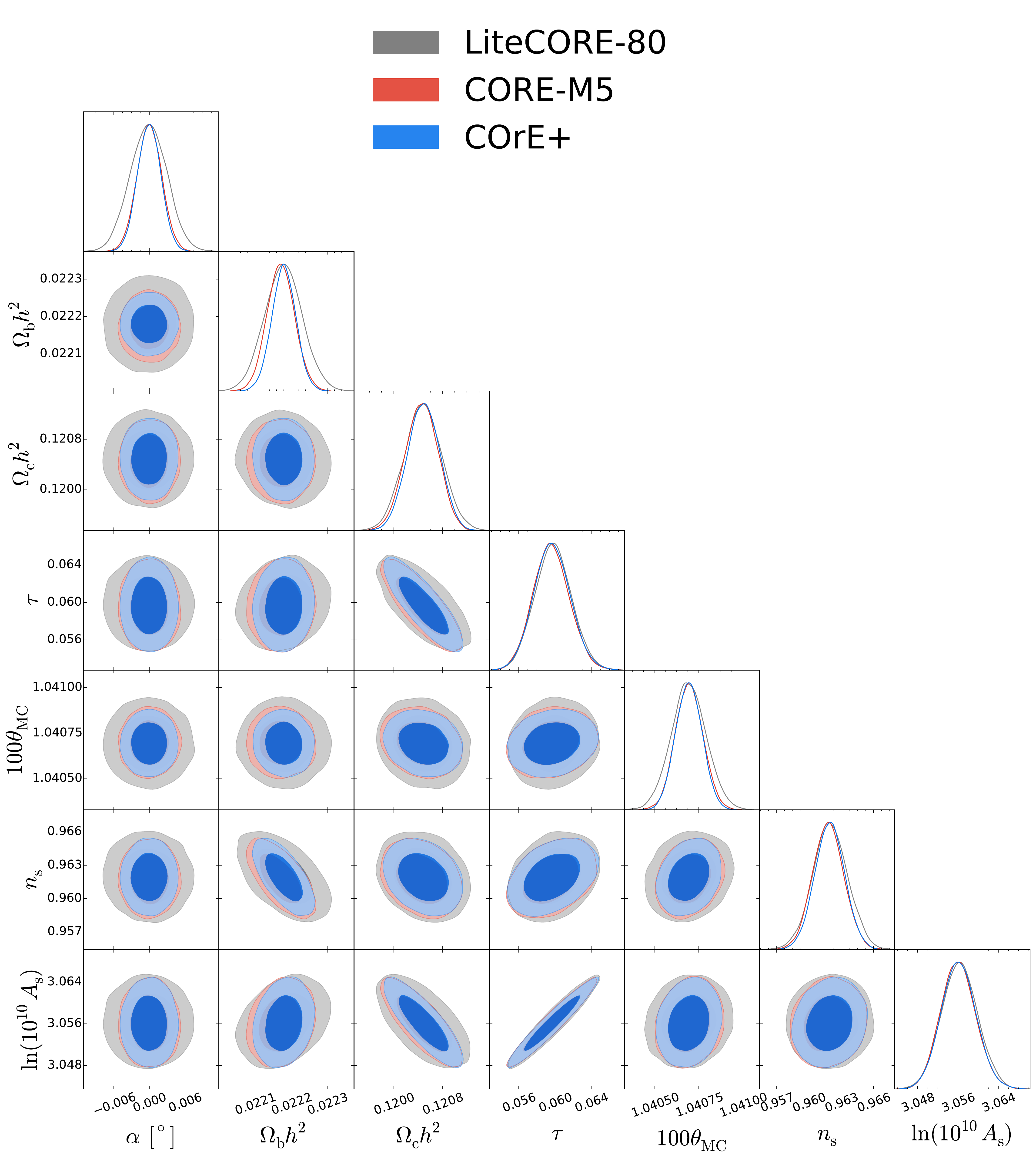}
\end{center}
\caption{Triangular plot showing the main degeneracies between the comological parameters listed in Tab.\ref{tab:biri}, for the three experimental configurations reported in this work.}\label{fig:biri}
\end{figure}

In Tab.~\ref{tab:biri}, we report the 68\% c.l. around the mean for the birefringence angle $\alpha$ and other cosmological parameters. We employ the full $TEB$ combination of power spectra for the four experimental configurations analyzed in this paper. 
In Fig.~\ref{fig:biri}, we report two-dimensional 68\% and 95\% probability contours for the same parameters listed in Tab.\ref{tab:biri} and for the same experimental configurations. The first column clearly shows no evidence of correlation between the birefringence angle and other cosmological parameters. One-dimensional posterior probability distributions for the same parameters are reported along the diagonal of Fig.~\ref{fig:biri}.

\

\begin{table}[h]
\scalebox{0.83}{\begin{tabular}{|c||c|c|c|c|}
\hline
Parameter                             & LiteCORE-80, TEB     & LiteCORE-120, TEB                         & CORE-M5, TEB	&COrE+, TEB\\
\hline
$ \alpha\,\mathrm{[^{\circ}]}$ & $ -0.0001\pm 0.0030                 $& $ 0.0000\pm 0.0026                 $	&$ 0.0000\pm 0.0021                 $&$ 0.0000\pm 0.0019                $\\
$ \Omega_b h^2          $ & $ 0.022179\pm 0.000052             $& $ 0.022181\pm 0.000041             $& $ 0.022172\pm 0.000038             $& $ 0.022180\pm 0.000033             $\\
$ \Omega_c h^2          $ & $ 0.12048\pm 0.00030               $& $ 0.12049\pm 0.00029               $& $ 0.12046\pm 0.00026               $& $ 0.12049\pm 0.00026               $\\
$ 100\theta_{MC}          $ & $ 1.040693\pm 0.000098             $& $ 1.040693\pm 0.000083             $& $ 1.040696\pm 0.000077             $& $ 1.040694\pm 0.000073             $\\
$ \tau                             $ & $ 0.0598\pm 0.0020                 $& $ 0.0597^{+0.0019}_{-0.0021}       $& $ 0.0596^{+0.0019}_{-0.0021}       $& $ 0.0597\pm 0.0020                 $\\
$ n_s                     $ & $ 0.9620\pm 0.0016                 $&$0.9619\pm 0.0015                 $& $ 0.9618\pm 0.0014                 $& $ 0.9619\pm 0.0014                 $\\
$ ln(10^{10} A_s)        $ & $ 3.0563\pm 0.0035                 $& $ 3.0562\pm 0.0035                 $& $ 3.0561\pm 0.0034                 $& $ 3.0562\pm 0.0034                 $\\
\hline
$ H_0                              $ [km/s/Mpc]& $ 66.95\pm 0.13                    $& $ 66.95\pm 0.12                    $& $ 66.96\pm 0.11                    $& $ 66.95\pm 0.10                    $\\
$ \sigma_8                         $ & $ 0.8173\pm 0.0012               $& $ 0.8173\pm 0.0011                 $& $ 0.81714\pm 0.00096               $& $ 0.81728\pm 0.00094                 $\\
\hline
\end{tabular}}
\caption{68\% CL for the birefringence angle $\alpha$ and other cosmological parameters for the four experimental configurations presented in this work and for the full $TEB$ field combination.}\label{tab:biri}

\end{table}

\section{Conclusions}

In this paper we forecasted the constraints on several cosmological parameters that can be achieved by the CORE-M5 satellite proposal.
Table \ref{tab:key} provides a summary of our main results. Assuming $\Lambda$CDM, the
 improvement with respect to Planck is extremely significant: CORE-M5 can simultaneously improve constraints on key parameters by
a factor $\sim 8$ ($\sigma_8$), $\sim 5.5$ ($H_0$, $\Omega_{cdm}h^2$), $4.5$ ($\Omega_bh^2$, $\tau$), and $3$ ($n_s$). 

Some of the parameters such as $\sigma_8$, $H_0$, and $\Omega_b h^2$ can be measured or derived independently
by galaxy surveys or luminosity distance measurements. Future comparisons with the CORE-M5 results will therefore
provide a crucial test for cosmology and the $\lcdm$ scenario and its extensions. 
The interest of such measurements by several means is exemplified by the current tensions between the Planck dataset and the local determination of the Hubble constant from \cite{riess2016} or measurements of weak lensing cosmic shear from surveys as CFHTLenS and KiDS-450 \cite{cfhtlens,kids}. These tensions may reveal either previously unknown systematic effects, or new physics. While these current tensions will likely be resolved by the time CORE flies, the large improvement brought by CORE on so many parameters will surely bring new opportunities for revealing tensions with whatever precision datasets will be available by then. These are opportunites for fundamental breakthoughs. 

\begin{table}[h]
\begin{center}\footnotesize
\scalebox{0.73}{
\begin{tabular}{|c|c|c|c|}
\hline
  Parameter & Description & Current results (Planck 2015+Lensing) & CORE expected uncertainties\\
\hline
    $\Lambda$CDM    &&&\\
\hline
    $\Omega_bh^2$ &Baryon Density & $\Omega_bh^2=0.02226 \pm 0.00016 \,$ (68 \% CL) \cite{planck2015}& $\sigma (\Omega_bh^2) = {\bf 0.000037}$ $\{4.3\}$\\
    $\Omega_ch^2$ &Cold Dark Matter Density & $\Omega_ch^2=0.1193 \pm 0.0014 \,$ (68 \% CL) \cite{planck2015}& $\sigma (\Omega_ch^2) = {\bf 0.00026}$ $\{5.4\}$\\
    $n_s$ &Scalar Spectral Index & $n_s=0.9653 \pm 0.0048 \,$ (68 \% CL) \cite{planck2015}& $\sigma (n_s) = {\bf 0.0014}$ $\{3.4\}$\\
    $\tau$ & Reionization Optical Depth & $0.063 \pm 0.014\,$ (68 \% CL) \cite{planck2015} & $\sigma (\tau) = {\bf 0.002}$ $\{7.0\}$\\
    $H_0$ [km/s/Mpc]& Hubble Constant & $H_0=67.51 \pm 0.64\,$ (68 \% CL) \cite{planck2015} & $\sigma (H_0) = {\bf 0.11}$ $\{5.8\}$ \\
    $\sigma_8$ & r.m.s. mass fluctuations & $\sigma_8=0.8150 \pm 0.0087\,$ (68 \% CL) \cite{planck2015} & $\sigma (\sigma_8) = {\bf 0.0011}$ $\{7.9\}$\\
\hline    
    Extensions    &&&\\
\hline
    $\Omega_{\rm k}$ &Curvature & $\Omega_{\rm k} = -0.0037^{+0.0083}_{-0.0069} \,$ (68 \% CL) \cite{planck2015} & $\sigma (\Omega_{\rm k}) = {\bf 0.0019}$ $\{4\}$\\
    $N_{\rm eff}$ &Relativistic Degrees of Freedom & $N_{\rm eff} = 2.94 \pm 0.20 \,$ (68 \% CL) \cite{planck2015}& $\sigma (N_{\rm eff}) = {\bf 0.041}$ $\{4.9\}$\\    
    $M_{\nu}$ & Total Neutrino Mass & $M_{\nu} < 0.315 \,$eV (68 \% CL) \cite{planck2015} & $\sigma (M_{\nu}) = {\bf 0.043}$ eV $\{7.3\}$\\
    $(m_s^{eff},N_s)$ & Sterile Neutrino Parameters& $(m_s ^{eff}< 0.33 eV, N_s < 3.24)\,$ (68 \% CL) \cite{planck2015} & $\sigma (m_s^{eff},N_s) = ({\bf 0.037} eV,{\bf 0.053})$ $\{8.9,4.5\}$ \\
    $Y_p$ & Primordial Helium abundance & $Y_p =0.247\pm0.014 \,$ (68 \% CL) \cite{planck2015} & $\sigma (Y_p) = {\bf 0.0029}$ $\{4.8\}$\\                        
    $Y_p$ & Primordial Helium (free $N_{eff}$)& $Y_p =0.259^{+0.020}_{-0.017} \,$ (68 \% CL) \cite{planck2015} & $\sigma (Y_p) = {\bf 0.0056}$ $\{3.2\}$\\
    $\tau_n$ [s]& Neutron Life Time & $\tau_n=908\pm69 \,$ (68 \% CL) \cite{Salvati:2015wxa} & $\sigma (\tau_n) = {\bf 13}$ $\{5.3\}$\\    

    $w$ & Dark Energy Eq. of State & $w=-1.42^{+0.25}_{-0.47} \,$ (68 \% CL) \cite{planck2015} & $\sigma (w) = {\bf 0.12}$ $\{3\}$\\    
    $T_0$ & CMB Temperature & Unconstrained \cite{planck2015} & $\sigma (T_0) = {\bf 0.018}$ K\\                        
    $p_{ann}$ & Dark Matter Annihilation & $p_{ann}< 3.4\times10^{-28} \,$ $cm^3/GeV/s$ (68 \% CL) \cite{planck2015} & $\sigma (p_{ann}) = {\bf 5.3\times10^{-29}}$ $cm^3/GeV/s$ $\{6.4\}$\\  
    $ g_\mathrm{eff}^4 $ & Neutrino self-interaction & $ g_\mathrm{eff}^4 < 0.22 \times 10^{-27} $ & $\sigma(g_\mathrm{eff}^4) = 0.34\times 10^{-28}$ $\{6.4\}$ \\                      
    $\alpha/\alpha_0$ & Fine Structure Constant & $\alpha/\alpha_0=0.9990\pm0.0034 \,$ (68 \% CL)  & $\sigma (\alpha/\alpha_0) = {\bf 0.0007}$ $\{4.8\}$\\                        
    $\Sigma_0-1$ & Modified Gravity & $\Sigma_0-1=0.10\pm0.11 \,$ (68 \% CL) \cite{planckmg} & $\sigma (\Sigma_0-1) = {\bf 0.044}$ $\{2.5\}$\\		                                                       
    $A_{2s1s}/8.2206$ & Recombination 2 photons rate& $A_{2s1s}/8.2206= 0.94\pm0.07\,$ (68 \% CL) \cite{planck2015} & $\sigma (A_{2s1s}/8.2206) = {\bf 0.015}$ $\{4.7\}$\\                        
    $\Delta (z_{reio})$ & Reionization Duration & $\Delta (z_{reio})< 2.26\,$ (68 \% CL) \cite{planck2016reio} & $\sigma (\Delta z_{reio}) = {\bf 0.58}$ $\{3.9\}$\\
\hline    
 \end{tabular}}
\end{center}
\caption{\small  Current limits from Planck 2015 and forecasted CORE-M5 uncertainties.
 The first $6$ rows assume a $\Lambda$CDM scenario while the following rows give the constraints on single parameter extensions. In the fourth column, numbers in curly brackets $\{...\}$ give the improvement in the parameter constraint when moving from Planck 2015 to CORE-M5, defined as the ratio of the uncertainties $\sigma^{Planck}/\sigma^{CORE}$.
 }
\label{tab:key}
\end{table}

In this paper, we have considered several possible extensions to the basic six parameters $\Lambda$CDM model. The forecasted constraints on these extra parameters
are summarized in the second section of Table \ref{tab:key}. As we can see, also on these extensions CORE-M5 can provide significantly more
stringent constraints than the current ones, with a factor of n improvement that ranges from $2$ up to more than $6$, clearly opening the window to new tests or discoveries for physics beyond the standard model. 

In particular, we found  that:

\begin{itemize}

\item  CORE-M5 alone could detect neutrino masses with an uncertainty of  $\sigma (M_{\nu}) = {0.043}$ eV, enough 
to rule out the inverted mass hierarchy at more than $95 \%$ c.l.. When combined with future galaxy clustering data as expected
from surveys as DESI or EUCLID, CORE-M5  will provide a guaranteed discovery for a neutrino mass. Other cosmological information from CORE-M5,
as clusters number counts (see the ECO companion paper \cite{ECOclusters}) could further reduce these uncertainties.

\item CORE-M5 could also provide extremely stringent constraints on the neutrino effective number $N_{\rm eff}$ with  $\sigma (N_{\rm eff}) = {0.041}$.
This uncertainty, that can be further reduced by combining the CORE-M5 data with clusters number counts data from CORE-M5 itself and/or
complementary galaxy surveys, will test the presence of extra light particles at recombination and the process of neutrino decoupling from the primordial
plasma at redshift $z\sim 10^9$. The nature of the neutrino background can be further tested by measuring its self-interactions.

\item The primordial Helium abundance $Y_p$ can be measured by CORE-M5 with an uncertainty of $\sigma (Y_p) = {0.0029}$ that is almost a factor
two better than current constraints from direct measurements from metal-poor extragalactic H II regions.

\item CORE-M5 will also significantly improve current constraints on curvature (by almost a factor $4$) and on the dark energy equation of state (by a factor $\sim 3$).
One key improvement will be the determination of the Hubble constant in these models: the possibility of an equation of state $w<-1$ to explain
current tensions on the values of $H_0$ can be significantly tested by CORE-M5.

\item By measuring the intermediate angular scale CMB polarization with unprecedented accuracy, CORE-M5 will scrutinize with the highest possible detail the process 
of recombination.
This will let CORE-M5 place bounds on known physical process as the amplitude of the recombination two photons rate (improving current
constraints by a factor $5$) but also to further improve the constraints on extra ionizing photons from dark matter annihilation and on variations of the fine structure constant.

\item Large angular scale polarization will also be measured by CORE, providing new constraints on the reionization process. It is here worthwhile to note that the
ability of CORE-M5 to measure polarization over a wide range of angular scales will provide a crucial test for the cosmological scenario. The constraints
on the optical depth $\tau$ from large angular scales, for example, can be only validated by a measurement of small angular scale polarization with results 
 consistent with the overall $\Lambda$CDM scenario.

\end{itemize}

\begin{table}[h]
\begin{center}\footnotesize
\scalebox{0.83}{\begin{tabular}{|c||c|c|c|c|}
\hline
 Parameter                     & CORE-M5         & CORE-M5     & CORE-M5                        	& CORE-M5\\
                     & vs Litebird         & vs LiteCORE-80     & vs LiteCORE-120                        	& vs COrE+\\

\hline
 $\Lambda$CDM                     &          &      &                         	&\\
\hline
$ \Omega_b h^2          $ & $ 3.5$& $ 1.4            $& $ 1.1              $& $ 0.9             $\\
$ \Omega_c h^2          $ & $ 2.3$& $ 1.3            $& $ 1.2              $& $ 1.0             $\\
$ 100\theta_{MC}         $ & $ 5.8$& $ 1.3            $& $ 1.1              $& $ 0.9            $\\
$ \tau                            $ & $ 1.0$& $ 1.0            $& $ 1.0              $& $ 1.0            $\\
$ n_s                            $ & $ 2.6$& $ 1.1            $& $ 1.1              $& $ 1.0            $\\
$ ln(10^{10} A_s)          $ & $ 1.2$& $ 1.1            $& $ 1.0              $& $ 1.0            $\\
$ H_0$[km/s/Mpc]           & $ 3.0$& $ 1.3            $& $ 1.1              $& $ 0.9            $\\
$ \sigma_8                    $ & $ 2.5$& $ 1.3            $& $ 1.1              $& $ 0.9            $\\
\hline
Extensions                     &          &      &                         	&\\
\hline
$ \Omega_k          $ & $ 1.8$& $ 1.1            $& $ 1.0              $& $ 1.0             $\\
$N_{\rm eff}$           & $ 4.8$& $ 1.5            $& $ 1.1              $& $ 0.9             $\\
$M_{\nu}$              & $ 1.6$& $ 1.1            $& $ 1.0              $& $ 1.0             $\\
$Y_p$                     & $ 3.0$& $ 1.4            $& $ 1.0              $& $ 0.9            $\\
\hline
\end{tabular}}
\end{center}
\caption{Improvements from CORE-M5 on cosmological parameters with respect to several proposed configuration defined as the ratio
of the forecasted $1$ $\sigma$ constraints, $\sigma/\sigma_{CORE-M5}$.}
\label{tab:improvements}
\end{table}

It is also interesting to summarize the constraints from different experimental configurations and to compare them.
We do this in Table~\ref{tab:improvements} where we report the ratio of the $1$-$\sigma$ forecasted error of a certain experimental
configuration over the expected $1$ $\sigma$ error from the proposed CORE-M5 setup. For generality, we also compare the constraints
with those expected from the JAXA Litebird proposal (\cite{litebird}) that is now in conceptual design phase (called ISAS Phase-A1) .
Litebird presents a significantly different experimental design with respect to the CORE configurations studied in this paper, with, for example, a 
smaller primary mirror of $60$ cm. As we can see from the results in Table~\ref{tab:improvements} any CORE configuration is expected
to constrain cosmological parameters with an improvement that ranges from a factor $2$ to $5$ respect to Litebird.
CORE-M5, for example, will constrain the neutrino effective number with a precision about $5$ times better than Litebird. 
It is clear from the results presented in the Table that CORE will have the possibility to probe new physics that will not be accessible by Litebird alone.  
However, constraints on the reionization optical depth will be comparable, since the imprint of reionization is mainly on large scale
polarization that can be equally measured by Litebird and CORE. 
Also from Table~\ref{tab:improvements} we see that CORE-M5 could produce constraints that are up to $50 \%$ better than those
expected from the cheaper LiteCORE-80 configuration. A significantly higher precision is indeed expected on key parameters
as the baryon abundance, the Hubble constant, the neutrino effective number and the primordial Helium abundance. 
On the other hand, the differences between CORE-M5 and LiteCORE-120 and COrE+ are expected to be of the order
of $\sim 10 \%$. From one side we can then consider the forecasts presented here for CORE-M5 as conservative: if the experimental 
sensitivity will be for some reason degraded to LiteCORE-120 we expect no significant variations in the constraints presented in this paper. 
On the other hand, the more expensive COrE+ configuration would only slightly improve the main parameter constraints and would not
present a decisive improvement in the specific scientific aspect of parameters recovery and model testing. Indeed the scientific driver for higher angular resolution is not the improvement in parameters accuracy.

To conclude, we have presented in this paper a large number of forecasts on cosmological parameters for the CORE-M5 proposed
mission. The expected improved constraints, presented in Table~\ref{tab:key} clearly calls for of a next CMB satellite mission as CORE.
CORE-M5 can probe new physics with unprecedented precision. We have compared the constraints with different experimental configurations
and found that the expected constraints are stable under a degradation of the experimental configuration to LiteCORE-120 that has a 
significantly smaller number of detectors.  Assuming the $\Lambda$CDM cosmological scenario, we also found that the CORE-M5 setup  
can produce constraints that are almost identical (at worst a $\sim 10 \%$ degradation) to the ones achievable by the larger 
aperture COrE+ configuration.

\appendix

\end{document}